%% file: ms.tex
\documentclass[a4paper,11pt]{article}
\pdfoutput=1 

\usepackage{docstyle} 

\usepackage[T1]{fontenc} 

\usepackage{lmodern}
\usepackage{enumitem}
\usepackage{slashed}
\usepackage{revsymb}

\title{\textcolor{mycolor}{Light Scalars in Light of UV/IR Mixing\\ \textit{\large{Classicalization via Synergy between Vainshtein \& Chameleon Screenings}}}}

\author{Florian Nortier}

\affiliation{Université Lyon 1, CNRS/IN2P3, IP2I Lyon, Villeurbanne, France}

\emailAdd{f.nortier@ip2i.in2p3.fr}

\abstract{Effective field theories featuring light scalar fields play a pivotal role in addressing fundamental questions in (astro)particle physics and cosmology. However, such theories often confront hierarchy problems in the absence of a symmetry.
Self-completion via classicalization offers a non-Wilsonian approach to ultraviolet (UV) completion, wherein new scalar self-interactions involving derivatives give rise to Vainshtein-like screening around energy-momentum sources. Rather than introducing new UV degrees of freedom to preserve unitarity at high energies, these theories reshuffle their infrared (IR) degrees of freedom by generating extended semiclassical objects---referred to as classicalons---which decay into a multitude of soft particles. This mechanism incorporates nonlocalizable fields, thereby realizing a form of UV/IR mixing that is analogous to the dynamics of black holes in gravitational theories.
In this article, having reviewed the fundamental principles of classicalization with a simple k-essence model, we then argue the necessity of maintaining a little hierarchy between the scalar mass and the scale of the first new resonances, thereby illustrating the impact of UV/IR mixing on hierarchy problems. Additionally, we investigate the effects of a scalar potential and couplings to fermions on the Vainshtein screening mechanism. We discuss that a chameleon-like screening mechanism must accompany the Vainshtein screening to preserve the integrity of classicalon solutions.
}

\keywords{Effective Field Theories, Hierarchy Problem, Nonperturbative Effects}

\arxivnumber{2511.01739}

\begin{document}

\maketitle
\flushbottom

\section{Introduction}
\label{introduction}
Relatively light scalar fields are ubiquitous in theoretical (astro)particle physics and cosmology. They serve as essential components in numerous models, including those involving spontaneous symmetry breaking (SSB) \cite{Peskin:1995ev}, particle dark matter \cite{Cirelli:2024ssz}, dynamical dark energy \cite{Joyce:2014kja}, and inflation \cite{Martin:2013tda}, to cite only a few representative reviews from the literature.

From an effective field theory (EFT) perspective \cite{Burgess:2020tbq}, light scalars face hierarchy problems \cite{Hebecker:2020aqr}: their masses require protection from radiative corrections induced by heavier particles. The principle of 't Hooft naturalness \cite{tHooft:1979rat} traditionally suggests that a symmetry should underlie their lightness, an approach that remains the most widely adopted. However, this symmetry-based explanation has been called into question by the discovery of the Higgs boson at the CERN LHC and ````nothing else'''' \cite{Cho:2007cb, Giudice:2017pzm, Craig:2020ojv}, motivating the search for alternative mechanisms to account for the lightness of scalar fields \cite{Craig:2022eqo, McCullough:2024evr}. In recent years, promising new approaches have emerged, including cosmological selection mechanisms (see Ref.~\cite{DAgnolo:2022mem} for a review), and the so-called `accidents' \cite{Brummer:2023znr, Brummer:2024ejc}. In this article, we explore a more radical possibility inspired by quantum gravity considerations: ultraviolet (UV)/infrared (IR) mixing.

The hierarchy problems associated with weakly coupled scalars originate from the Wilsonian perspective of quantum field theories (QFTs) as EFTs \cite{Wilson:1970ag}, where the decoupling of scales is of paramount importance. It is crucial to recognize that the principles of decoupling and the axioms of textbook QFT \cite{Duncan:2012aja} are deeply intertwined. Nevertheless, it has been established that the locality axiom, typically enforced through microcausality \cite{Duncan:2012aja}, is overly restrictive when addressing nonperturbative phenomena in gravity, such as black holes (BHs) \cite{Aharony:1998tt,Giddings:2001pt,Giddings:2004ud,Giddings:2005id,Giddings:2006vu,Giddings:2006sj,Giddings:2006be,Giddings:2007ie,Giddings:2007pj,Giddings:2021qas,Buoninfante:2023dyd,Giddings:2024qcf}. This suggests that the realization of locality in the fundamental theory of nature may differ substantially from conventional wisdom. If a certain degree of nonlocality is permitted within QFT, one can envision correlations between short- and long-range physics---referred to as UV/IR mixing---which could explain the apparent violation of naturalness in the EFT from a naive Wilsonian perspective \cite{Cohen:1998zx,Dienes:2001se,Cheung:2014vva,Ibanez:2017kvh,Ibanez:2017oqr,Lust:2017wrl,Craig:2019fdy,Craig:2019zbn,Castellano:2021mmx,Craig:2022eqo,McCullough:2024evr,Cribiori:2025oek}.

\begin{figure}[t]
\begin{center}
\includegraphics[height=8.5cm]{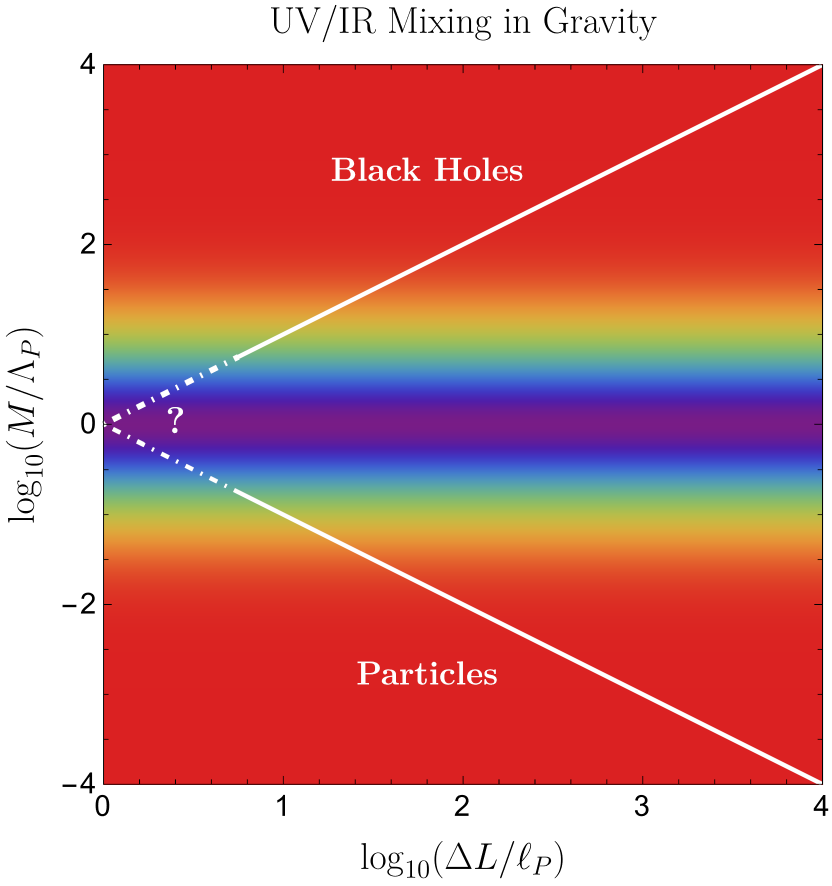}
\end{center}
\caption{Schematic representation of UV/IR mixing in gravity upon crossing the Planck scale, \(\Lambda_P\). A pointlike object with mass \(M \ll \Lambda_P\) is accurately described as a quantum particle, for which pair creation becomes significant at distances below the quantum wavelength \(\Delta L \sim 1/M\) (with gravitational effects remaining negligible). Conversely, for \(M \gg \Lambda_P\), the object is most appropriately characterized as a classical black hole, where gravitational effects dominate within the gravitational radius \(\Delta L \sim \ell_P^2 M\) (with quantum fluctuations being negligible). At the boundary where \(M \sim \Lambda_P\), the length scale satisfies \(\Delta L \sim \ell_P\), and both quantum fluctuations and gravitational effects become equally significant. The theoretical description of such a state remains an open question. The rainbow background in the figure highlights this UV/IR mixing, with the inversion of the relationship between mass and length scale when crossing $\Lambda_P$.
}
\label{rainbow}
\end{figure}

A concrete example of UV/IR mixing\footnote{In string theory, UV/IR mixing is realized through modular invariance, the significance of which has been recently revisited in Refs.~\cite{Abel:2021tyt,Abel:2023hkk,Abel:2024twz}.} in well-established theories emerges from the relationship between energy-momentum and spacetime resolution. For a quantum particle described by QFT, wave-particle duality dictates \(\lambdabar_C \equiv 1/M\), where \(M\) is the particle mass and \(\lambdabar_C\) its (reduced) Compton wavelength. In contrast, for a BH described by General Relativity (GR), its size is determined by the Schwarzschild radius \(R_S \sim \ell_P^2 M\), with \(\ell_P\) denoting the Planck length and \(M\) the BH mass. The boundary between the particle and BH regimes is set by the (reduced) Planck scale:
\begin{equation}
\Lambda_P \equiv 1/\ell_P \equiv \sqrt{\frac{1}{8\pi G_N}} \sim 10^{18} \ \text{GeV},
\end{equation}
at which nonperturbative effects of quantum gravity become significant. Consequently, strong-field gravity inverts the conventional relationship between energy-momentum and spacetime resolution, as depicted in Fig.~\ref{rainbow}.

This inversion leads to the hypothesis known as `asymptotic darkness' \cite{Aharony:1998tt,Shomer:2007vq,Dvali:2010bf,Dvali:2010ue}: an ultra-Planckian scattering process between 2 particles, with a center-of-mass energy \(\sqrt{s} \gg \Lambda_P\) and an impact parameter \(b \ll R_S \sim \ell_P^2 \sqrt{s}\), should result in the formation of a BH of mass \(M \sim \sqrt{s}\). This BH subsequently evaporates---via Hawking radiation \cite{Hawking:1974rv,Hawking:1975vcx}---into \(N_\circledast \sim R_S \sqrt{s} \gg 1\) soft particles, each with energy \(\omega \sim 1/R_S\). This scenario is supported by the \(S\)-matrix program in gravity \cite{Giddings:2009gj,Giddings:2011xs}, both in GR and string theory \cite{Amati:1987wq,Gross:1987kza,tHooft:1987vrq,Gross:1987ar,Amati:1987uf,Amati:1988tn,Mende:1989wt,Amati:1990xe,Amati:1993tb,Banks:1999gd,Eardley:2002re,Kohlprath:2002yh,Giddings:2004xy,Giddings:2007bw,Giddings:2007qq,Dvali:2014ila,Addazi:2016ksu}.
As a result, \(\ell_P\) emerges as the smallest length scale accessible in any experiment, rendering the introduction of new degrees of freedom with mass \(M_\text{new} \gg \Lambda_P\) meaningless, as they cannot be distinguished from a BH of equivalent mass \cite{Dvali:2010bf,Dvali:2010ue}. It follows that enforcing locality as a strict microcausality condition lacks physical meaning in a theory characterized by a minimal length scale \cite{Hossenfelder:2012jw}, where light cones inherently appear `fuzzy' \cite{Marshakov:2002ff,Boos:2020qgg}.

A corollary of asymptotic darkness is that pure gravity in GR, when treated as a QFT \cite{Donoghue:1994dn, Burgess:2003jk, Donoghue:2017pgk,Donoghue:2022eay}, may be self-complete in the ultra-Planckian regime \cite{Dvali:2010bf, Dvali:2010ue, Dvali:2011th}. The transition amplitude for \(2 \to 2\) ultra-Planckian hard scattering processes of gravitons---processes that violate tree-level unitarity---is suppressed by a factor of \(e^{-N_\circledast}\), since a BH-like object is produced and preferentially decays into \(N_\circledast \gg 1\) soft particles \cite{Dvali:2011th,Dvali:2014ila,Addazi:2016ksu}; unitarity is thus preserved nonperturbatively. To reconcile a quantum description of BHs with unitarity, a natural framework---proposed in Refs.~\cite{Dvali:2011aa,Dvali:2012rt,Dvali:2012gb,Dvali:2012uq,Dvali:2012wq,Dvali:2013vxa}---is to interpret them as coherent states composed of \(N_\circledast \gg 1\) gravitons\footnote{For comprehensive reviews and additional references, see Refs.~\cite{Casadio:2015lis, Giusti:2019wdx}. Subsequent developments are discussed in Refs.~\cite{Buoninfante:2019swn,Dvali:2020wft,Casadio:2021eio,Dvali:2021ooc,Dvali:2021tez,Dvali:2021ofp,Dvali:2023qlk,Raj:2023irr,Feng:2025nai,Dvali:2025gvd,Dvali:2025sog}. Although not widely adopted as a quantum description of BHs, this framework simply follows the textbook treatment of a semiclassical background for a bosonic field in QFT: a coherent state with a large occupation number \cite{Zhang:1999is,Duncan:2012aja}.}. This corpuscular description of gravity has also been extended to other gravitational backgrounds of cosmological significance, yielding important implications for the cosmological constant problem \cite{Binetruy:2012kx,Dvali:2013eja,Dvali:2014gua,Casadio:2015xva,Kuhnel:2015yka,Dvali:2017eba,Dvali:2018jhn,Dvali:2020etd,Giusti:2021shf,Berezhiani:2021zst,Berezhiani:2024boz}.


Should new states be necessary to preserve unitarity at energies \(\sqrt{s} \sim \Lambda_P\), they must consequently emerge at or below the scale\footnote{If extra dimensions of space exist, this is the higher-dimensional Planck scale.} \(\Lambda_P\), where gravitational interactions are expected to become strongly coupled. Around the Planck scale, new resonances---sometimes dubbed quantum BHs \cite{Meade:2007sz, Calmet:2008dg, Gingrich:2009hj, Calmet:2012fv}, Planckions \cite{Treder:1985kb,Dvali:2016ovn}, or BH precursors \cite{Calmet:2014gya, Calmet:2017omb}---should appear, the properties of which demand a UV-complete theory of quantum gravity for their description. One may then hypothesize that these new states could represent composite resonances of gravitons, potentially forming their own composite gravitational strings before transitioning into the ultra-Planckian BH regime \cite{Dvali:2010bf}. This scenario is analogous to the regulation of IR behavior in quantum chromodynamics (QCD), where the strong interaction between quarks and gluons generates hadronic strings \cite{Narison:2002woh}.  While this hypothesis is not logically inconsistent, its investigation requires nonperturbative techniques in quantum gravity, such as lattice approaches\footnote{The modern approach to lattice gravity is known as causal dynamical triangulations (CDT) \cite{Loll:2019rdj}.}.

\begin{figure}[t]
\begin{center}
\includegraphics[width=15.3cm]{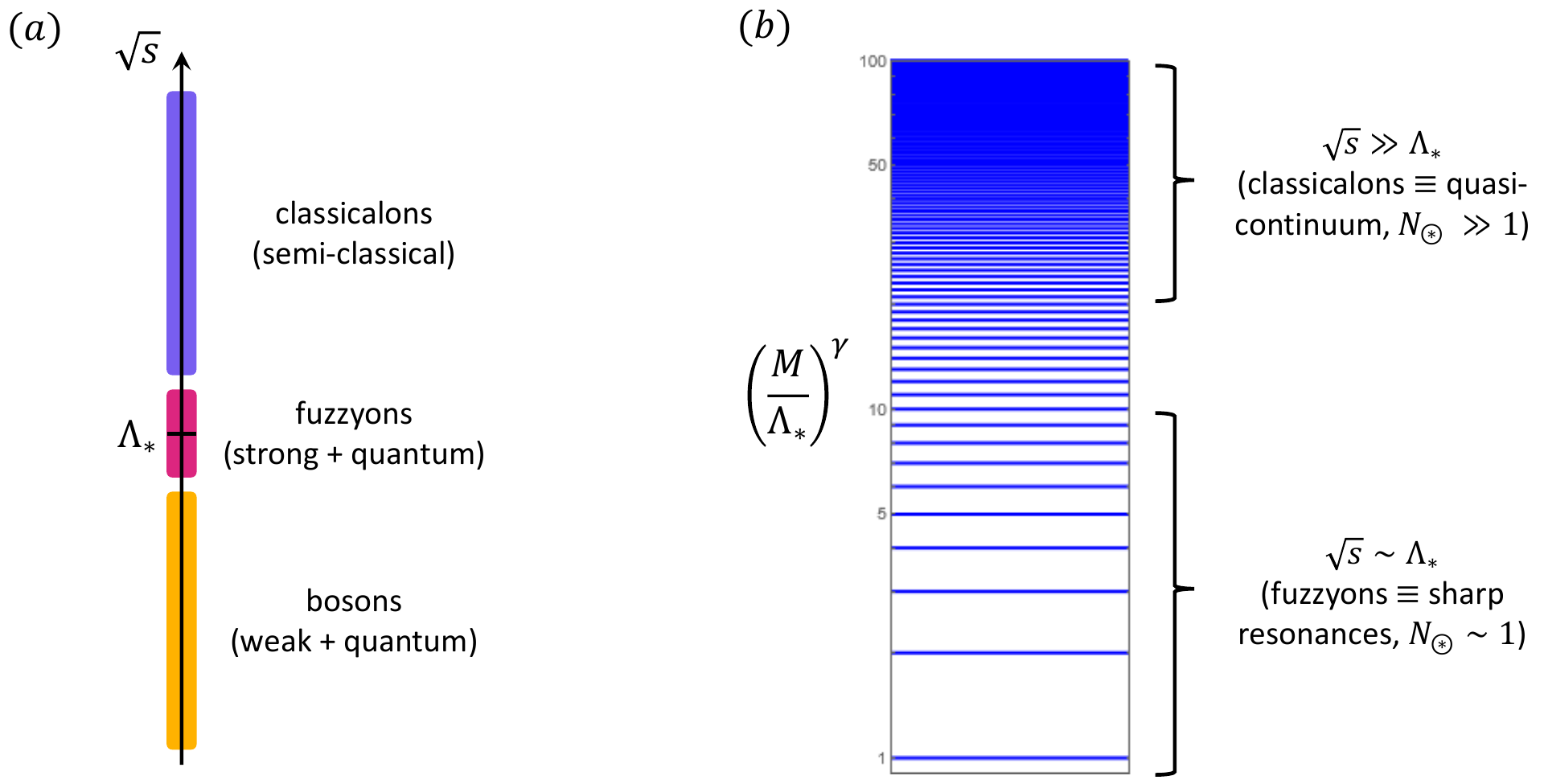}
\end{center}
\caption{Panel $(a)$: Classicalization is governed by an interaction scale \(\Lambda_\ast\).
For a scattering process at center-of-mass energy $\sqrt{s}$ and impact parameter $b \sim 1/\sqrt{s}$, 3 distinct regimes can be identified:
$(i)$ the production of \(\mathcal{O}(1)\) weakly interacting bosons when \(\sqrt{s} \ll \Lambda_\ast\);
$(ii)$ the production of a narrow, strongly-coupled resonance (a fuzzyon) decaying into \(N_\circledast \sim 1\) bosons when \(\sqrt{s} \sim \Lambda_\ast\);
$(iii)\) the production of a semiclassical state (a classicalon) decaying into \(N_\circledast \gg 1\) soft bosons when \(\sqrt{s} \gg \Lambda_\ast\).
Panel $(b)$: Schematic representation of the mass spectrum of composite states of $N_\circledast$ bosons in a theory exhibiting classicalization (logarithmic scale). As discussed in Ref.~\cite{Dvali:2010gv,Dvali:2011nh}, this spectrum is expected to be quantized as a function of the interaction scale \(\Lambda_\ast\) and a real parameter \(\gamma>1\), which depends on the operator responsible for triggering classicalization. In a collider experiment with a center-of-mass energy \(\sqrt{s} \sim \Lambda_\ast\), only quantum composite states (fuzzyons) within the strongly-coupled regime can be probed $(N_\circledast \sim 1)$. For \(\sqrt{s} \gg \Lambda_\ast\), however, the states (classicalons) form a quasi-continuum that is effectively described by a semiclassical approach $(N_\circledast \gg 1)$.}
\label{spectrum}
\end{figure}

Inspired by the foregoing considerations, the concept of `classicalization'---first introduced in Ref.~\cite{Dvali:2010jz} and subsequently developed in Refs.~\cite{Dvali:2010ns,Dvali:2011nj,Bajc:2011ey,Dvali:2011th,Dvali:2011nh,Brouzakis:2011zs,Grojean:2011bq,Rizos:2011wj,Dvali:2012zc,Dvali:2012mx,Rizos:2012qs,Alberte:2012is,Vikman:2012bx,Berkhahn:2013woa,Brouzakis:2014bwa,Asimakis:2014bza,Dvali:2015ywa,Dvali:2016ovn,Dvali:2018xoc,Dvali:2021ooc,Dvali:2022vzz}---posits that certain EFTs may not admit a conventional Wilsonian UV completion. Instead, they might achieve self-UV completion\footnote{The `self-healing' mechanism, as discussed in Ref.~\cite{Aydemir:2012nz}, reveals that the scale at which tree-level unitarity is violated does not necessarily align with the emergence of new physics. This mechanism preserves unitarity by incorporating bound states within the EFT framework, thereby obviating the need for additional degrees of freedom.} by unitarizing through the nonperturbative formation of extended semiclassical objects termed `classicalons' (analogous to BHs in gravity). This mechanism involves transitioning through a strongly-coupled intermediate regime of composite states---referred to as `fuzzyons\footnote{The term `fuzzyons' alludes to the minimal length scale \(\ell_\ast\) at which the classicalizer field makes new types of composite states.}' in Ref.~\cite{Chattopadhyay:2023nbj}---at a characteristic scale \(\Lambda_\ast \equiv 1/\ell_\ast\), and the quasi-continuum of classicalon states emerges in the deep-UV region of the spectrum \cite{Dvali:2011nh} (see Fig.~\ref{spectrum}). Thus, UV/IR mixing constitutes the `DNA' of classicalization. Classicalizing theories have been identified as nonlocalizable\footnote{Based on the construction in Ref.~\cite{Nortier:2023dkq}, where nonlocal form factors appear even at tree level (unlike in the present work), the model of Ref.~\cite{Chattopadhyay:2023nbj} has been argued to exhibit classicalization. For alternative constructions of this type with a classicalizing behavior, see also Refs.~\cite{Addazi:2015ppa,Buoninfante:2018gce,Buoninfante:2019swn,Addazi:2020nkm}.} QFTs, in which microcausality is superseded by macrocausality \cite{Keltner:2015qqb,Keltner:2015xda,Buoninfante:2023dyd}.

The remarkable phenomenon of classicalization is not exclusive to gravitons. For instance, in modified gravity, it is common to encounter additional forces mediated by new scalar fields with derivative self-couplings, which have significant implications for cosmology (see Ref.~\cite{Joyce:2014kja} for a comprehensive review). These models exhibit the Vainshtein\footnote{The literature on dark energy theories lacks a standardized nomenclature for screening mechanisms.
In this article, we adopt the terminology of Ref.~\cite{Brax:2012jr}, wherein mechanisms relying on nonlinearities in the field derivatives are referred to as Vainshtein-like screening, and those based on nonlinearities in the fields themselves are termed chameleon-like screening. For brevity, we omit the suffix ``like''.} screening mechanism \cite{Vainshtein:1972sx,Nicolis:2004qq} around astrophysical bodies, enabling them to remain consistent with the stringent constraints on new long-range forces within the solar system while still being active at cosmological scales.
Notable examples include k-essence models~\cite{Armendariz-Picon:1999hyi,Chiba:1999ka,Armendariz-Picon:2000nqq,Armendariz-Picon:2000ulo,Brax:2014wla} and Galileon theories~\cite{Luty:2003vm,Nicolis:2004qq,Nicolis:2008in,Deffayet:2009wt,Deffayet:2009mn}.
Vainshtein screening appears fundamentally incompatible with a Wilsonian UV completion \cite{Adams:2006sv,Kovner:2012yi,Kaloper:2014vqa,Burrage:2020bxp}, and classicalization remains the only known possibility for a UV completion\footnote{Unfortunately, the large length scales involved in dark energy models led the authors of Ref.~\cite{Padilla:2017wth} to guess that such scenarios should be excluded by cross-section measurements at hadron colliders. This criticism does not pertain to other applications featuring a sufficiently large classicalizing energy scale.} \cite{Dvali:2007kt,Dvali:2012zc,Keltner:2015qqb,Keltner:2015xda,Brax:2016jjt,Padilla:2017wth}. This outcome is not unexpected, as Vainshtein screening inherently exhibits UV/IR mixing \cite{Keltner:2015qqb,Keltner:2015xda}.

In this article, we examine the field-theoretic aspects of classicalization for gauge singlet scalars, independent of its specific applications in (astro)particle physics or cosmology.
In Section~\ref{K-essence_Model}, we review the interplay between Vainshtein screening and classicalization within a unified framework, adopting the perspective of self-UV completion rather than the conventional motivation of screening scalar forces in dark energy theories.
We also summarize the criticisms in the literature concerning classicalization, along with their counterarguments, and elucidate the fundamental distinction between the standard EFT interpretation of nonrenormalizable interactions and their interpretation within the framework of classicalization.
In Section~\ref{from_kessence_kchameleon}, we investigate the constraints imposed by introducing a potential that does not disrupt the screening mechanism---a consideration overlooked in previous studies.
In particular, we provide a detailed analysis of the claim in Ref.~\cite{Dvali:2010jz} that the mediator must be lighter than the classicalizing scale $\Lambda_\ast$, thereby suggesting the existence of light scalars with a necessary but little hierarchy.
Then, we explore, for the first time, the consequences of coupling the classicalizing scalar to other fields, with a focus on fermions. In particular, we demonstrate the necessity of incorporating a variant of the chameleon screening mechanism---well-known in dark energy theories \cite{Joyce:2014kja}---to preserve an active Vainshtein screening within the theory when a scalar potential and/or a direct coupling to matter are present.
In Section~\ref{conclusion}, we summarize our results and discuss potential avenues for future research. N.B.: Our conventions follow those adopted in the QFT textbook by Peskin and Schroeder \cite{Peskin:1995ev}.

\section{K-essence as Classicalizer}
\label{K-essence_Model}
For pedagogical purposes, we begin with a brief review of the classicalization proposal. We then proceed to examine a k-essence model involving a massless scalar field characterized by a simple kinetic self-interaction. This section establishes the foundational results upon which the original analysis in Section~\ref{from_kessence_kchameleon} is based.

\subsection{Classicalization in a Nutshell}
\label{nutshell}
The classicalization phenomenon \cite{Dvali:2010jz} relies on a bosonic `classicalizer' field that generalizes the role of gravity by generating extended classical field configurations, known as classicalons. For this mechanism to operate, the classicalizer must couple to an operator that becomes strongly interacting at short distances---such as in a scattering process---thereby inducing the formation of a classicalon. When derived exclusively from self-sourcing---that is, in the absence of external sources---classicalons are nontopological solitons~\cite{Lee:1991ax} belonging to the universal class of objects known as `saturons,' whose defining characteristic is the saturation of unitarity and entropy bounds \cite{Dvali:2019jjw,Dvali:2019ulr,Dvali:2020wqi,Dvali:2021rlf,Dvali:2021tez,Dvali:2023xfz,Dvali:2023qlk,Contri:2025eod}.

To illustrate this concept, consider a scalar classicalizer field \(\phi(x)\) of mass \(m\) defined on a \(3+1\)-dimensional Minkowski spacetime \(\mathbb{R}^{1,3}\). A prototypical classicalizing operator---referred to as `UV screener'---takes the form:
\begin{equation}
\frac{\phi}{\Lambda_\ast^{d-4}} \, \mathcal{O}_S^{(d-1)},
\label{class_scalar_op}
\end{equation}
where \(\mathcal{O}_S^{(d-1)}\) is a scalar composite operator\footnote{One could also consider an operator of the form:
\begin{equation}
\frac{\partial_\mu\phi}{\Lambda_\ast^{d-4}} \, \mathcal{O}_V^{\mu\, (d-2)},
\end{equation}
in which the classicalizer field couples to a vector operator \(\mathcal{O}_V^{\mu\, (d-2)}\) of dimension \(d-2\). However, this operator is equivalent to the scalar case in Eq.~\eqref{class_scalar_op}, since integration by parts yields
\begin{equation}
\int d^4x \ \frac{\partial_\mu\phi}{\Lambda_\ast^{d-4}} \, \mathcal{O}_V^{\mu\, (d-2)}
= \int d^4x \ \frac{\phi}{\Lambda_\ast^{d-4}} \, \mathcal{O}_S^{ (d-1)}, \quad \text{with} \quad \mathcal{O}_S^{ (d-1)} = -\partial_\mu\mathcal{O}_V^{\mu\, (d-2)},
\end{equation}
assuming the boundary term vanishes.} of dimension \(d-1 \geq 3\). This operator encodes a nonrenormalizable interaction involving \(\phi(x)\) and/or other fields, with the interaction scale governed by \(\Lambda_\ast \equiv 1/\ell_\ast\).

\begin{figure}[t]
\begin{center}
\includegraphics[height=6cm]{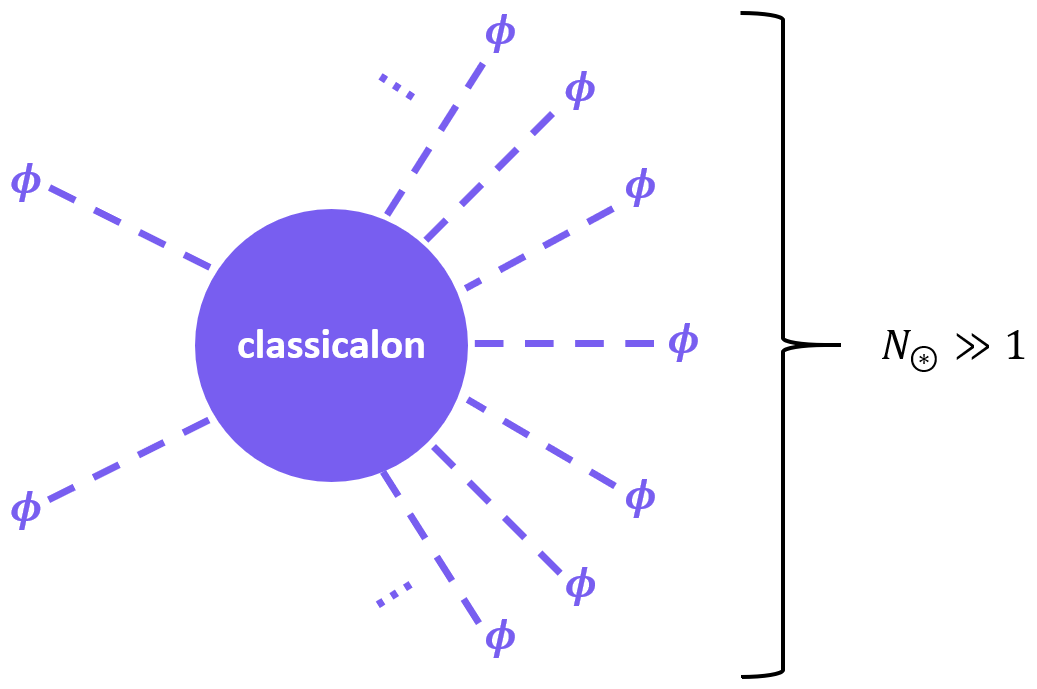}
\end{center}
\caption{Feynman diagram of a nonperturbative process, \(2 \to N_\circledast \gg 1\), mediated by the production and evaporation of a classicalon involving a k-essence field \(\phi\).
}
\label{Feyn1}
\end{figure}

From a conventional Wilsonian perspective, the Lagrangian density resembles that of an EFT, where the tree-level \(\phi\phi \to \phi\phi\) scattering amplitude \(\mathcal{M}(s,t)\)---expressed in terms of the Mandelstam variables \(s\) and \(t\)---violates unitarity in high-energy scattering processes when \(s \sim -t \gtrsim \Lambda_\ast^2\). Traditionally, this violation would necessitate the introduction of new degrees of freedom to restore theoretical consistency. Classicalization, however, offers a distinct, non-Wilsonian UV completion where unitarity is preserved nonperturbatively \cite{Dvali:2010jz,Dvali:2016ovn}. Specifically, the formation of a classicalon of mass $M_\circledast \sim \sqrt{s}$ and radius
\begin{equation}
R_\circledast = \ell_\ast \left( \frac{M_\circledast}{4\pi \Lambda_\ast} \right)^{\gamma-1}, \quad \gamma > 1,
\end{equation}
which grows with a model-dependent exponent $\gamma-1$ of the center-of-mass energy \(\sqrt{s} \gg \Lambda_\ast\), prevents the localization of \(\phi\) particles within a distance shorter than \(R_\circledast(s) \gg \ell_\ast\). This classicalon can be described as a coherent state \cite{Grojean:2011bq,Alberte:2012is,Berkhahn:2013woa,Dvali:2022vzz} of
\begin{equation}
N_\circledast = \frac{M_\circledast}{\omega} \gg 1
\end{equation}
weakly interacting \(\phi\) bosons confined within a region of size \(2R_\circledast\). The energy of each constituent boson is typically\footnote{A Bose-Einstein statistical model of the classicalon is developed in Ref.~\cite{Grojean:2011bq}, and its results align with the semiclassical coherent-state approach described in Ref.~\cite{Alberte:2012is}.}:
\begin{equation}
\omega = \sqrt{\left( \frac{\pi}{2 R_\circledast} \right)^2 + m^2} > m.
\label{varepsilon}
\end{equation}
Like BHs undergoing Hawking radiation, the classicalon evaporates thermally into \(N_\circledast \gg 1\) quanta (see Fig.~\ref{Feyn1}).

Starting from an \(\mathcal{O}(1)\) number of bosons in the initial state ($\sqrt{s} \gg \Lambda_\ast$), the amplitude for producing a specific \(N_\circledast\)-boson microstate of the classicalon \(\mathfrak{S}\) is exponentially suppressed as \(\sim e^{-N_\circledast}\). However, from the perspective of the \(S\)-matrix, individual microstates are indistinguishable. Consequently, one must perform an inclusive sum over the vast number of \(\sim e^{+N_\circledast}\) microstates of \(\mathfrak{S}\), which compensates for the exponential suppression \cite{Dvali:2010jz,Dvali:2011th,Grojean:2011bq,Alberte:2012is,Dvali:2020wqi,Dvali:2022vzz}. Unitarity is preserved in the \(2 \to 2\) process because it proceeds via an intermediate classicalon state, \(2 \to \mathfrak{S} \to 2\). The probability for any single microstate to decay back into 2 particles is suppressed as \(\sim e^{-N_\circledast}\).

The core idea of classicalization is to transform a \(2 \to 2\) hard scattering process---one that would otherwise violate tree-level unitarity---into a soft \(2 \to N_\circledast \gg 1\) process\footnote{The idea that a UV completion could manifest as the production of many soft particles has also been explored recently in other scenarios \cite{Hook:2023pba,Cheung:2024wme}, independent of UV/IR mixing and classicalization.}, thereby realizing UV/IR mixing. This can be interpreted as a collection of feeble elementary interactions among the $N_\circledast$ classicalon's constituents.
If these bosons are massive, their average individual energy \(\omega > m\) must be hierarchically smaller than the interaction scale \(\Lambda_\ast\); otherwise, the \(N_\circledast\) bosons would not reside in the feebly interacting regime of the EFT.
This requirement explains why UV/IR mixing necessitates a little hierarchy \(m \ll \Lambda_\ast\), as previously suggested in Ref.~\cite{Dvali:2010jz}.

\subsection{Classicalization by Vainshtein Screening}
\label{Class_Vainsh}
\subsubsection{Prototype Model}
Consider the following Lagrangian density \(\mathcal{L}_X\) for a real scalar field \(\phi(x)\) of mass dimension 1 (the classicalizer), defined on \(\mathbb{R}^{1,3}\):
\begin{equation}
\mathcal{L}_X \equiv \Lambda_\ast^4 \, \mathcal{K}(X),
\quad \text{where} \quad
X \equiv \dfrac{\partial^\mu \phi \partial_\mu \phi}{2 \Lambda_\ast^4}.
\label{S_X}
\end{equation}
This Lagrangian density depends solely on the kinetic variable \(X\). For simplicity, we adopt the following kinetic function:
\begin{equation}
\mathcal{K}(X) = X + c_2 X^2,
\quad \text{with} \quad
c_2 \equiv \pm 1,
\label{kin_funct}
\end{equation}
which yields the Lagrangian density,
\begin{equation}
\mathcal{L}_X = \dfrac{1}{2} \, \partial^\mu \phi \partial_\mu \phi + \dfrac{c_2}{4 \Lambda_\ast^4} \left( \partial^\mu \phi \partial_\mu \phi \right)^2,
\label{L_kessence}
\end{equation}
of a prototype model of k-essence. Here, the second term represents a nonrenormalizable kinetic self-interaction. The Lagrangian density \(\mathcal{L}_X\) exhibits both a shift symmetry \(\phi(x) \mapsto \phi(x) + \phi_c\) (where \(\phi_c \in \mathbb{R}\) is a constant) and a \(\mathbb{Z}_2\) symmetry \(\phi(x) \mapsto -\phi(x)\).

At first glance, this theory is formulated as an EFT, truncated at the leading self-interaction term permitted by these symmetries, with a naive perturbative cutoff \(\Lambda_\ast \equiv 1/\ell_\ast\). However, if this theory self-completes through classicalization, it implies that it is secretly UV complete, despite its initial appearance as an EFT. Our objective is to review the plausibility of this assertion.

To determine whether this theory can lead to classicalization\footnote{In the original article~\cite{Dvali:2010jz}, the authors also considered the classicalization of theories described by the action:
\begin{equation}
\int d^4x \ F^2(\phi) \, \partial^\mu \phi \partial_\mu \phi,
\end{equation}
where $F(\phi)$ is a function of $\phi$ satisfying $F^2(0) = 1/2$.
However, as discussed in Ref.~\cite{Cheung:2007st}, such a Lagrangian can be reformulated as the kinetic term of a free theory for a field $\widetilde{\phi}$ through the field redefinition $\widetilde{\phi}(\phi)$, defined by $d\widetilde{\phi}/d\phi \equiv F(\phi)$. Therefore, such a theory is secretly a noninteracting theory and does not exhibit classicalization.}, we must verify the following conditions: \((i)\) the existence of a classicalon solution in the presence of a sharply localized source; \((ii)\) the stability of this classicalon solution against quantum corrections; and \((iii)\) whether the nonlinearities arising from the kinetic self-coupling prevent the localization of \(\phi\) quanta within the classicalization radius.

The final ingredient for achieving $(ii)$ is thus the coupling of the field $\phi(x)$ to an external classical source $J(x)$, which is a spurion, namely a Lorentz scalar that transforms as $J(x) \mapsto -J(x)$ under the $\mathbb{Z}_2$ symmetry. The corresponding source term in the Lagrangian density takes the form:
\begin{equation}
\mathcal{L}_J = \frac{\phi}{\Lambda_\ast} \, J.
\label{source_term}
\end{equation}

\subsubsection{Classicalon Solution from Classical Vainshtein Screening}
\label{sec_background}
To determine whether the kinetic self-interaction in \(\mathcal{L}_X\) can induce classicalization, we must first verify the existence of a classicalon solution within a semiclassical framework. To enable an analytic derivation of this solution, we adopt the approach outlined in Ref.~\cite{Dvali:2010jz} and investigate the response of \(\phi(x)\) to an external, localized energy-momentum source \(J(x)\). In its rest frame, we define the source as
\begin{equation}
J = - \mathcal{E} \, \delta^{(3)}(\mathbf{r}),
\quad \text{and} \quad
\int d^3 \mathbf{r} \ \delta^{(3)}(\mathbf{r}) = 1
\end{equation}
which represents a pointlike source\footnote{In dark energy theories, the magnitude \(\mathcal{E}\) of the localized source is represented by the product of a scalar coupling and the mass of the astrophysical object (e.g., a star) around which Vainshtein screening takes place \cite{Joyce:2014kja}. Naturally, the pointlike approximation should be relaxed when required.} of magnitude \(\mathcal{E} > 0\). The key insight is that if a classicalon can be generated by a pointlike external source, it necessarily also arises from the kinetic self-interaction when attempting to sharply localize a wave packet of \(\phi\) bosons. Notably, when the source $J(x)$ is expressed in its rest frame and the sign of $\mathcal{E}$ is fixed, $\mathcal{L}_J$ breaks both the Lorentz-Poincaré and $\mathbb{Z}_2$ symmetries. The shift transformation, by contrast, merely produces an extra $\phi$-independent term in $\mathcal{L}_J$ that does not contribute to the Euler-Lagrange equation \eqref{ELE0}, leaving it invariant.

The first step in a semiclassical analysis is to derive the classical background solution for the field. The Euler-Lagrange equation for the Lagrangian density \(\mathcal{L}_X + \mathcal{L}_J\) is given by
\begin{equation}
\square \phi + \dfrac{c_2}{\Lambda_\ast^4} \, \partial^\mu \left( \partial^\nu \phi \partial_\nu \phi \partial_\mu \phi \right) = - \dfrac{\mathcal{E}}{\Lambda_\ast} \, \delta^{(3)}(\mathbf{r}),
\label{ELE0}
\end{equation}
which, for a static source, simplifies to
\begin{equation}
\overrightarrow{\nabla} \cdot \left[ \overrightarrow{\nabla} \phi - \dfrac{c_2}{\Lambda_\ast^4} \left( \overrightarrow{\nabla} \phi \right)^2 \overrightarrow{\nabla} \phi \right] = \dfrac{\mathcal{E}}{\Lambda_\ast} \, \delta^{(3)}(\mathbf{r}).
\end{equation}
Given the spherical symmetry of the source, we adopt spherical coordinates \(\mathbf{r} \equiv (r, \theta, \varphi)\). Applying the divergence theorem\footnote{Also referred to as Gauss's theorem or the Gauss-Ostrogradsky theorem.}, we integrate the above equation to obtain:
\begin{equation}
\phi' - \dfrac{c_2}{\Lambda_\ast^4} \, \phi^{\prime \, 3} = \frac{1}{\Omega r^2} \cdot \dfrac{\mathcal{E}}{\Lambda_\ast},
\end{equation}
where \(\phi' \equiv d\phi/dr\), and \(\Omega \equiv 4\pi\) denotes the solid angle. This cubic equation in \(\phi'\) admits a solution that, while not particularly illuminating in its full form \cite{Joyce:2014kja}, allows us to identify 3 distinct regimes: $(i)$ the \textit{linear regime}, characterized by the hierarchy \(\phi' \gg \ell_\ast^4 \, \phi^{\prime \, 3}\); $(ii)$ the \textit{nonlinear regime}, characterized by the opposite hierarchy \(\phi' \ll \ell_\ast^4 \, \phi^{\prime \, 3}\); and $(iii)$ the \textit{transition regime}, marking the onset of nonlinearities at the length scale:
\begin{equation}
R_V \equiv \ell_\ast \sqrt{ \frac{\mathcal{E}}{\Omega \Lambda_\ast} },
\label{V_rad}
\end{equation}
known as the Vainshtein radius.

\begin{figure}[t]
\begin{center}
\includegraphics[height=8.5cm]{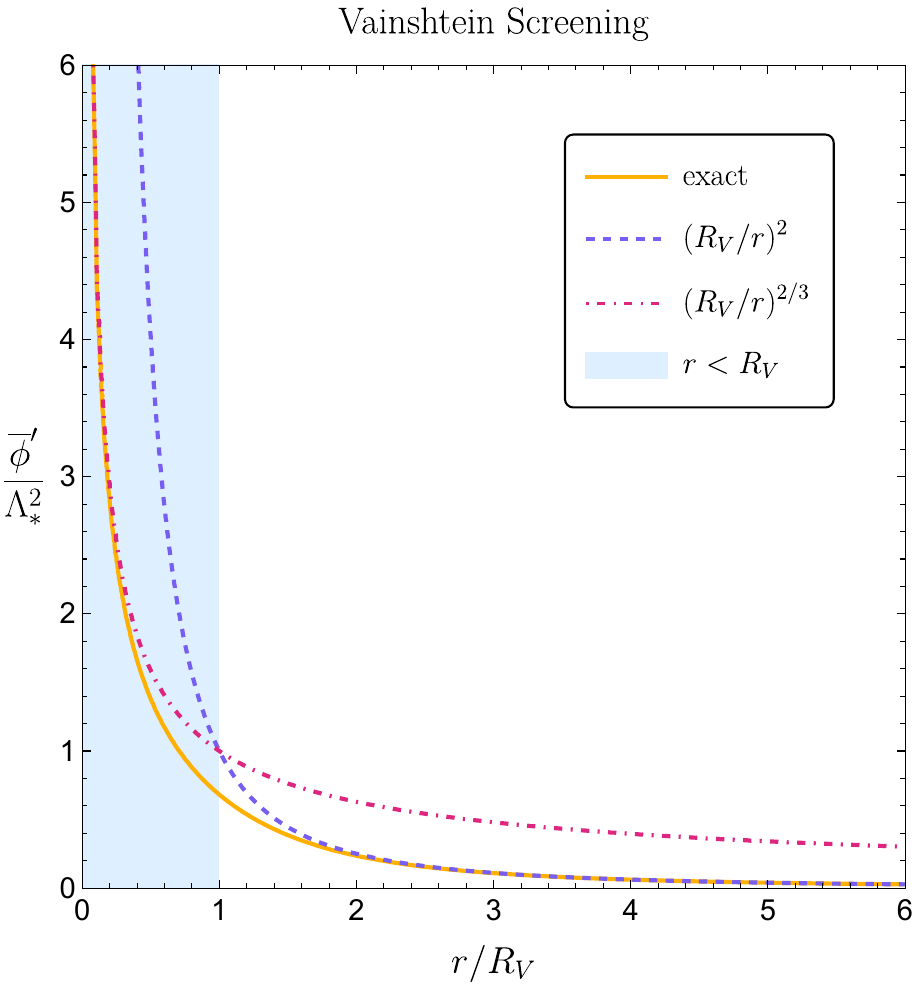}
\end{center}
\caption{Comparison of the exact background solution \(\overline{\phi}^\prime(r)\) with the asymptotic solutions in both the linear and nonlinear regimes, demarcated by the Vainshtein radius \(R_V\), at which \(\overline{\phi}^\prime(r) \sim \Lambda_\ast^2\), the classicalizing scale.}
\label{plot_Vainshtein_screening}
\end{figure}

To ensure a real solution that smoothly interpolates between the linear and nonlinear regimes, the background solution \(\overline{\phi}'(r)\) takes the form
\begin{equation}
\dfrac{\overline{\phi}'(r)}{\Lambda_\ast^2} \simeq
\left\{
\begin{array}{cllll}
\left( \dfrac{R_V}{r} \right)^2 & \ll 1 & \text{for} & r \gg R_V & \text{(linear regime)}, \\[0.5em]
\mathcal{O}(1) & & \text{for} & r \sim R_V & \text{(transition regime)}, \\[0.5em]
(-c_2)^{1/3} \left( \dfrac{R_V}{r} \right)^{2/3} & \gg 1 & \text{for} & r \ll R_V & \text{(nonlinear regime)},
\end{array}
\right.
\label{background_sol}
\end{equation}
which admits a real solution only when \(c_2 = -1\). Notably, for all regimes,
\begin{equation}
\left|\dfrac{\overline{\phi}(r)}{r \, \overline{\phi}^{\, \prime} (r)}\right| = \mathcal{O}(1).
\end{equation}
Here, the nonlinearities implement the classical version of the Vainshtein screening mechanism: the ratio of the background solutions \(\overline{\phi}^{\, \prime} (r)\), with and without the kinetic self-interaction term, scales as \((r/R_V)^{4/3} \ll 1\) at distances \(r \ll R_V\) from the source at \(r=0\). Consequently, when nonlinearities dominate, they suppress the scalar force in the region surrounding the source, see Fig.~\ref{plot_Vainshtein_screening}.  

The preceding discussion yields several key observations:
\begin{itemize}[label=$\spadesuit$]
\item The Vainshtein radius, defined in Eq.~\eqref{V_rad}, represents the transition scale between the linear and nonlinear regimes, growing with \(\mathcal{E}\). For \(\mathcal{E} \ll \Lambda_\ast\), the radius satisfies \(R_V \ll \ell_\ast\), and nonlinearities remain negligible within the valid regime of the EFT. In contrast, for \(\mathcal{E} \gg \Lambda_\ast\), the radius becomes \(R_V \gg \ell_\ast\), and the source is enclosed by a spherical region of radius \(R_V\) where nonlinearities dominate (the `Vainshtein core'). A notable puzzle in this k-essence model is that, although \(R_V \gg \ell_\ast\) lies within the EFT's valid length scale, the condition \(\phi'(r \ll R_V) \gg \Lambda_\ast\) appears to exceed the regime of validity in terms of the background amplitude \(\overline{\phi}^{\, \prime}(r)\) \cite{Keltner:2015xda}. Assessing the radiative stability of this background is therefore of paramount importance to validate the reliability of this semiclassical analysis.

\item The requirement \(c_2 = -1\) violates positivity bounds \cite{Pham:1985cr,Adams:2006sv,Dvali:2012zc,Bellazzini:2020cot,Arkani-Hamed:2020blm}---a point to which we will return in Section~\ref{positivity_bounds}---whereas \(c_2 = +1\) guarantees the existence of a Wilsonian UV completion. This suggests that Vainshtein screening occurs only when the UV completion is non-Wilsonian.
\end{itemize}
Taken together, these characteristics serve as signatures of the UV/IR mixing induced by the response of the derivative self-sourcing term to a localized source with \(\mathcal{E} \gg \Lambda_\ast\).

\subsubsection{Radiative Stability from Quantum Vainshtein Screening}
\label{radiative_stability}
The second stage of the semiclassical analysis involves examining the propagation and interactions of the quantum fluctuations \(\delta \phi(x)\) on top of the background \(\overline{\phi}(x)\) in the regime where \(\mathcal{E} \gg \Lambda_\ast\). It is essential to verify that these quantum fluctuations do not spoil the semiclassical approximation, analogous to the treatment of classical potentials in inflationary models \cite{Cheung:2007st,Weinberg:2008hq,Burgess:2009ea}. We will discuss that quantum corrections are naturally suppressed by the quantum version of Vainshtein screening \cite{Nicolis:2004qq,Nicolis:2008in,Vikman:2012bx,Brouzakis:2014bwa,deRham:2014wfa,Asimakis:2014bza,Brax:2016jjt,Padilla:2017wth}.

For convenience, we decompose the field as \(\phi(x) = \overline{\phi}(x) + \delta \phi(x)\), and similarly decompose the kinetic variable \(X = \overline{X} + \delta X\), where
\begin{equation}
\overline{X} \equiv \frac{\partial^\mu \overline{\phi} \partial_\mu \overline{\phi}}{2\Lambda_\ast^4},
\quad \text{and} \quad
\delta X \equiv \frac{\partial^\mu \delta \phi \, \partial_\mu \delta \phi + 2 \partial^\mu \overline{\phi} \partial_\mu \delta \phi}{2\Lambda_\ast^4}.
\end{equation}
The Lagrangian density \(\mathcal{L}_X\) in Eq.~\eqref{S_X} can then be expressed in terms of the fluctuations \(\delta \phi(x)\). Expanding the kinetic function in a Taylor series yields
\begin{equation}
\mathcal{K}(\overline{X} + \delta X) \simeq \mathcal{K}(\overline{X}) + \mathcal{K}^{(1)}(\overline{X}) \, \delta X + \frac{\mathcal{K}^{(2)}(\overline{X})}{2} \, \delta X^2 + \mathcal{O} \left( \delta X^3 \right),
\end{equation}
where \(\mathcal{K}^{(n)}(X) \equiv d^n \mathcal{K}(X) / dX^n\). Given our specific choice of \(\mathcal{K}(X)\) in Eq.~\eqref{kin_funct}, this second-order expansion is exact, since
\begin{equation}
\mathcal{K}^{(1)}(X) = 1 + 2 c_2 X, \quad
\mathcal{K}^{(2)}(X) = 2 c_2, \quad
\text{and} \quad
\forall n > 2, \, \mathcal{K}^{(n)}(X) = 0.
\end{equation}

\subparagraph{Kinetic Term:}
From the original Lagrangian density \(\mathcal{L}_X\) in Eq.~\eqref{S_X}, we derive the kinetic term for the massless fluctuations \(\delta \phi(x)\), which takes the form:
\begin{equation}
\mathcal{L}_\mathrm{kin} = \frac{\mathcal{Z}_\phi^{\mu\nu}}{2} \, \partial_\mu \delta \phi \partial_\nu \delta \phi,
\quad \text{where} \quad
\mathcal{Z}_\phi^{\mu\nu} \equiv \mathcal{K}^{(1)}(\overline{X}) \, \eta^{\mu\nu} + \frac{\mathcal{K}^{(2)}(\overline{X})}{\Lambda_\ast^4} \, \partial^\mu \overline{\phi} \partial^\nu \overline{\phi},
\end{equation}
with \(\eta_{\mu\nu}\) denoting the Minkowski metric. Owing to the spherical symmetry of the background \([\overline{\phi}(x) = \overline{\phi}(r)]\), the kinetic term can be recast as
\begin{equation}
\mathcal{L}_\mathrm{kin} = \frac{Z_\phi(x)}{2} \left[ (\partial_t \delta \phi)^2 - (\partial_\Omega \delta \phi)^2 - B(x) \cdot (\partial_r \delta \phi)^2 \right],
\end{equation}
where the angular derivative term is given by
\begin{equation}
(\partial_\Omega \delta \phi)^2 \equiv \left( \frac{\partial_\theta \delta \phi}{r} \right)^2 + \left( \frac{\partial_\varphi \delta \phi}{r \sin \theta} \right)^2,
\end{equation}
and the introduced nonvanishing functions are defined as
\begin{equation}
Z_\phi(x) \equiv \mathcal{K}^{(1)}(\overline{X}), \quad
B(x) \equiv \dfrac{\mathcal{K}^{(1)}(\overline{X}) + 2 \overline{X} \, \mathcal{K}^{(2)}(\overline{X})}{\mathcal{K}^{(1)}(\overline{X})},
\end{equation}
with \(Z_\phi > 0\) required to avoid ghost instabilities \cite{Joyce:2014kja}. It is important to note that the kinetic term is generally manifestly anisotropic. This anisotropy is acceptable, as the presence of the external source $J(x)$ in its rest frame inherently breaks the manifest Lorentz invariance of the original Lagrangian density. Below, we discuss the distinct regimes of the background solution in Eq.~\eqref{background_sol}, based on our choice of kinetic function in Eq.~\eqref{kin_funct}.

In the Vainshtein core (\(r \ll R_V\)), we find \(B(r) \to 3\), and the ghost-free condition imposes
\begin{equation}
Z_\phi(r) \underset{|\overline{\phi}^\prime| \gg \Lambda_\ast^2}{\simeq} -c_2 \left[ \frac{\overline{\phi}^{\, \prime}(r)}{\Lambda_\ast^2} \right]^2 > 0
\quad \Longrightarrow \quad c_2 = -1,
\label{Z_phi}
\end{equation}
thus recovering the same condition on \(c_2\) required for the existence of \(\overline{\phi}(r \ll R_V)\). Failure to satisfy this condition would indicate an internal inconsistency within this k-essence model. Zooming in on a specific point \(\mathbf{r_0} \equiv (r_0, \theta_0, \varphi_0)\) within the Vainshtein core, with a resolution of \(\Delta \mathbf{r} \equiv (\Delta r, \Delta \theta, \Delta \varphi)\), we obtain
\begin{equation}
\sqrt{\frac{1}{Z_\phi(r_0+\Delta r)}}
\underset{\Delta r \, \ll \, r_0}{\simeq}
\sqrt{\frac{1}{Z_\phi(r_0)}}
+ \Delta r \cdot \left. \frac{d}{dr} \left[ \sqrt{\frac{1}{Z_\phi(r)}} \right] \right|_{r=r_0}
+ \mathcal{O} \left( \frac{\Delta r^2}{r_0^2} \right),
\end{equation}
where
\begin{equation}
\Delta r \cdot  \left. \frac{d}{dr} \left[ \sqrt{\frac{1}{Z_\phi(r)}} \right] \right|_{r=r_0}
\simeq \left( \frac{r_0}{R_V} \right)^\frac{2}{3} \frac{\Delta r}{r_0} \ll 1,
\end{equation}
such that only the leading-order term is retained. Consequently, we can perform the field-strength renormalization of the fluctuation field to absorb the isotropic factor \(Z_\phi(r_0)\) in the kinetic term, leading to a Taylor expansion that yields
\begin{equation}
\mathcal{L}_\mathrm{kin} \simeq \frac{1}{2} \left[ (\partial_t \delta \phi_Z)^2 - (\partial_\Omega \delta \phi_Z)^2 - 3 (\partial_r \delta \phi_Z)^2 \right],
\quad \text{with} \quad
\delta \phi_Z(x) \equiv \sqrt{Z_\phi(r_0)} \, \delta\phi(x).
\label{L_2}
\end{equation}
This field-strength renormalization is essential for determining the regime of validity of the EFT for the fluctuations, as we will discuss when analyzing the interaction terms.

In the linear (\(r \gg R_V\)) and transition (\(r \sim R_V\)) regimes, we have \(Z_\phi(r) \sim B(r) \sim 1\), and the ghost-free condition remains consistent with \(c_2 = -1\). While it is possible to introduce the renormalized field \(\delta \phi_Z(x) \equiv \sqrt{Z_\phi} \, \delta\phi(x)\), this field-strength renormalization does not significantly modify the interaction scale, as we will demonstrate.

\subparagraph{Interaction Terms:}
The cubic and quartic interaction terms for $\delta \phi(x)$ are given by
\begin{align}
\mathcal{L}_\mathrm{int} = -\frac{\partial^\nu \overline{\phi} \partial_\nu \delta\phi \, \partial^\mu \delta \phi \, \partial_\mu \delta \phi}{\Lambda_\ast^4} -\frac{\left( \partial^\mu \delta \phi \, \partial_\mu \delta\phi \right)^2}{4 \Lambda_\ast^4},
\label{L_3}
\end{align}
respectively. It is important to emphasize that the shift symmetry remains preserved by the Lagrangian density of the fluctuations $(\delta \phi \mapsto \delta \phi + \phi_c)$. We now focus on the distinct regimes of the background solution, as outlined in Eq.~\eqref{background_sol}, using our chosen kinetic function in Eq.~\eqref{kin_funct}.

In the nonlinear regime (\(r \ll R_V\)), the interaction terms exhibit the following asymptotic forms:
\begin{align}
\mathcal{L}_\mathrm{int} \simeq \dfrac{\partial_r \delta \phi_Z \, \partial^\mu \delta \phi_Z \, \partial_\mu \delta \phi_Z}{\overline{\Lambda_\ast}^2} -\dfrac{\left( \partial^\mu \delta \phi_Z \, \partial_\mu \delta \phi_Z \right)^2}{4 \overline{\Lambda_\ast}^4},
\label{L_4}
\end{align}
where, after field-strength renormalization, the effective interaction scale at which the coupling becomes of $\mathcal{O}(1)$ is
\begin{equation}
 \overline{\Lambda_\ast}(r_0) \equiv \sqrt{Z_\phi(r_0)} \, \Lambda_\ast,
\end{equation}
which depends explicitly on \(r_0\). Based on the background solution in the Vainshtein core, as provided in Eq.~\eqref{background_sol}, and the expression for the renormalization factor \(Z_\phi(r_0)\) in Eq.~\eqref{Z_phi}, we can draw the following conclusions:

\begin{itemize}[label=$\spadesuit$]
\item Deep within the Vainshtein core, \( \overline{\Lambda_\ast}(r_0) \gg \Lambda_\ast\). This indicates that the perturbative cutoff \( \overline{\Lambda_\ast}\) is blueshifted relative to the original (naive) cutoff \(\Lambda_\ast\) in \(\mathcal{L}_X\). Consequently, the regime of validity of the EFT is `redressed' by Vainshtein screening: the interaction scale for fluctuations about the classicalon background becomes \(\overline{\Lambda_\ast}(r_0) \gg \Lambda_\ast\), rather than \(\Lambda_\ast\). The stronger the blueshift effect, the more suppressed the self-interactions in Eq.~\eqref{L_4}. This suppression represents another manifestation of Vainshtein screening: self-interactions among quantum fluctuations are weakened. As a corollary, loop corrections are also suppressed, as demonstrated for k-essences in Refs.~\cite{Vikman:2012bx,deRham:2014wfa,Asimakis:2014bza,Brax:2016jjt,Padilla:2017wth} and for Galileons in Refs.~\cite{Nicolis:2004qq,Nicolis:2008in,Brouzakis:2014bwa}. In essence, quantum effects do not disrupt the classical background in the Vainshtein core due to the screening mechanism.

\item For a fixed \(\mathbf{r_0}\), the factor \(Z_\phi(r_0)\) increases with \(\mathcal{E}\), enhancing the blueshift effect for larger values of \(\mathcal{E}\). It is worth noting that the volume of the Vainshtein core also scales as \(R_V^3\).

\item For a given value of \(\mathcal{E}\), the factor \(Z_\phi(r_0)\) increases as \(r_0 \to 0\), strengthening the blueshift effect as one approaches the source. Notably, \(\overline{\phi}^{\, \prime}(r)\) diverges for \(r_0 \to 0\), reflecting the pointlike nature of the external source $J(x)$. Moreover, we need not assume the validity of our semiclassical treatment---or, consequently, of the background solution in Eq.~\eqref{background_sol}---at radii $r \lesssim r_\star \equiv 1/\mathcal{E}$, the length scale associated with the source of magnitude $\mathcal{E} \gg \Lambda_\ast$.
\end{itemize}

Applying the same methodology, it is evident that in the linear (\(r \gg R_V\)) and transition (\(r \sim R_V\)) regimes, the interaction scale for fluctuations about the classicalon background remains \(\Lambda_\ast\), since the fluctuation fields undergo negligible field-strength renormalization (\(Z_\phi \sim 1\)). Fortunately, quantum fluctuations are expected to remain under EFT control in this case, which is why the blueshift effect is critical only within the Vainshtein core.

In summary, Vainshtein screening corresponds to the dominance of the kinetic self-interaction over the kinetic term of the k-essence field. The Vainshtein core thus defines a spatial region in which \(\phi\) bosons cannot propagate as weakly interacting particles, as nonlinear interactions prevail \cite{Vainshtein:1972sx,Padilla:2017wth}. From the perspective of background fluctuations $\delta \phi(x)$, their kinetic term acquires a large renormalization due to the background for \(r \ll R_V\); the Vainshtein mechanism is therefore a form of screening by inertia \cite{Joyce:2014kja}. After field strength renormalization, this corresponds to fluctuations $\delta \phi_Z(x)$ that behave effectively as quasi-free particles in the limit \(r_0 \to 0\) via the blueshift of the interaction scale $\Lambda_\ast$.

\subsubsection{Classicalization from Kinetic Self-interactions}
\label{class_kin_self}
The foregoing analysis demonstrates that classicalon solutions arise in response to a sharply localized external energy-momentum source. This ensures that any attempt to localize quanta coupled to \(\phi(x)\) within \(R_V\) will generate a Vainshtein core surrounding the source, thereby preventing the probing of shorter length scales through the emission of a hard k-essence boson.

Consider a scattering process in the toy model of k-essence bosons; in this case, there is no external source: the localized energy-momentum source is provided by the momentum exchange of the scattered \(\phi\) bosons themselves, at center-of-mass energy $\sqrt{s} \gg \Lambda_\ast$; the kinetic self-interaction then acts as a self-sourcing term, preventing the probing of length scales below the classicalization radius \(R_\circledast\). We now examine how the preceding discussion is modified.

\paragraph{Self-sourcing:}
As discussed in Ref.~\cite{Dvali:2010jz}, the radius \(R_\circledast\) can be obtained by examining the static scenario in which a field configuration \(\phi_0(r)\), characterized by a mass energy \(M_\circledast \sim \sqrt{s}\), is confined within a sphere of radius \(R_\circledast\), the latter representing the typical scale of variation for \(\phi_0(r)\). A dimensional analysis yields
\begin{equation}
M_\circledast \sim \int_{r < R_\circledast} d^3 \mathbf{r} \left( \partial \phi_0 \right)^2 \sim \left. \Omega r \phi_0^2 \right|_{r=R_\circledast}.
\end{equation}
The localized field configuration $\phi_0(r)$ of size \(R_\circledast\) then serves as a source for the field \(\phi(x)\) itself, with an integrated value parametrically given by
\begin{equation}
\left| \frac{\mathcal{E}_0}{\Lambda_\ast} \right| \sim \left| \int_{r < R_\circledast} d^3 \mathbf{r} \ \frac{\delta}{\delta \phi_0} \left[ \frac{\left( \partial \phi_0 \right)^4}{\Lambda_\ast^4} \right] \right|
\sim \left. \left| \frac{\Omega}{r} \cdot \frac{\phi_0^3}{\Lambda_\ast^4} \right| \right|_{r=R_\circledast}.
\end{equation}
An analysis analogous to that presented in Section~\ref{sec_background}, in which the pointlike source is replaced by this extended source, yields
\begin{equation}
\forall \, r \gg R_\circledast \, , \quad \phi(r) \sim - \frac{1}{\Omega r} \cdot \frac{\mathcal{E}_0}{\Lambda_\ast},
\end{equation}
which exhibits a Coulomb-like tail outside the source. This leads to the classicalization radius $R_\circledast$ for the k-essence field $\phi(x)$, where the kinetic self-interaction begins to dominate the kinetic term:
\begin{equation}
R_\circledast \equiv \ell_\ast \left( \frac{M_\circledast}{\Omega \Lambda_\ast} \right)^\frac{1}{3} \gg \ell_\ast,
\quad \text{with} \quad M_\circledast \gg \Lambda_\ast.
\label{C_rad_1}
\end{equation}
This result has been explicitly confirmed through the study of spherical wavepacket collapse, both analytically \cite{Dvali:2010ns,Dvali:2011th} and numerically \cite{Brouzakis:2011zs,Rizos:2011wj}. These studies provide evidence that the wavepacket propagates freely until reaching the distance \(R_\circledast\), at which point scattering effects prevent the resolution of smaller length scales.

\begin{figure}[t]
\begin{center}
\includegraphics[height=8.5cm]{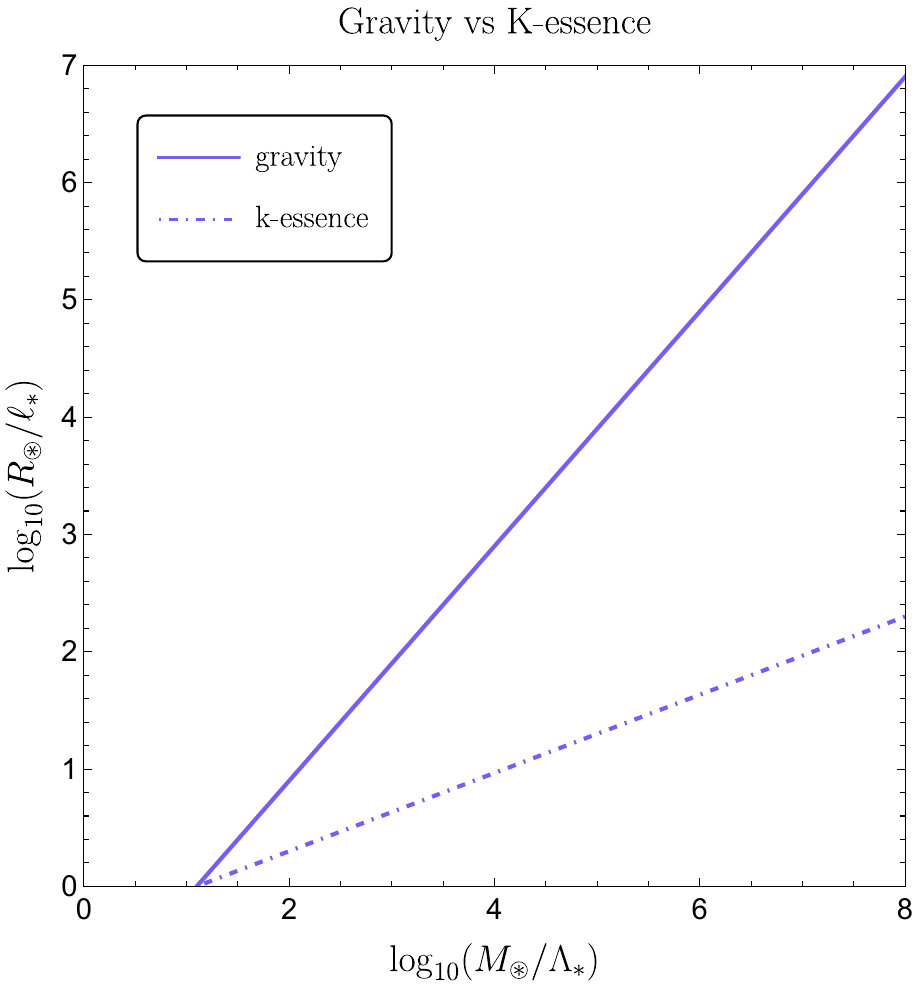}
\end{center}
\caption{A comparison between gravity and massless k-essence with respect to the growth rate of the classicalon radius $R_\circledast$ as a function of its mass $M_\circledast$. To ensure a meaningful comparison, the classicalization length $\ell_\ast \equiv 1/\Lambda_\ast$ is fixed to the Planck length $\ell_P$ in both scenarios. The plot indicates that $R_\circledast$ increases more rapidly with increasing $M_\circledast$ in gravity than in k-essence.}
\label{Gravity_vs_Kessence}
\end{figure}

From this discussion, it becomes clear how k-essence achieves self-UV completion through classicalization during a scattering process with \(\sqrt{s} \gg \Lambda_\ast\) and $b \ll R_\circledast$: a classicalon of radius \(R_\circledast\) and mass \(M_\circledast \sim \sqrt{s}\) is formed. The greater the energy injected by the scattering process, the larger the classicalon and the stronger the blueshift effect within the Vainshtein core for the UV modes. The behavior of $R_\circledast$ is analogous to that of the Schwarzschild radius for a BH of mass $M_\circledast \sim \sqrt{s}$,
\begin{equation}
R_S \equiv \ell_P \left( \dfrac{M_\circledast}{\Omega \Lambda_P} \right) \gg \ell_P,
\quad \text{where} \quad M_\circledast \gg \Lambda_P,
\label{S_rad}
\end{equation}
beyond which gravitational nonlinearities become significant. In Fig.~\ref{Gravity_vs_Kessence}, we compare the efficiency of the classicalization phenomenon between gravity and k-essence for \(\ell_\ast = \ell_P\). As discussed in Ref.~\cite{Dvali:2011th}, gravity represents the most efficient classicalizer, as the Schwarzschild radius in Eq.~\eqref{S_rad} grows more rapidly than \(R_\circledast\) with increasing \(M_\circledast\).

\paragraph{Dimensional Analysis:}
It is illuminating to demonstrate that the parametric form of \(R_\circledast\) can be derived through dimensional analysis alone \cite{Dvali:2010jz}. To this end, we adopt a system of units \cite{Giudice:2016yja} that restores the \(\hbar\)-dependence while retaining \(c = 1\), thereby distinguishing units of energy \((E)\) and length \((L)\). Our focus is on an inclusive scattering process involving two \(\phi\) bosons with a center-of-mass energy \(\sqrt{s}\) and an impact parameter \(b \sim 1/\sqrt{s}\). The action for the field \(\phi(x)\) is expressed as
\begin{equation}
S_X = \int d^4x \ \mathcal{L}_X, \quad
\mathcal{L}_X = \frac{1}{2} \, \partial^\mu \phi \partial_\mu \phi + \frac{G_\phi}{4} \left( \partial^\mu \phi \partial_\mu \phi \right)^2.
\end{equation}
The dimensionalities of the relevant quantities are as follows:
\begin{align}
&[\partial] = L^{-1}, \quad
[d^4x] = L^4, \quad
[S_X] = [\hbar] = EL, \quad
[\mathcal{L}_X] = EL^{-3}, \nonumber \\
&[\phi] = E^{\frac{1}{2}} L^{-\frac{1}{2}}, \quad
[G_\phi] = E^{-1} L^3, \quad
[\sqrt{s}] = E,
\end{align}
where the dimensions in natural units \((\hbar = 1)\) are recovered by setting \(E = L^{-1}\).

Relationships among physical quantities should depend solely on the parameters \(\hbar\), \(G_\phi\), and \(\sqrt{s}\). From these, one can construct: \((i)\) a unique quantum length associated exclusively with the system's kinematics,
\begin{equation}
\lambdabar_s \equiv \frac{\hbar}{\sqrt{s}},
\end{equation}
which corresponds to the de Broglie wavelength of the source (i.e., the incoming particles); \((ii)\) a unique quantum length associated solely with the kinetic self-coupling,
\begin{equation}
\ell_\ast \equiv \left| \hbar G_\phi \right|^{\frac{1}{4}},
\end{equation}
which defines the length scale at which this interaction becomes strongly coupled; and \((iii)\) a unique classical length,
\begin{equation}
r_c \equiv \left| \sqrt{s} G_\phi \right|^{\frac{1}{3}},
\end{equation}
which sets the length scale defining the scattering cross-section \(\sigma \sim r_c^2\) \cite{Dvali:2010ns}. In the classical limit \(\hbar \to 0\), only \(r_c\) remains finite, while \(\lambdabar_s, \ell_\ast \to 0\), as expected. In natural units \((\hbar = 1)\), the correspondence with our previous notations is obtained by identifying \(G_\phi = c_2/\Lambda_\ast^4\) and \(r_c = \Omega^{1/3} R_\circledast\). Thus, dimensional analysis enables us to accurately determine the parametric dependence of the system's relevant length scales, particularly the classicalization radius\footnote{This analysis holds for \(\sqrt{s} \gg \Lambda_\ast\) only under the assumption that classicalization indeed occurs---a condition that dimensional analysis alone cannot confirm.}.

\paragraph{Quantum Criticality:}
As a semiclassical state, a classicalon can be characterized as a coherent state---a quantum state with a large occupation number $N_\circledast \gg 1$ of k-essence bosons.
By definition,
\begin{equation}
N_\circledast \equiv \frac{M_\circledast}{\omega}
\sim \left( \frac{M_\circledast}{\Lambda_\ast} \right)^\frac{4}{3},
\end{equation}
where $\omega$ is the average energy of a constituent boson, see Eqs.~\eqref{varepsilon} and \eqref{C_rad_1}.
These soft bosons interact through a quantum coupling $\alpha$, which can be derived from the Lagrangian density $\mathcal{L}_X$ in Eq.~\eqref{L_kessence}:
\begin{equation}
\alpha \equiv \left( \frac{\omega}{\Lambda_\ast} \right)^4 \sim \left( \frac{M_\circledast}{\Lambda_\ast} \right)^{-\frac{4}{3}}.
\end{equation}
One can then introduce a collective coupling for the classicalon state,
\begin{equation}
\alpha_c \equiv \alpha N_\circledast \sim 1,
\end{equation}
a hallmark of a saturon~\cite{Dvali:2020wqi,Dvali:2021tez}.
This result reflects the quantum criticality of the $N_\circledast$-boson system, marking a quantum critical point at which collective interactions among the classicalon constituents become significant.
Despite the weak coupling between individual bosons ($\alpha \ll 1$), this critical point signals the onset of nonperturbative collective effects \cite{Berkhahn:2013woa,Dvali:2015ywa}. Therefore, classicalization can be interpreted as the attainment of a fixed point for the collective coupling\footnote{This holds provided that a mass term does not prevent the classicalization radius from exceeding the corresponding Compton wavelength (see Section~\ref{vanishing_VEV} for an example).}---exhibiting critical behavior---rather than for the individual couplings between the constituents.

\subsection{New Landscape Islands in Terra Incognita}
\label{no-go}
Self-UV completion by classicalization opens the possibility of a new Landscape of QFTs, potentially building a bridge to quantum gravity. However, many open issues certainly remain. In particular, this topic does not possess the same degree of maturity as textbook QFT \cite{Peskin:1995ev, Duncan:2012aja}, and the reader may have several concerns about the viability of such a proposal. In the following, we review several criticisms that have been raised in the community, and their proposed resolutions, without claiming any originality. The objective is to briefly summarize and coherently present the relevant literature, which is often dispersed across various topics and communities.

\subsubsection{Trouble with Asymptotic States? Classicalons vs Hadrons}
The concept of asymptotic darkness---that is, the dominance of BH states in ultra-Planckian particle scatterings at impact parameter $b \ll R_S$---has occasionally been criticized on the grounds of the definition of asymptotic states. Since classicalization generalizes this phenomenon to interactions beyond gravity, such criticisms are equally relevant. For example, in Section 2 of Ref.~\cite{Buoninfante:2024yth}, it is argued that the notion of gravitons as asymptotic states in the ultra-Planckian regime may be ill-defined due to nonperturbative graviton self-interactions\footnote{This criticism is unrelated to the legitimate discussions concerning IR divergences and the soft dressing of asymptotic states. Such subtleties arise even in the absence of classicalization, and we have no further insights to contribute on this matter. See Ref.~\cite{Addazi:2016ksu} for soft dressing in the context of BH production in a highly inelastic scattering process.}, analogous to how quarks and gluons cannot be defined as asymptotic states in QCD because they are confined within hadrons.

Despite the fact that both BH production in GR and hadronization in QCD arise from nonperturbative dynamics, the comparison between these 2 phenomena is not appropriate for critiquing classicalization. To clarify this distinction, let us consider the scattering of 2 massive particles, each with mass \(M \ll \Lambda_P\), in the context of GR, rather than massless particles. For an ultra-Planckian scattering characterized by \(\sqrt{s} \gg \Lambda_P\) and $b \ll R_S$, one expects the production of a BH with mass \(M_\circledast \sim \sqrt{s}\). A naive question arises: Why are the initial particle states well-defined as particles (rather than BHs) if they themselves possess ultra-Planckian energies \(E \gg \Lambda_P\), similar to how quarks cannot be defined as asymptotic states in QCD? The key point \cite{Arkani-Hamed:1998sfv} is that energy and momentum are not Lorentz-invariant quantities; one can perform a Lorentz boost to study the particle in its rest frame, where \(E \ll \Lambda_P\). The Lorentz-invariant quantity for an asymptotic particle is its mass \(M \ll \Lambda_P\). While such a particle can emit soft gravitons in its rest frame, this emission occurs within the weak-field regime, where BH formation does not take place.

In the case of scattering between 2 particles of mass \(M \ll \Lambda_P\), 2 Lorentz-invariant and independent kinematic quantities can be defined: the Mandelstam variables \(s\) and \(t\). If \(\sqrt{s} \gg \Lambda_P\) but \(\sqrt{-t} \ll \Lambda_P\), with the impact parameter \(b \gg R_S \sim \ell_P^2\sqrt{s}\), the system does not form a BH; instead, it can be analyzed through the eikonal resummation of soft graviton exchange \cite{Giddings:2009gj,Giddings:2011xs}. BH formation is expected only when \(\sqrt{s} \gg \Lambda_P\) and $b \ll R_S$.

In contrast, for light quarks in QCD, hadronization results from strong coupling in the IR regime: in its rest frame, an `asymptotic' quark is strongly coupled to the gluon field and will inevitably form a hadron. Therefore, although nonperturbative effects are fundamental in both GR and QCD, their underlying mechanisms are fundamentally distinct. This discussion can be extended to massless particles as well (which lack a rest frame), with the critical observation being that energy and momentum are not Lorentz-invariant quantities.

\subsubsection{Beyond Spherical Symmetry: Classicalization vs Eikonal Regimes}
In the previous sections, our analysis has focused exclusively on classicalization involving perfectly spherical sources. However, Ref.~\cite{Akhoury:2011en} demonstrates that classicalon formation fails to occur in scattering configurations that deviate substantially from spherical symmetry, thereby calling into question the viability of UV completing an EFT via this mechanism. The efficiency of classicalization is maximized for a perfectly spherical source, a feature associated with the diminished effectiveness of Vainshtein screening in nonspherical configurations \cite{Brax:2011sv,Bloomfield:2014zfa}.

In fact, classicalization is the phenomenon responsible for unitarizing a scattering process with $\sqrt{s} \gg \Lambda_\ast$ and an impact parameter \(b \ll R_\circledast\)  \cite{Dvali:2011th}. This scenario represents the configuration closest to spherical symmetry. A significant deviation from spherical symmetry, characterized by \(b \gg R_\circledast\), implies that unitarization is instead expected to occur through the eikonal resummation of the large number of soft mediators exchanged in the \(t\)-channel between the scattered particles. This mechanism is well documented in the literature on the gravitational \(S\)-matrix \cite{Giddings:2009gj,Giddings:2011xs} and is distinct from Vainshtein screening, which can also be investigated via the resummation of Feynman diagrams \cite{Davis:2021oce}. Consequently, depending on the hierarchy between \(b\) and \(R_\circledast\), the nonperturbative mechanism underlying unitarity should exhibit a smooth transition ($b \sim R_\circledast$) between the eikonal and classicalization regimes \cite{Dvali:2011th}.

\subsubsection{Causality and Positivity Bounds: The Fate of the Fate}
\label{positivity_bounds}
As discussed in Ref.~\cite{Adams:2006sv}, certain nongravitational EFTs with apparently local and Lorentz-invariant actions can give rise to superluminal propagation on specific semiclassical backgrounds, even within their regime of validity. It is important to emphasize that superluminality, in itself, does not constitute a pathology. However, it has been argued that, within such EFTs, it is possible to construct closed timelike curves (CTCs) at the classical level, thereby challenging the conventional notion of causality. The underlying reasoning is that a UV completion consistent with the standard axioms of QFT---including Lorentz invariance, unitarity, analyticity, and locality---imposes positivity bounds on certain coefficients of the EFT operators to prevent such issues \cite{Adams:2006sv}. These bounds can also be derived from dispersion relations \cite{Pham:1985cr,Adams:2006sv,Bellazzini:2020cot,Arkani-Hamed:2020blm}. It is well-established, however, that the k-essence theory with \(c_2 = -1\) \cite{Pham:1985cr,Adams:2006sv,Dvali:2012zc,Bellazzini:2020cot,Arkani-Hamed:2020blm} and the Galileons theories \cite{Tolley:2020gtv,Caron-Huot:2020cmc,Bellazzini:2021oaj,Serra:2023nrn} violate these positivity bounds, thereby raising doubts about the validity of classicalization. The issue of superluminality and CTCs for k-essence and Galileon fields has been highlighted in Refs.~\cite{Bonvin:2006vc,Bonvin:2007mw, Hinterbichler:2009kq, Goon:2010xh, Evslin:2011vh, Evslin:2011rj}.

The critique based on positivity bounds can be challenged by examining its 2 foundational pillars:

\subparagraph{$(i)$ Superluminality:} Drawing an analogy with chronology protection arguments in GR \cite{Hawking:1991nk,Kim:1991mc}, several studies for both k-essence and Galileon theories \cite{Babichev:2007dw,Burrage:2011cr,Kaplan:2024qtf} have argued that strong quantum backreaction renders CTCs sensitive to UV physics, thereby indicating that such a pathological background lies beyond the regime of control within the EFT. This observation aligns with expectations for theories that achieve self-UV completion through classicalization, as discussed in Ref.~\cite{Dvali:2012zc}. Qualitatively, classicalization introduces the concept of a fundamental limit on spacetime resolution associated with a given field.
As a result, light cones appear fuzzy, and the notion of a pointlike event loses its meaning.
It is therefore unsurprising that the traditional concept of microcausality must be reconsidered. Further exploration of the distinction between micro- and macro-causality in theories exhibiting Vainshtein screening can be found in Refs.~\cite{Keltner:2015qqb,Keltner:2015xda,Cintia:2025fzn}.
The conclusion, based on these studies, is that k-essence and Galileon theories do not result in dramatic violations of causality, such as the possible existence of CTCs.

\subparagraph{$(ii)$ Dispersion Relations:} Classicalization embodies the notion of a minimal length scale that an interaction can probe and is thus nonperturbatively nonlocal, despite the apparent locality of the k-essence Lagrangian density in Eq.~\eqref{L_kessence}. This can be formalized within the axiomatic framework of QFT using Jaffe’s classification of strictly localizable, quasi-local, and nonlocalizable fields, specifically by examining the UV behavior of the spectral density functions \cite{Meiman:1964vmk, Jaffe:1966an, Jaffe:1967nb, Keltner:2015xda, Tokuda:2019nqb}. The Wightman formulation, causality, and the standard properties of the $S$-matrix for theories with nonlocalizable fields have been studied in Refs.~\cite{Iofa:1969fj, Iofa:1969ex, Steinmann:1970cm, Fainberg:1977wp, Fainberg:1978cc, Solovev:1980tle, Fainberg:1992jt, Soloviev:1999rv, Soloviev:2001qe, Soloviev:2001qd, Bruning:2004jr, Soloviev:2005qd, Soloviev:2006ah, Soloviev:2009cy, Soloviev:2010bg}. Positivity bounds are typically derived from dispersion relations for textbook QFTs, which deal with tempered localizable fields, a subclass of strictly localizable fields \cite{Tokuda:2019nqb}, whereas classicalizing QFTs involve nonlocalizable fields \cite{Keltner:2015xda,Buoninfante:2023dyd}. The status of positivity bounds on QFTs with nonlocalizable fields is discussed in Refs.~\cite{Fainberg:1971ia, Tokuda:2019nqb, Buoninfante:2023dyd, Buoninfante:2024ibt}, and the results for classicalizing theories are consistent with the violation of the bounds obtained by assuming tempered localizable fields \cite{Tokuda:2019nqb, Buoninfante:2023dyd}, e.g., \( c_2 = -1 \). The properties of nonlocalizable fields, whose spectral density functions grow faster than those of strictly localizable fields, explain why classicalization appears at odds with any interpretation in a Wilsonian perspective \cite{Kovner:2012yi}. We also stress that this is the reason the classicalization proposal for scalars is fundamentally different from the controversial Higgsplosion scenario \cite{Khoze:2017tjt, Khoze:2017ifq, Khoze:2017lft, Khoze:2017uga, Belyaev:2018mtd} in the framework of the standard electroweak theory, involving only strictly localizable fields \cite{Monin:2018cbi, Khoze:2018qhz}, while both deal with multi-scalar boson production at high energies.\\

The conclusion of this discussion is that the nonlocal features of classicalization invalidate the usual mathematical hypotheses used to derive positivity bounds from dispersion relations in local QFTs, while exhibiting a self-protection mechanism against the ability to observe violations of causality.

\subsubsection{Nonrenormalizable Operators Demystified}
In the models under consideration, classicalization emerges through operators---referred to as UV screeners---that belong to the infinite class of nonrenormalizable operators within the conventional EFT framework.
Since this phenomenon becomes apparent in scattering processes with \(\sqrt{s}  \gg \Lambda_\circledast\) and $b \ll R_\circledast$, practitioners of EFTs may find the interpretation of these nonrenormalizable operators within the classicalization paradigm unclear.
To clarify this point, it is worthwhile to first review the UV origin of nonrenormalizable operators in a standard EFT \cite{Burgess:2020tbq}.

\paragraph{Ultraviolet Remnants:}
The action of an EFT can be expressed as
\begin{equation}
S_{\mathrm{EFT}} = \int d^4x \left( \mathcal{L}_\mathrm{R} + \mathcal{L}_\mathrm{NR} \right),
\end{equation}
where \(\mathcal{L}_\mathrm{R}\) and \(\mathcal{L}_\mathrm{NR}\) denote the renormalizable and nonrenormalizable parts of the Lagrangian density, respectively.
Within the Wilsonian EFT framework, the latter are termed `irrelevant' because, within the regime of validity of the EFT, they are suppressed by powers of a heavy interaction scale \(\Lambda\) according to
\begin{equation}
\mathcal{L}_\mathrm{NR} = \sum_{k=1}^{+\infty} \dfrac{c_k}{\Lambda^{d_k-4}} \, \mathcal{O}^{(k, \, d_k)}_\mathrm{NR},
\end{equation}
where $\mathcal{O}^{(k, d_k)}_\mathrm{NR}$ represents an operator of dimension \(d_k>4\), and the coefficients \(c_k\) are naturally of \(\mathcal{O}(1)\) unless suppressed by selection rules from (approximate) symmetries.
In the Wilsonian approach, this infinite tower of operators arises from integrating out heavy fields.

To illustrate, consider a toy model featuring 1 real scalar field \(\Phi(x)\) of mass \(M_\Phi\) and 1 massless Dirac fermion \(\psi(x)\). The Lagrangian density, including a Yukawa coupling $y_\psi$, is given by
\begin{equation}
\mathcal{L}_\mathrm{UV} [\Phi, \bar{\psi}, \psi]
\equiv
i \bar{\psi} \gamma^\mu \overleftrightarrow{\partial_\mu} \psi
- \dfrac{1}{2} \, \Phi ( \square + M_\Phi^2 ) \Phi
+ y_\psi \, \bar{\psi} \psi \Phi,
\end{equation}
with the Dirac matrices \(\gamma^\mu\), and the derivative operator defined as \(\overleftrightarrow{\partial_\mu} = \overrightarrow{\partial_\mu} - \overleftarrow{\partial_\mu}\).
An EFT involving only the light field \(\psi(x)\) can be obtained by integrating out the heavy field \(\Phi(x)\). The effective action \(S_\mathrm{EFT}\) is formally nonperturbatively defined as
\begin{equation}
e^{iS_\mathrm{EFT}} = \int [\mathcal{D}\Phi] \ e^{iS_\mathrm{UV}[\Phi, \, \bar{\psi}, \, \psi]},
\end{equation}
where \(S_\mathrm{UV}\) is the action associated with \(\mathcal{L}_\mathrm{UV}\).
At tree level, \(\Phi\) can be integrated out using its Euler-Lagrange equation, which amounts to the substitution
\begin{equation}
\Phi \mapsto y_\psi \left( \dfrac{1}{\square + M_\Phi^2} \right) \bar{\psi} \psi,
\end{equation}
yielding a nonlocal effective Lagrangian density:
\begin{equation}
\mathcal{L}_\mathrm{EFT}^\mathrm{Nloc}[\bar{\psi}, \psi]
=
i \bar{\psi} \gamma^\mu \overleftrightarrow{\partial_\mu} \psi
+ \frac{y_\psi^2}{2} \, \bar{\psi}\psi \left( \dfrac{1}{\square + M_\Phi^2} \right) \bar{\psi}\psi.
\end{equation}
This EFT is equivalent to the original theory only at tree level\footnote{More generally, a systematic method for integrating out fields in perturbation theory is the background field method~\cite{Zhang:2016pja}.} for the \(\Phi\) field, as the nonlocal form factor in the quartic term encodes the tree-level propagation of the massive \(\Phi\) particle.
For Euclidean momenta \(p_E^2 \ll M_\Phi^2\), a Taylor expansion of this form factor---known as the operator product expansion (OPE)---produces a local EFT Lagrangian density:
\begin{equation}
\mathcal{L}_\mathrm{EFT}^\mathrm{loc}[\bar{\psi}, \psi]
\simeq
i \bar{\psi} \gamma^\mu \overleftrightarrow{\partial_\mu} \psi
+ \frac{y_\psi^2}{2}
\left[ \frac{\bar{\psi}\psi\bar{\psi}\psi}{M_\Phi^2}
- \frac{\bar{\psi}\psi \square \bar{\psi}\psi}{M_\Phi^4}
+ \mathcal{O} \left( \frac{\square^2}{M_\Phi^6} \right)
\right],
\end{equation}
where the tower of effective operators is truncated at the appropriate order in \(\square/M^2\) via power counting.
These nonrenormalizable operators\footnote{While this terminology is standard, it is not accurate, as the EFT remains perturbatively renormalizable at any given order in perturbation theory.} thus represent the low-energy remnants of the \(\Phi\) propagator, with the interaction scale \(\Lambda\) set by the heavy mass \(M_\Phi\) if $y_\psi \sim 1$.

The role of nonrenormalizable operators---referred to as `UV remnants'---is evident from this example: they provide an approximate description of the virtual propagations of heavy particles and are not fundamental.
Crucially, these operators are treated perturbatively and must remain so, even when seeking semiclassical solutions within an EFT (which are nonperturbative in couplings but perturbative in the EFT expansion).
This perturbative treatment ensures that EFTs remain free of ghost instabilities, in contrast to generic higher-derivative theories~\cite{Simon:1990ic, Weinberg:2008hq, Burgess:2014lwa, Solomon:2017nlh}.

\paragraph{Ultraviolet Screeners:}
In the context of classicalization, the nonrenormalizable operators (UV screeners) responsible for this phenomenon must be treated nonperturbatively.
To maintain consistency with the discussion about UV remnants, this implies that their fundamental origin cannot be attributed to the OPE of form factors coming from integrating out UV degrees of freedom \cite{Dvali:2011nj,Dvali:2012zc,Dvali:2016ovn,Kaloper:2014vqa}.
To clarify this distinction, the nonrenormalizable operators in the EFT can be decomposed as follows:
\begin{equation}
\mathcal{L}_\mathrm{NR} =
\sum_{i \geq 1} \dfrac{\kappa_i}{\Lambda_\ast^{d_i-4}} \, \mathcal{O}^{(i, \, d_i)}_\mathrm{scr}
+
\sum_{j = 1}^{+\infty} \dfrac{c_j}{M^{d_j-4}} \, \mathcal{O}^{(j, \, d_j)}_\mathrm{rem},
\end{equation}
where the operators \(\mathcal{O}^{(i, d_i)}_\mathrm{scr}\) and \(\mathcal{O}^{(j, d_j)}_\mathrm{rem}\) represent UV screeners and UV remnants of dimension $d_i>4$ and $d_j>4$, respectively.
Here, \(\kappa_i\) and \(c_j\) are dimensionless couplings, \(\Lambda_\ast\) denotes the classicalization scale, and \(M\) is the mass scale associated with the next layer of UV degrees of freedom that have been integrated out.
Given that classicalization is intended to UV complete the theory within the energy range between \(\Lambda_\ast\) and \(M\), the hierarchy \(\Lambda_\ast \ll M\) must hold. Note that the number of UV screeners may be infinite if they are defined through arbitrary functions of the fields and their derivatives, like $\mathcal{K}(X)$ in Eq.~\eqref{S_X}.

For energies \(E \ll \Lambda_\ast\), the nonperturbative dynamics of UV screeners---including fuzzyons and the infinite tower of classicalons---can be integrated out.
This procedure yields a nonlocal EFT Lagrangian density, which can then be expanded into a tower of UV remnants via the OPE.
Thus, below the scale \(\Lambda_\ast\), an EFT can be constructed using conventional methods, despite an apparent violation of traditional positivity bounds.

Since UV screeners govern the classicalization dynamics, they must be treated nonperturbatively, and their selection must ensure the absence of physical ghost-like degrees of freedom in the spectrum.
This challenge is well documented in the modified gravity literature~\cite{Joyce:2014kja}, where k-essence and Galileon theories are specifically constructed to avoid such pathologies\footnote{For a broader discussion of these constructions (in the context of dark energy theories), see the review~\cite{Langlois:2018dxi}.}.
This requirement represents a fundamental distinction from UV remnants.

As a final remark on the interpretation of UV screeners within the EFT framework, the reader may question how the values of the couplings \(\kappa_i\) align with the renormalization program. A detailed analysis of the perturbative renormalization of the EFT governing the fluctuations around the classicalon background lies beyond the scope of this article (see Ref.~\cite{Brax:2016jjt} for an attempt in this direction).
Here, we offer only preliminary considerations on the subject.
Within the classicalon radius, the running of these couplings is effectively frozen, as the dynamics in this region is predominantly classical, as demonstrated in Ref.~\cite{Padilla:2017wth}.
Since quantum corrections are suppressed\footnote{In certain cases, nonrenormalization theorems guarantee that a coupling \(\kappa_i\) does not run \cite{Goon:2016ihr,Heisenberg:2020cyi}.}, no large logarithms should require resummation.
Consequently, one should be able to choose the renormalized values of the couplings \(\kappa_i\) to match those of the classical Lagrangian $(\hbar \to 0)$, with, e.g., a subtraction point specified by the background amplitude \(\overline{\phi_0}^\prime \equiv \Lambda_\ast^2\). The counterterms should then be employed to absorb the UV divergences, as usual in the background field method \cite{Peskin:1995ev}. This also implies that if an operator vanishes classically, it is incorporated with a renormalized coupling that is zero in the Lagrangian density, yet accompanied by a nonvanishing counterterm to address the UV divergences \cite{Brax:2016jjt}.

\section{From K-essence to K-chameleon}
\label{from_kessence_kchameleon}
Let us revisit the massless k-essence model analyzed in Section~\ref{Class_Vainsh}.
For physical applications, scalar fields are typically introduced with a potential $V(\phi)$, which explicitly breaks the shift symmetry. In the presence of a localized source, the Euler-Lagrange equation~\eqref{ELE0} takes the modified form
\begin{equation}
\square \phi + \frac{c_2}{\Lambda_\ast^4} \, \partial^\mu \left( \partial^\nu \phi \, \partial_\nu \phi \, \partial_\mu \phi \right) + \frac{dV}{d\phi}
= -\dfrac{\mathcal{E}}{\Lambda_\ast} \, \delta^{(3)}(\mathbf{r}),
\label{ELE_V}
\end{equation}
where \(c_2 = -1\) and the left-hand side is no longer a total divergence. Our objective is to study how this potential affects Vainshtein screening---and, consequently, the conditions for UV completion through classicalization.

\subsection{Massive K-essence}
We begin by examining a quadratic potential, specifically a nontachyonic mass term:
\begin{equation}
V(\phi) = V_m(\phi) \equiv \frac{m^2}{2} \, \phi^2,
\end{equation}
with $m > 0$.
A light k-essence boson, characterized by \(m \ll \Lambda_\ast\), is 't Hooft natural, as the limit \(m \to 0\) restores the shift symmetry of the Lagrangian density, thereby ensuring the radiative stability of the light scalar. As already mentioned in Section~\ref{nutshell}, UV completion via classicalization renders a little hierarchy \(m \ll \Lambda_\ast\) a necessary consistency condition for the k-essence model.

\paragraph{Background:}
One can establish the conditions under which the potential dominates the other terms in the Lagrangian density across the various regimes discussed in Section~\ref{sec_background} for massless k-essence, with \(\mathcal{E} \gg \Lambda_\ast\). To this end, it is convenient to introduce the (reduced) Compton wavelength \(\lambdabar_C \equiv 1/m\) of the k-essence boson.

\subparagraph{Linear Regime \((r \gg R_V)\):}
In this regime, the kinetic term dominates over the kinetic self-interaction term. Consequently, the mass term should be compared to the kinetic term:
\begin{equation}
\left| \dfrac{ m^2 \, \phi^2 }{ \partial^\mu \phi \partial_\mu \phi} \right|
\underset{\phi(x) \, \equiv \, \overline{\phi}(r)}{\sim}
\left( \frac{r}{\lambdabar_C} \right)^2.
\end{equation}
This ratio remains small when \(r \ll \lambdabar_C\). However, for \(r \gg \lambdabar_C\), the mass term prevails over the kinetic term, rendering the solution in Eq.~\eqref{background_sol} invalid.

\subparagraph{Transition Regime \((r \sim R_V)\):}
Here, the kinetic term and the kinetic self-interaction term are of comparable magnitude. The comparison yields:
\begin{equation}
\left| \dfrac{ m^2 \, \phi^2 }{ \partial^\mu \phi \partial_\mu \phi} \right|
\sim \left| \dfrac{ m^2 \, \phi^2 }{ \ell_\ast^4 (\partial^\mu \phi \partial_\mu \phi)^2} \right|
\underset{\phi(x) \, \equiv \, \overline{\phi}(r)}{\sim}
\left( \frac{r}{\lambdabar_C} \right)^2,
\end{equation}
leading to the same conclusions as in the linear regime.

\subparagraph{Nonlinear Regime \((r \ll R_V)\):}
In this regime, the kinetic self-interaction term dominates over the kinetic term. Thus, the mass term should be compared to the former:
\begin{equation}
\left| \dfrac{ m^2 \, \phi^2 }{ \ell_\ast^4 (\partial^\mu \phi \partial_\mu \phi)^2} \right|
\underset{\phi(x) \, \equiv \, \overline{\phi}(r)}{\sim}
\left( \frac{r}{\lambdabar_C} \right)^2 \left( \frac{r}{R_V} \right)^\frac{4}{3},
\end{equation}
which remains small provided that \(r \ll \lambdabar_C\).

We now compare these scaling arguments with the exact solution for massive k-essence in the presence of a localized source. The Euler-Lagrange equation is provided by Eq.~\eqref{ELE_V}, with the potential specified as \(V(\phi) = V_m(\phi)\).
In the linear regime, the background solution takes the form
\begin{equation}
\overline{\phi}(r) \simeq - \left(\frac{R_V}{\ell_\ast} \right)^2 \dfrac{e^{-\frac{r}{\lambdabar_C}}}{r}
\quad \xrightarrow{r \ll \lambdabar_C} \quad
-  \left(\frac{R_V}{\ell_\ast} \right)^2 \dfrac{1}{r},
\label{phi-background}
\end{equation}
which corresponds to the Yukawa potential for a massive boson in relativistic quantum mechanics. This potential reduces to the Coulomb potential in the limit \(r \ll \lambdabar_C\). The mass term thus introduces the well-known `mass screening' effect at the foundation of Yukawa's theory of nuclear interactions~\cite{Yukawa:1935xg}, i.e., a massive boson cannot mediate interactions beyond its Compton wavelength.

\begin{figure}[t]
\begin{center}
\includegraphics[height=7.9cm]{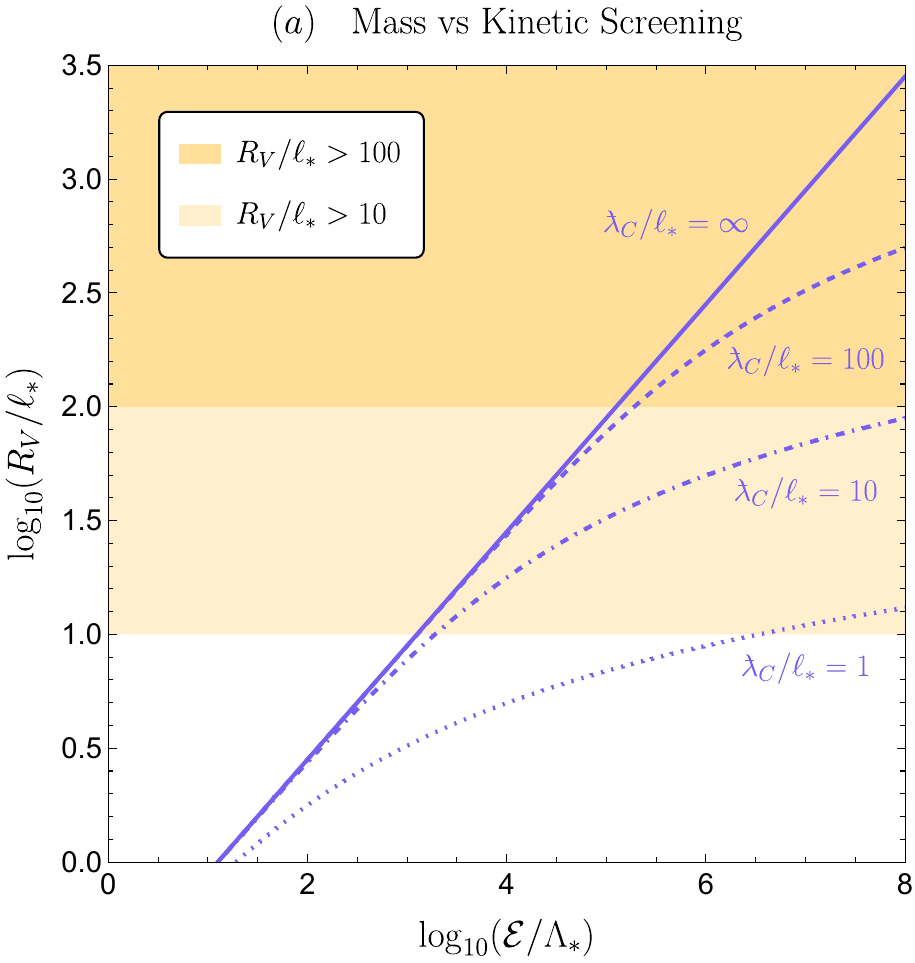}\hfill
\includegraphics[height=7.9cm]{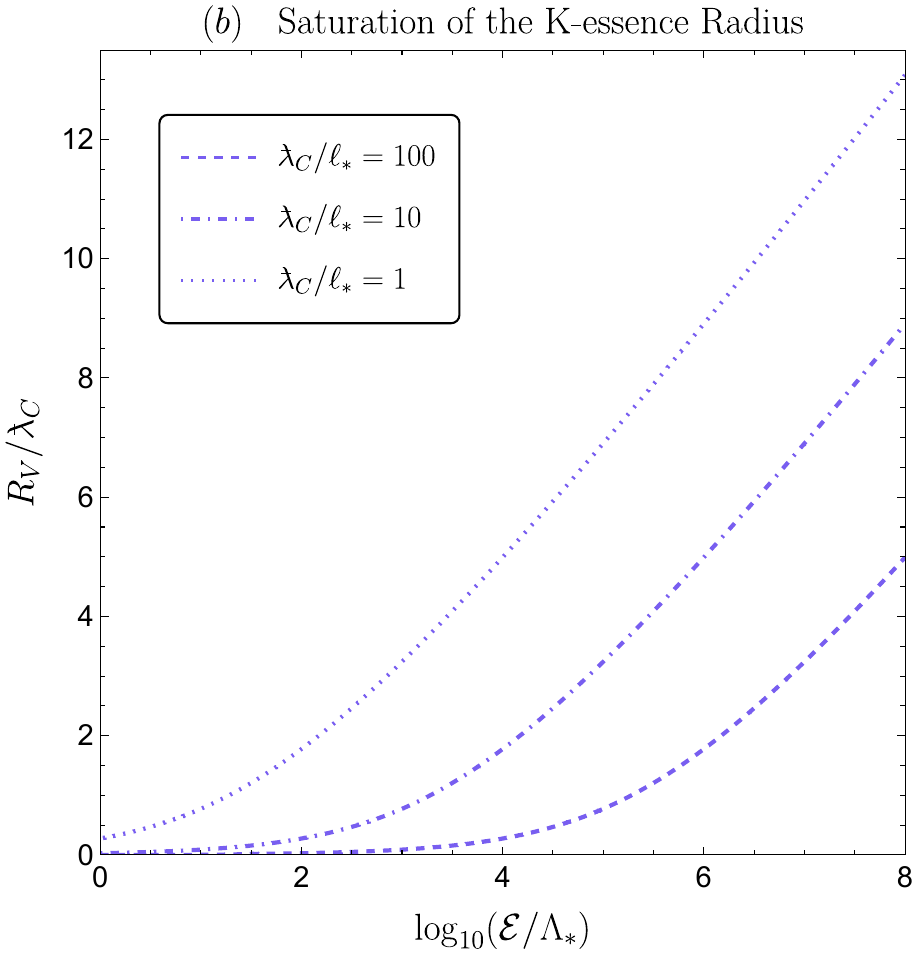}\hfill
\end{center}
\caption{We investigate the impact of mass screening beyond the Compton wavelength \(\lambdabar_C\) on the Vainshtein radius \(R_V\) as the source magnitude \(\mathcal{E}\) exceeds the interaction scale \(\Lambda_\ast \equiv 1/\ell_\ast\). In panel \((a)\), we observe that, for massive k-essence, \(R_V\) departs from the massless case upon reaching \(\lambdabar_C\). Panel \((c)\) demonstrates that \(R_V\) saturates once it attains \(\lambdabar_C\). Notably, even when $\mathcal{E}/\Lambda_\ast \gg 1$, the ratio \(R_\circledast / \lambdabar_C\) remains of order \(\mathcal{O}(10)\).
}
\label{massive_Kessence}
\end{figure}

\begin{figure}[t]
\begin{center}
\includegraphics[height=7.8cm]{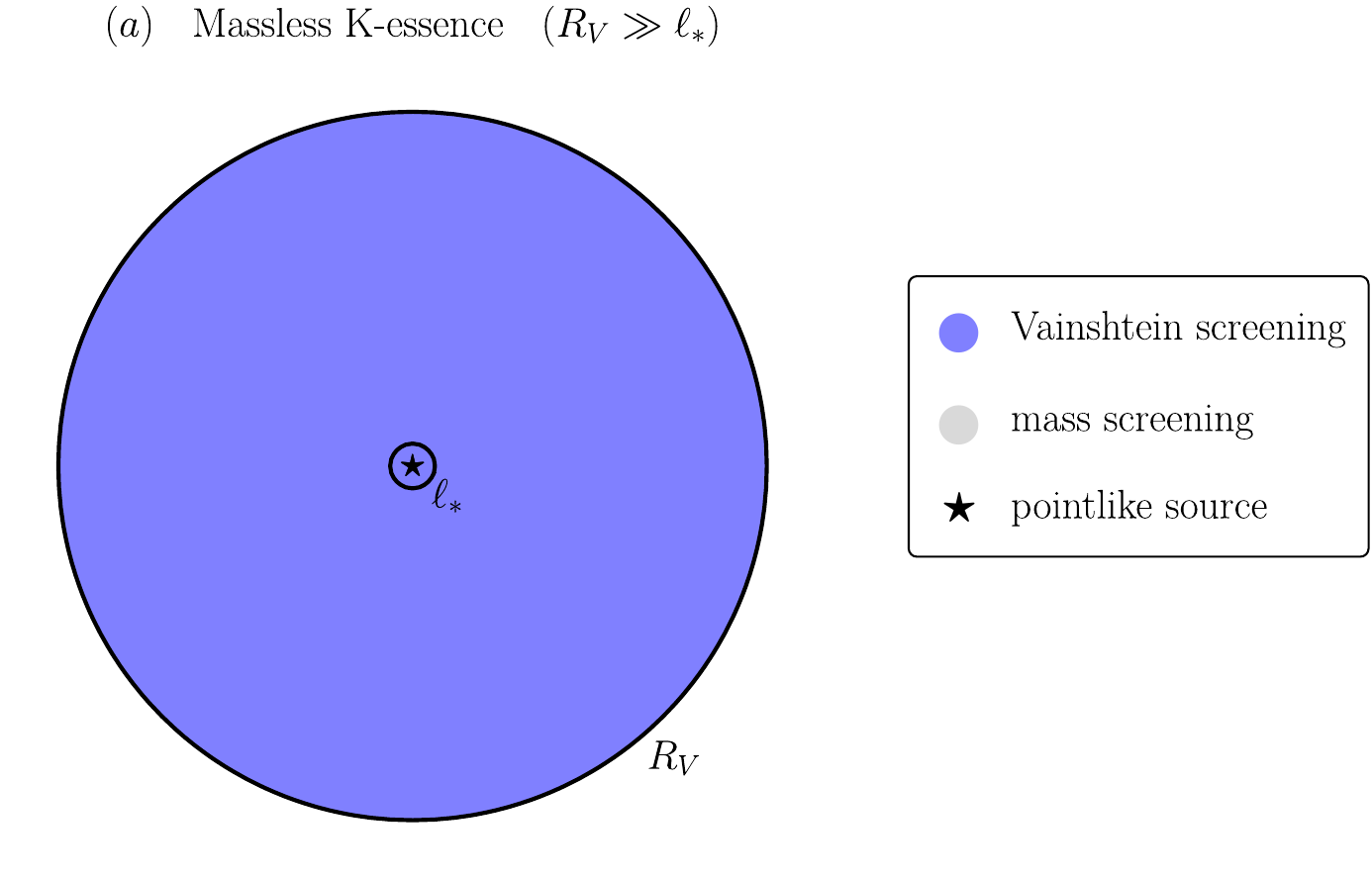}\\
\includegraphics[height=7.8cm]{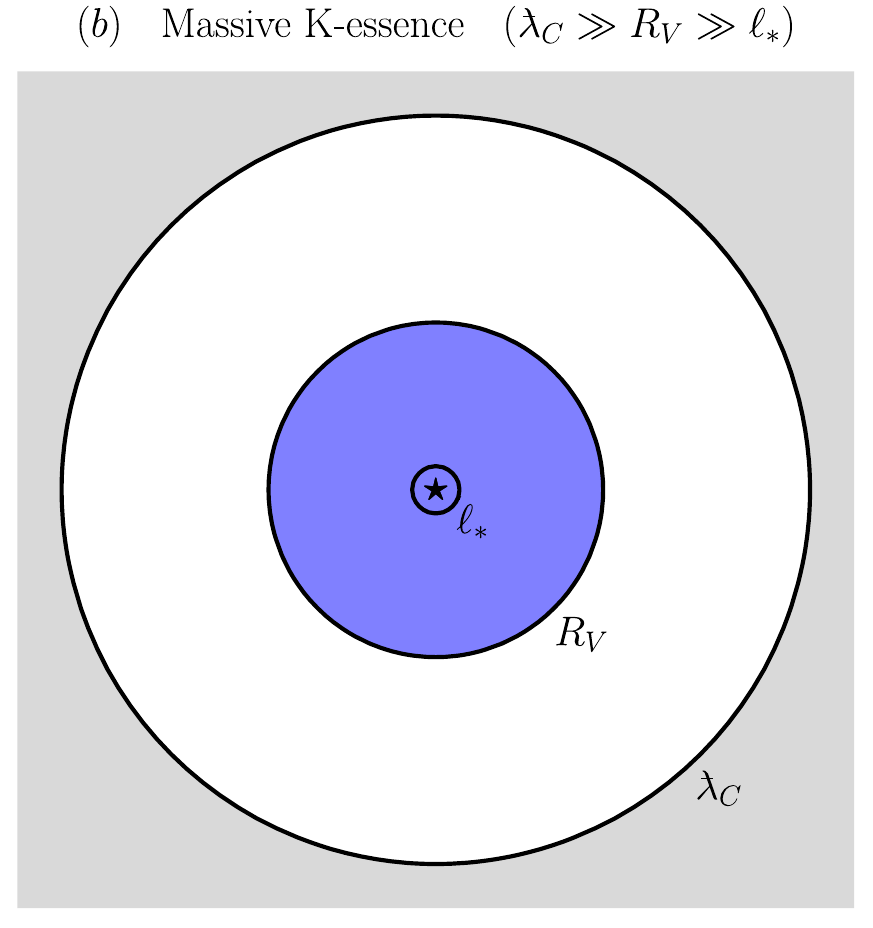}
\includegraphics[height=7.8cm]{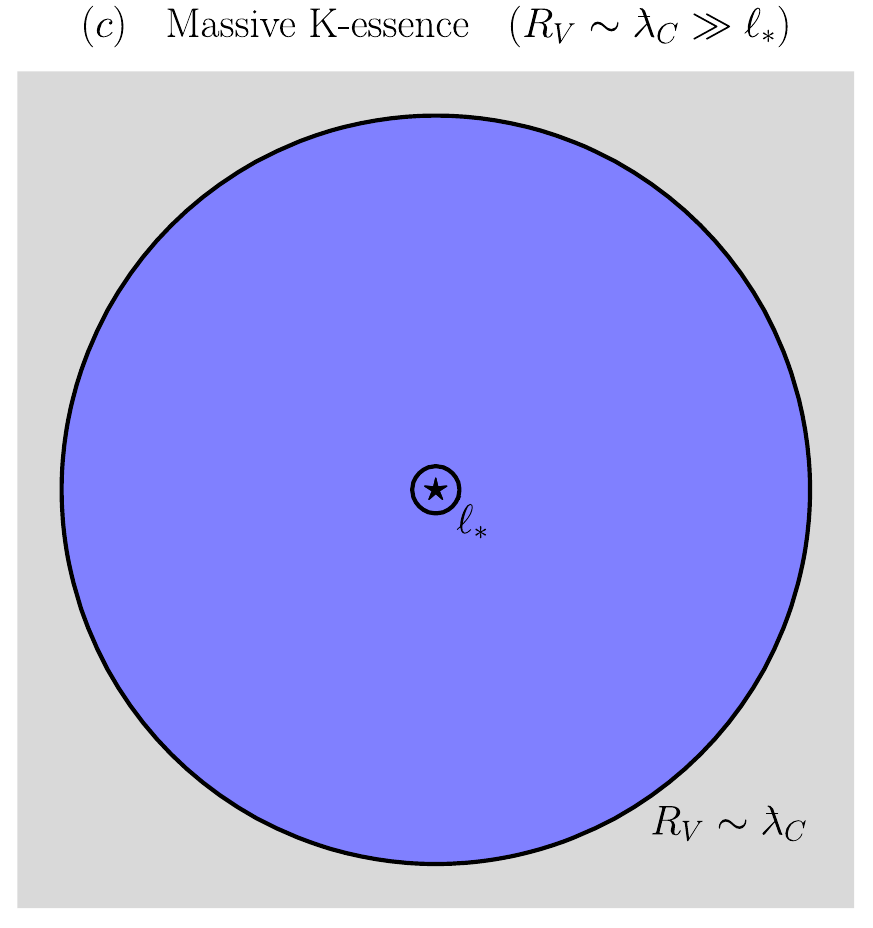}
\end{center}
\caption{Schematic representation of the distinct screening regions surrounding a pointlike source for massless versus massive k-essence, interacting solely through kinetic self-interactions. The fundamental length scale \(\ell_\ast \equiv 1/\Lambda_\ast\) remains the smallest in the system.
Panel \((a)\): 
For massless k-essence, the boson exhibits an infinite interaction range, allowing the Vainshtein radius \(R_V\) to grow without bound as the source magnitude \(\mathcal{E} \gg \Lambda_\ast\) increases. This radius defines the Vainshtein core.
Panels \((b)\) and \((c)\):
For massive k-essence, the Compton wavelength \(\lambdabar_C\) determines the range of self-interaction; beyond this range, mass screening occurs.
Panel \((b)\):
When \(R_V \ll \lambdabar_C\), the Vainshtein core expands unimpeded by mass screening.
Panel \((c)\):
Once the Vainshtein core fills the entire sphere of radius \(\lambdabar_C\), the Vainshtein radius saturates at \(R_V \sim \lambdabar_C\). Beyond this radius, Vainshtein screening cannot extend further.
}
\label{circles1}
\end{figure}

From the preceding discussion, if \(\lambdabar_C \gg R_V\), the background solution for \(r \ll \lambdabar_C\)---where mass screening is negligible---is well approximated by the massless k-essence solution in Eq.~\eqref{background_sol}. However, when \(R_V \sim \lambdabar_C\), it is possible to determine the conditions under which nonlinearities become significant. As shown in Fig.~\ref{massive_Kessence}, nonlinear effects fail to develop for \(r \gg \lambdabar_C\), even when \(\mathcal{E} / \Lambda_\ast \sim 10^{8}\). Consequently, the Vainshtein core cannot extend significantly beyond the sphere of radius \(R_V \sim \lambdabar_C\), as illustrated in Fig.~\ref{circles1}. This behavior can be summarized by the following expression:
\begin{equation}
R_V \sim \ell_\ast \sqrt{\frac{\mathcal{E}}{\Omega \Lambda_\ast}} \cdot
\theta \left[ \lambdabar_C - \ell_\ast \sqrt{\frac{\mathcal{E}}{\Omega \Lambda_\ast}} \right]
+ \lambdabar_C \cdot \theta \left[ \ell_\ast \sqrt{\frac{\mathcal{E}}{\Omega\Lambda_\ast}} - \lambdabar_C \right] \lesssim \lambdabar_C,
\end{equation}
from Eq.~\eqref{V_rad}, with the Heaviside step function $\theta(z)$.

\paragraph{Perturbations:}
For the fluctuation \(\delta \phi(x)\) on top of the background $\overline{\phi}(r)$, the quadratic potential introduces a mass term as follows:
\begin{equation}
V_m ( \overline{\phi} + \delta \phi )
\supset \frac{m^2}{2} \, \delta \phi^2
= \frac{\overline{m}^2}{2} \, \delta \phi_Z^2,
\quad
\overline{m}(r_0) \equiv \frac{m}{\sqrt{Z_\phi(r_0)}}.
\end{equation}
Here, we use the renormalized field \(\delta \phi_Z(x)\) from Eq.~\eqref{L_2} to identify the effective mass \(\overline{m}(r_0)\) of the fluctuations of the background \(\overline{\phi}(r)\). Deep within the Vainshtein core, where \(Z_\phi(r_0) \gg 1\), the mass undergoes a redshift, yielding \(\overline{m}(r_0) \ll m\). Consequently, the mass term does not disturb the screening mechanism, and the background solution is perturbatively stable.

\paragraph{Classicalization:}
Within the framework of self-UV completion via classicalization, unitarity is preserved for \(\sqrt{s} \gg \Lambda_\ast\) through the formation of classicalons, the radii of which grow with their masses \(M_\circledast \sim \sqrt{s}\). However, for massive k-essence, our analysis---illustrated in Fig.~\ref{massive_Kessence} with an external source---demonstrates that \(R_\circledast\) saturates at the Compton wavelength \(\lambdabar_C\), leading to the following behavior:
\begin{equation}
R_\circledast \sim \ell_\ast \left(\frac{M_\circledast}{\Omega\Lambda_\ast}\right)^\frac{1}{3} \cdot
\theta \left[ \lambdabar_C - \ell_\ast \left(\frac{M_\circledast}{\Omega\Lambda_\ast}\right)^\frac{1}{3} \right]
+ \lambdabar_C \cdot
\theta \left[ \ell_\ast \left(\frac{M_\circledast}{\Omega\Lambda_\ast}\right)^\frac{1}{3} - \lambdabar_C \right] \lesssim \lambdabar_C,
\end{equation}
as derived from Eq.~\eqref{C_rad_1}. When \(\ell_\ast \sim \lambdabar_C\), the nonperturbative regime fails to transition into a semiclassical regime over a wide range of \(\ell_\ast / \lambdabar_s\). This limitation arises because the \(N_\circledast \gg 1\) bosons within the coherent state must remain feebly interacting---a condition that is never satisfied when \(\ell_\ast \sim \lambdabar_C\), given that the kinetic self-interaction becomes strongly coupled at energies \(\omega > m \sim \Lambda_\ast\) per boson. In the absence of a classicalon regime to implement UV/IR mixing, the exponential suppression of \(2 \to 2\) scattering amplitudes cannot be invoked to preserve unitarity, so one needs the condition \(\lambdabar_C \gg \ell_\ast\). Furthermore, the mass parameter \(m\) must correspond to that of the low-energy EFT, as the latter provides an accurate description of the dynamics for \(r \gg R_\circledast\). To ensure a reliable, weakly coupled EFT regime outside of the Vainshtein core---where the semiclassical approach remains valid---it is also necessary that \(\lambdabar_C \gg \ell_\ast\) in the linear regime.

We have thereby substantiated the conjecture presented in Ref.~\cite{Dvali:2010jz}: UV completion through classicalization requires at least a little hierarchy, specifically \(m \ll \Lambda_\ast\). In the absence of such a hierarchy, the theory encounters the same strong-coupling problem as any EFT pushed beyond its strong-coupling scale: an infinite tower of higher-dimensional operators comes to dominate, and one loses predictivity, rendering the EFT ineffective. This outcome exemplifies the concrete impact of UV/IR mixing on the hierarchy of scales.

\subsection{K-chameleon: Kinetically Catalyzed Chameleon Screening}
\label{double-screening}
We have observed that a quadratic potential for the k-essence field $\phi(x)$, i.e., a mass term, imposes a little hierarchy for Vainshtein screening to take effect. It is therefore instructive to examine more complex potentials. Motivated by SSB, we focus on potentials invariant under the $\mathbb{Z}_2$ symmetry $\phi(x) \mapsto -\phi(x)$; however, the same methodology can be applied to study more general potentials.

\subsubsection{Symmetric Vacuum}
\label{vanishing_VEV}
Consider the previously discussed massive k-essence model with a perturbative \(\phi^4\) interaction, where the potential exhibits a vanishing vacuum expectation value (VEV):
\begin{equation}
V(\phi) = V_\oplus(\phi) \equiv \frac{m^2}{2} \, \phi^2 + \frac{\lambda}{4!} \, \phi^4,
\label{V_plus}
\end{equation}
with $m, \ \lambda > 0$.
In this case, the mass term is not protected by a shift symmetry, rendering a hierarchy \(m \ll \Lambda_\ast\) unnatural in the 't Hooft sense, analogous to the Higgs boson mass in the standard electroweak theory \cite{Hebecker:2020aqr}.
Nonetheless, as our analysis of the quadratic potential suggests, such a modest hierarchy is essential for classicalization to occur.

\paragraph{Vainshtein Screening:}
In this section, we focus on the regime \(r \ll \lambdabar_C\), where the quadratic term in \(V(\phi)\) becomes negligible.
This allows us to isolate the quartic term and examine how this new self-interaction modifies the background solution, which remains well-approximated by Eq.~\eqref{background_sol} for \(r \ll \lambdabar_C\).

\subparagraph{Linear Regime \((r \gg R_V)\):}
In this regime, the kinetic term dominates over the kinetic self-interaction.
To assess the relative importance of the \(\phi^4\) interaction, we compare it to the kinetic term:
\begin{equation}
\left| \dfrac{\lambda \, \phi^4}{\partial^\mu \phi \partial_\mu \phi} \right| 
\underset{\phi(x) \equiv \overline{\phi}(r)}{\sim}
\lambda \left( \frac{R_V}{\ell_\ast} \right)^4,
\end{equation}
which is \(\ll 1\) when \(R_V \ll \ell_\ast\), i.e., within the standard EFT regime where $\mathcal{E}$ and \(\phi(x) \ll \Lambda_\ast\).
However, even for a moderately weak coupling (e.g., \(\lambda \sim 0.1\)), the ratio becomes \(\gg 1\) when \(\mathcal{E} \gg \Lambda_\ast\).
In this case, the potential dominates over the linear term, and the background solution \(\overline{\phi}(r)\) approaches the VEV of \(V(\phi)\).

\subparagraph{Transition Regime \((r \sim R_V)\):}
Here, the kinetic term and the kinetic self-interaction are of comparable magnitude, yielding:
\begin{equation}
\left| \dfrac{\lambda \, \phi^4}{\partial^\mu \phi \partial_\mu \phi} \right|
\sim \left| \dfrac{\lambda \, \phi^4}{\ell_\ast^4(\partial^\mu \phi \partial_\mu \phi)^2} \right|
\underset{\phi(x) \equiv \overline{\phi}(r)}{\sim}
\lambda \left( \frac{R_V}{\ell_\ast} \right)^4,
\end{equation}
and thus the conclusions remain consistent with those of the linear regime.

\subparagraph{Nonlinear Regime \((r \ll R_V)\):}
In this regime, with \(\mathcal{E} \gg \Lambda_\ast\), the kinetic self-interaction dominates the kinetic term, so we compare the \(\phi^4\) interaction to the kinetic self-interaction:
\begin{equation}
\left| \dfrac{\lambda \, \phi^4}{\ell_\ast^4(\partial^\mu \phi \partial_\mu \phi)^2} \right|
\underset{\phi(x) \equiv \overline{\phi}(r)}{\sim}
\lambda \left( \dfrac{r}{\ell_\ast} \right)^4
\lesssim \lambda \left( \dfrac{\lambdabar_C}{\ell_\ast} \right)^4.
\end{equation}
This scaling estimate reveals that a Vainshtein core for \(r \gg \ell_\ast\) can only exist if \(\lambda \ll 1\) to sufficiently suppress the growth of \((r/\ell_\ast)^4 \gg 1\).
Consequently, for applications involving a moderately weak coupling, a clear tension arises with Vainshtein screening.

\begin{figure}[t]
\begin{center}
\includegraphics[height=7.1cm]{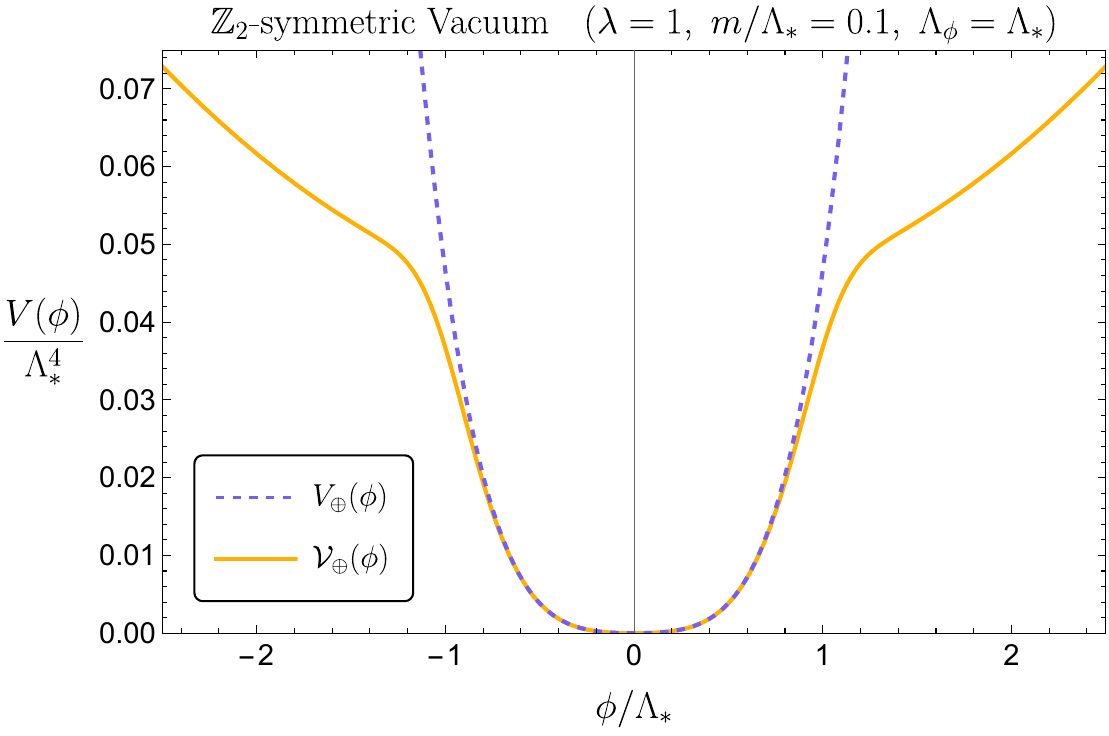}
\end{center}
\caption{Comparison of the scalar potentials \(V_\oplus(\phi)\) in Eq.~\eqref{V_plus} and \(\mathcal{V}_\oplus(\phi)\) in Eq.~\eqref{V_curl_plus} with a \(\mathbb{Z}_2\)-symmetric vacuum.
The 2 potentials begin to deviate at \(|\phi| \sim \Lambda_\phi = \Lambda_\ast\), the threshold at which chameleon screening of the \(\phi^4\) interaction becomes effective.}
\label{plot_symmetric}
\end{figure}

\paragraph{Chameleon Screening:}
To resolve this tension between a quartic potential and Vainshtein screening, we seek a minimal modification of the model that satisfies the following 2 criteria:
\((i)\) a moderately weak $\phi^4$ coupling (e.g., \(\lambda \sim 0.1\) or even 1) when \(\partial \phi \ll \Lambda_\ast^2\);
\((ii)\) a potential that approaches a mass term as \(\partial \phi \gg \Lambda_\ast^2\).
A solution is to recognize that, within the Vainshtein core ($R_V \gg r \gg r_\star$), both \(|\overline{\phi}^{\, \prime}(r)| \gg \Lambda_\ast^2\) and \(|\overline{\phi}(r)| \gg \Lambda_\ast\) hold.
We therefore modify the potential as follows:
\begin{equation}
V(\phi) = \mathcal{V}_\oplus (\phi) \equiv \frac{m^2}{2} \, \phi^2 + \frac{\lambda}{4!} \, \Lambda_\phi^4 \tanh \left[ \left( \frac{\phi}{\Lambda_\phi} \right)^4\right],
\label{V_curl_plus}
\end{equation}
where, following Dirac naturalness, we set\footnote{A hierarchy \(\Lambda_\phi \ll \Lambda_\ast\) is typically excluded, since the EFT would become strongly coupled for hard scattering processes at energies \(\sqrt{s} \sim \Lambda_\phi\), well below the classicalization scale \(\Lambda_\ast\).
At such energies, the kinetic self-coupling responsible for classicalization has not yet become operative to UV complete the theory.} the new scale \(\Lambda_\phi \equiv 1/\ell_\phi \sim \Lambda_\ast\).
This example of $V(\phi)$ meets the 2 criteria outlined previously, with the following asymptotic limits (see Fig.~\ref{plot_symmetric}):
\begin{equation}
\Lambda_\phi^4 \tanh \left[\left( \frac{\phi}{\Lambda_\phi} \right)^4\right] \simeq
\begin{cases}
\phi^4 & \text{for } |\phi| \ll \Lambda_\phi, \\
\Lambda_\phi^4 & \text{for } |\phi| \gg \Lambda_\phi.
\end{cases}
\end{equation}
The second condition guarantees that the mass term governs the potential within the Vainshtein core.

Light scalar fields whose effective potential exhibits strong environmental dependence---commonly referred to as `chameleons'---are well-documented in the dark energy literature~\cite{Khoury:2003aq,Khoury:2003rn}.
In these models, the potential is effectively modified in dense environments, resulting in a large effective mass for the field in regions where its amplitude $|\phi(x)|$ exceeds a critical threshold\footnote{Another approach involves achieving an effectively suppressed coupling to matter: see, e.g., the dilaton screening \cite{Damour:1994zq, Brax:2011ja} or the `symmetron' mechanism \cite{Olive:2007aj, Hinterbichler:2010es, Hinterbichler:2011ca}.} (a phenomenon known as `chameleon screening').
Chameleon fields can be incorporated into a k-essence model, a hybrid framework referred to as `k-chameleons'~\cite{Wei:2004rw}.

\begin{figure}[t]
\begin{center}
\includegraphics[height=7.8cm]{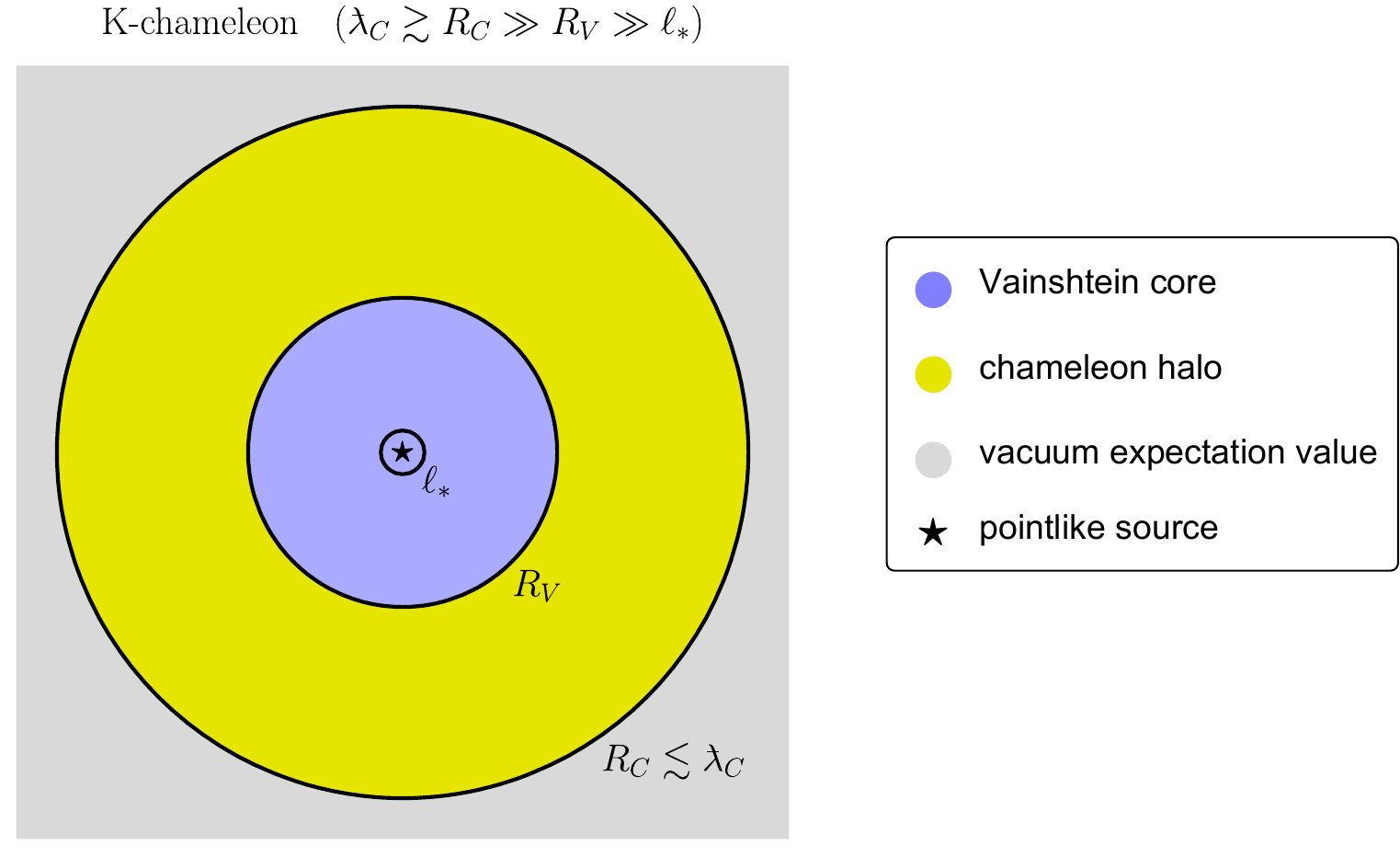}
\end{center}
\caption{Schematic representation of the distinct screening regions surrounding a pointlike source for a k-chameleon, which interacts through both kinetic and potential self-interactions.
The Compton wavelength, \(\lambdabar_C\), defines the maximum range of these self-interactions, beyond which a classicalon cannot extend.
The Vainshtein radius, \(R_V\), and the chameleon radius, \(R_C\), consistently satisfy \(R_C \gtrsim R_V\).
The fundamental length scale, \(\ell_\ast\), remains the smallest in the system.
When \(R_C \ll \lambdabar_C\), both the Vainshtein and chameleon screening regions expand with the source magnitude \(\mathcal{E} \gg \Lambda_\ast\).
The chameleon halo continues to grow until it occupies the entire sphere of radius \(\lambdabar_C\).
However, when \(R_C \sim R_V \sim \lambdabar_C\), both the chameleon and Vainshtein screening effects fill the entire sphere of radius \(\lambdabar_C\) and cannot extend further (see Panel~$(c)$ of Fig.~\ref{circles1}). For radii \(r \gg R_C\), the classical background of the field is determined by the vacuum expectation value of its potential.}
\label{circles2}
\end{figure}

In the present framework, Vainshtein screening leads to large field values, \(|\overline{\phi}(r)| \gg \Lambda_\phi\), such that the self-interactions in the scalar field potential are strongly suppressed within the Vainshtein core---a phenomenon we continue to term `chameleon screening'.
Notably, the region in which this new screening mechanism operates extends beyond the Vainshtein core.
Specifically, the departure from a quartic potential becomes significant when \(|\overline{\phi}(r)| \sim \Lambda_\phi\), which occurs at the `chameleon radius':
\begin{equation}
R_C \equiv \ell_\phi \left( \frac{R_V}{\ell_\ast} \right)^2 = \ell_\phi \left( \frac{\mathcal{E}}{\Omega \Lambda_\ast} \right) \gg R_V \gg \ell_\ast \sim \ell_\phi,
\end{equation}
if $\mathcal{E} \gg \Lambda_\ast$ and $R_C \ll \lambdabar_C$, as derived from the background solution in Eq.~\eqref{phi-background}. When \(R_C \to \lambdabar_C\), its growth ceases.
The internal structure of the classicalon is illustrated in Fig.~\ref{circles2} and exhibits the following features:
\begin{itemize}[label=$\spadesuit$]
\item \(R_V\) defines the boundary of the Vainshtein core. Within this region, kinetic self-interactions dominate the dynamics;
\item \(R_C\) marks the outer limit of the region where chameleon screening takes effect (designated as the `chameleon halo', as it envelops the Vainshtein core);
\item these screening regions are restricted to length scales not exceeding \(\lambdabar_C\) due to the effects of mass screening.
\end{itemize}

\subsubsection{Tachyon Condensation}
\label{tachyon_cond}
\begin{figure}[t]
\begin{center}
\includegraphics[height=7.1cm]{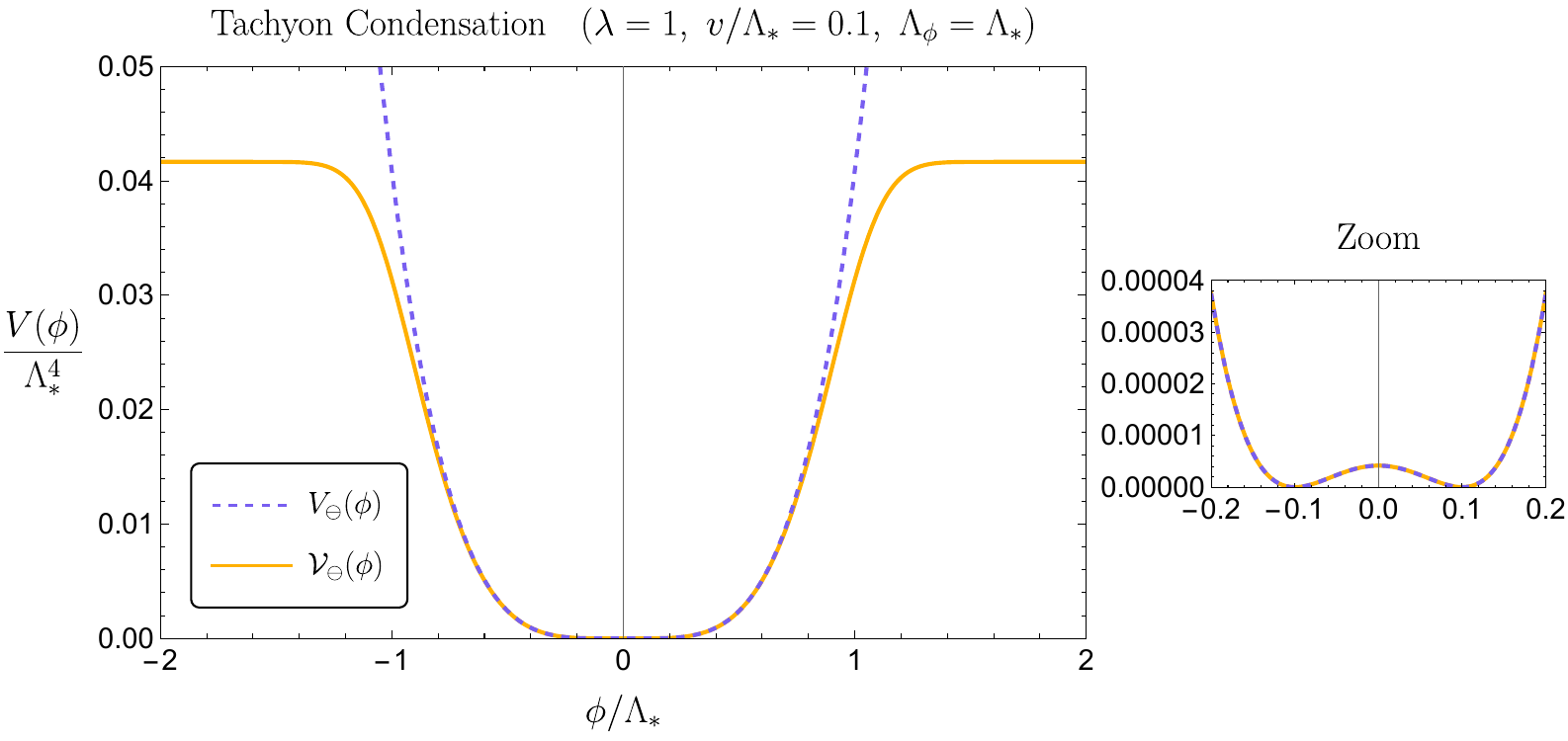}
\end{center}
\caption{Comparison of the scalar potentials \(V_\ominus(\phi)\) in Eq.~\eqref{V_minus} and \(\mathcal{V}_\ominus(\phi)\) in Eq.~\eqref{V_curl_minus} with tachyon condensation.
The 2 potentials begin to deviate at \(|\phi| \sim \Lambda_\phi = \Lambda_\ast\), the threshold at which chameleon screening becomes effective.}
\label{plot_tachyon}
\end{figure}

The canonical example of the SSB of the \(\mathbb{Z}_2\) symmetry through tachyon condensation is provided by the following renormalizable potential:
\begin{align}
V_\ominus (\phi) &\equiv \frac{\lambda}{4!} \left( \phi^2 - v^2 \right)^2,
\label{V_minus} \\
&= - \frac{1}{2} \left( \frac{\lambda v^2}{6} \right) \phi^2 + \frac{\lambda}{4!} \, \phi^4 + \frac{\lambda v^4}{24},
\end{align}
where $\lambda > 0$, and $v$ is the VEV of $\phi(x)$.
As discussed in Section~\ref{vanishing_VEV}, a quartic term must be modified to accommodate Vainshtein screening.
However, since the quadratic term is tachyonic, it cannot dominate for \(|\phi| \gg \Lambda_\phi\) without compromising the stability of the system.
To address these challenges, we consider the following potential:
\begin{equation}
V(\phi) = \mathcal{V}_\ominus \equiv \frac{\lambda}{4!} \, \Lambda_\phi^4 \tanh \left[ \dfrac{\left( \phi^2 - v^2 \right)^2}{\Lambda_\phi^4} \right],
\label{V_curl_minus}
\end{equation}
with the Dirac natural choice \(\Lambda_\phi \sim \Lambda_\ast\).
This potential satisfies the asymptotic behavior (see Fig.~\ref{plot_tachyon}):
\begin{equation}
\mathcal{V}_\ominus \simeq
\begin{cases}
V_\ominus(\phi) & \text{for } |\phi| \ll \Lambda_\phi, \\
\frac{\lambda}{4!} \Lambda_\phi^4 & \text{for } |\phi| \gg \Lambda_\phi.
\end{cases}
\end{equation}
A notable feature of this model is the restoration of shift symmetry for large field values, \(|\phi| \gg \Lambda_\phi\), as the potential approaches a constant in this regime.

In this scenario, \(R_C\) can grow without being constrained by \(\lambdabar_C\), the Compton wavelength of fluctuations around the asymptotic VEV, $v$, since the quadratic term of \(V(\phi)\) is also subject to chameleon screening.
For \(r \ll R_C\), the dynamics is entirely governed by terms that respect the shift symmetry, which is thus restored within both the Vainshtein core and the chameleon halo.
Although mass screening no longer influences the development of nonlinearities, the requirement of a little hierarchy, \(\sqrt{\lambda} v \ll \Lambda_\ast\), remains essential.
This condition ensures the existence of a reliable field theory with a light boson for \(r \gg R_C\), thereby enabling the definition of a semiclassical regime in that region.

\subsubsection{Radiative Stability and Classicalization}
\paragraph{Quantum Stability:}
The models presented in Sections~\ref{vanishing_VEV} and \ref{tachyon_cond} feature potentials with a specific asymptotic form for \(|\phi(x)| \gg \Lambda_\phi\):
\begin{equation}
\tanh \left[ \left( \frac{\phi}{\Lambda_\phi} \right)^4 \right]
\underset{|\phi| \gg \Lambda_\phi}{\simeq}
1 - 2 e^{-2 \left(\frac{\phi}{\Lambda_\phi} \right)^4},
\end{equation}
which must remain stable under quantum fluctuations of the background, \(\delta \phi(x) \equiv \phi(x) - \overline{\phi}(r)\).
To demonstrate this stability, we observe that this asymptotic form can be expressed as
\begin{equation}
e^{-2 \left(\frac{\overline{\phi} + \delta \phi}{\Lambda_\phi} \right)^4}
\simeq
e^{-2 \left(\frac{\overline{\phi}}{\Lambda_\phi} \right)^4}
 \sum_{n=0}^{N_0} \left( \frac{\delta \phi}{\overline{\phi}} \right)^n P_n \left( \frac{\overline{\phi}}{\Lambda_\phi} \right)
+ o \left[ \left( \frac{\delta \phi}{\overline{\phi}} \right)^{N_0} \right],
\label{Taylor_phi}
\end{equation}
with polynomials\footnote{We emphasize that the index $n$ should not be confused with the degree of the polynomial $P_n(z)$.} $P_n(z)$.
For \(r \ll R_C\), the background field satisfies \(|\overline{\phi}(r)| \gg \Lambda_\phi\), ensuring that all the $\delta \phi(x)$ interaction terms arising from the potential are suppressed by the exponential prefactor.
This suppression persists even after field strength renormalization, \(\delta \phi(x) \to \delta \phi_Z(x)\), as defined in Eq.~\eqref{L_2}.

From this discussion, it is evident\footnote{For \(|\overline{\phi}| \gg \Lambda_\phi\), the exponential factor in Eq.~\eqref{Taylor_phi} always dominates over both polynomial and logarithmic factors of \(|\overline{\phi}| / \Lambda_\phi\).} that radiative corrections cannot alter the form of the potential for \(|\phi| \gg \Lambda_\phi\) in a manner that would induce vacuum instability\footnote{
It is important to note, in general, that fluctuations of the background may exhibit tachyonic instabilities.
Instead of ghosts, this is not pathological \cite{Joyce:2014kja}: classicalon solutions do not correspond to stable field configurations; instead, they decay into a large number of soft quanta.
} (see Ref.~\cite{Devoto:2022qen} for a review on false vacuum decay).
The combination of Vainshtein and chameleon screenings can therefore serve as a stabilization mechanism for the vacuum in theories involving scalar fields\footnote{For discussions on vacuum stability in another class of nonlocal theories, see Refs.~\cite{Ghoshal:2017egr,Ghoshal:2022mnj}.}.

\paragraph{Classicalization:}
We have examined the conditions under which a classicalon solution can form in the presence of a pointlike external source.
In a scattering process, the self-interactions of the field serve as the effective source.
As demonstrated in Ref.~\cite{Dvali:2010ns}, only self-interaction terms involving derivatives give rise to classicalization, in contrast to those arising from a potential. The origin lies in the underlying Vainshtein screening mechanism, which emerges at a substantial distance from the source and thus necessitates derivative interactions~\cite{Joyce:2014kja}.
As a result, the classicalization radius \(R_\circledast\) for the k-chameleon---below which the k-chameleon field loses the ability to resolve smaller length scales---continues to be governed by the kinetic self-interaction and is defined by Eq.~\eqref{C_rad_1}.

\paragraph{Hierarchy Problem:}
Upon integrating out the fuzzyons at the scale \(\Lambda_\ast\), along with the infinite tower of classicalons, one may question the implications for the hierarchy problem with respect to the mass $m$ of the fluctuations around the VEV of the k-chameleon field.
Due to the exponential suppression \(\sim e^{-N_\circledast}\) of an effective vertex involving a classicalon $(N_\circledast \gg 1)$ and $\mathcal{O}(1)$ particles---as discussed in Section~\ref{nutshell}---the contributions of classicalons to the threshold corrections \(\delta m^2\) to $m^2$ are likewise exponentially suppressed.
The radiative corrections \(\delta m^2\) are therefore dominated by the heavier fuzzyons---lying at the scale \(\Lambda_\mathrm{fuzz}\)---since \(N_\circledast \sim 1\) for them. The scale \(\Lambda_\mathrm{fuzz}\) can be defined as the mass \(M_\circledast\) for which \(R_\circledast = \ell_\ast\) in Eq.~\eqref{C_rad_1}, yielding \(\Lambda_\mathrm{fuzz} = \Omega \Lambda_\ast\).
This relationship leads to the following scaling behavior:
\begin{equation}
\delta m^2 \sim \left( \frac{\Lambda_\mathrm{fuzz}}{\Omega} \right)^2 = \Lambda_\ast^2.
\end{equation}
From naturalness alone, this suggests that \(m \sim \Lambda_\ast\).
However, classicalization unitarizes scattering amplitudes ($\sqrt{s} \gg \Lambda_\ast$, $b \sim 1/\sqrt{s}$) by converting the exchange of a few hard bosons (strong coupling) into the production of \(N_\circledast \gg 1\) soft bosons (weak coupling).
Consequently, the k-chameleon bosons must be sufficiently light to be described within the weakly interacting EFT below the scale $\Lambda_\ast$, thereby satisfying the little hierarchy \(m \ll \Lambda_\ast\). 
This result shows how UV/IR mixing is realized in the language of the \(S\)-matrix.

\subsection{Interaction with Matter}
\label{interactions_matter}
In this section, we address the coupling of a k-chameleon boson to matter, with a focus on its interaction with a fermion field.
Given the prevalence of Yukawa couplings in particle physics, our discussion will center on this specific interaction. The objective is to determine how such an interaction with matter can maintain the integrity of Vainshtein screening.

\subsubsection{Yukawa Coupling vs Vainshtein Screening}
\label{Yukawa_coupling}
Consider a model featuring 1 real scalar field \(\phi(x)\) and 1 Dirac fermion \(\Psi(x)\), described by the Lagrangian density \( \mathcal{L}_X - V(\phi) + \mathcal{L}_\Psi\), where
\begin{equation}
\mathcal{L}_\Psi =
\frac{i}{2} \, \overline{\Psi} \gamma^\mu \overleftrightarrow{\partial_\mu} \Psi
- y \left[ (\Psi_R^\dagger \cdot \phi) \Psi_L + \Psi_L^\dagger (\phi \cdot \Psi_R) \right],
\end{equation}
with a real Yukawa coupling $y$.
Here, the Dirac matrices \(\gamma^\mu\) are expressed in a chiral basis.
Motivated by a model with a \(\mathbb{Z}_2\) symmetry---whether spontaneously broken by the potential \(V(\phi)\) or not---, we assign distinct parities to the Weyl spinors:
\begin{equation}
\mathbb{Z}_2: \quad
\phi(x) \mapsto - \phi(x), \quad
\Psi_L(x) \mapsto + \Psi_L(x), \quad
\Psi_R(x) \mapsto - \Psi_R(x).
\end{equation}
This assignment ensures that the Yukawa interactions in \(\mathcal{L}_\Psi\) are the only terms invariant under the \(\mathbb{Z}_2\) symmetry, which introduces an additional contribution to the breaking of shift symmetry, already violated by \(V(\phi)\).

By selecting a potential \(V(\phi)\) compatible with Vainshtein screening, the background solution $\overline{\phi}(r)$ for \(r \ll R_C\) is provided by Eq.~\eqref{background_sol}, and we add the perturbations \(\delta \phi(x) \equiv \phi(x) - \overline{\phi}(r_0)\).
The fermionic Lagrangian density is then written as
\begin{equation}
\mathcal{L}_\Psi =
\frac{i}{2} \, \overline{\Psi} \gamma^\mu \overleftrightarrow{\partial_\mu} \Psi
-  \overline{M_\Psi}(r_0) \, \overline{\Psi}\Psi
-  \overline{y}(r_0) \, \delta \phi_Z \overline{\Psi}\Psi,
\label{L_eff_coupl}
\end{equation}
where \(\delta \phi_Z\) is the renormalized field defined in Eq.~\eqref{Z_phi}, and
\begin{equation}
\overline{M_\Psi}(r_0) \equiv y \, \overline{\phi}(r_0),
\quad
\overline{y}(r_0) \equiv \frac{y}{\sqrt{Z_\phi(r_0)}}.
\label{eff_coupl}
\end{equation}
As a result, the fermion acquires an effective mass \(\overline{M_\Psi}(r_0)\) within the classicalon, which increases as one approaches the source $(r_0 \to 0)$, while the effective Yukawa coupling \(\overline{y}(r_0)\) decreases.
A natural question arises regarding the impact of the Yukawa coupling on Vainshtein screening.
This issue becomes particularly clear when examining the threshold corrections to the mass of the perturbations \(\delta \phi_Z(x)\)---after field strength renormalization in Eq.~\eqref{L_2}---through fermion loop corrections inside the Vainshtein core:
\begin{equation}
\delta \overline{m}^2 \sim \left( \dfrac{\overline{y}}{\Omega} \,  \overline{M_\Psi} \right)^2
\quad \Longrightarrow \quad
\frac{\delta  \overline{m}^2(r_0)}{\Lambda_\ast^2}
\sim \frac{y^4}{\Omega^2} \left( \frac{r_0}{\ell_\ast} \right)^2 \gg 1,
\label{delta_m_eff}
\end{equation}
for a moderately weak Yukawa coupling, and $R_V \gg r_0 \gg \ell_\ast$.
The fluctuation mass consequently receives substantial quantum corrections from the Yukawa coupling, which destabilizes the classicalon solution. The issue becomes even more pronounced within the chameleon halo, where \(Z_\phi \sim 1\).
Thus, the model requires modification.

\subsubsection{Conformal-like Coupling and Chameleon Screening}
\label{conformal_coupling}
The first modification to the model of Section~\ref{Yukawa_coupling} introduces a conformal-like coupling between the k-chameleon and the fermion\footnote{For conformal couplings in the context of chameleon dark energy, see, e.g., Refs.~\cite{Khoury:2003aq,Khoury:2003rn}. Here, we employ the term ``conformal-like coupling'' by analogy with the (true) conformal coupling of chameleon fields to fermions in modified gravity \cite{Barreira:2014gwa}.}. The Lagrangian density \(\mathcal{L}_\Psi\) is accordingly modified to\footnote{For simplicity, we adopt an identical conformal-like coupling for both chiralities.}
\begin{equation}
\mathcal{L}_\Psi =
\frac{i}{2} \, e^{\left(\frac{\phi}{\Lambda_\Psi}\right)^2} \overline{\Psi} \gamma^\mu \overleftrightarrow{\partial_\mu} \Psi
- y \, \phi \overline{\Psi}\Psi,
\end{equation}
with the scale \(\Lambda_\Psi \equiv 1/\ell_\Psi \sim \Lambda_\ast\), which is Dirac natural. A hierarchy \(\Lambda_\Psi \ll \Lambda_\ast\) poses a strong-coupling problem for the EFT at the scale $\Lambda_\Psi$, since classicalization does not set in until \(\Lambda_\ast\), the scale needed to UV complete the theory.

To elucidate the effect of the conformal-like factor on the Yukawa coupling, we perform the field redefinition:
\begin{equation}
\Psi(x) \mapsto e^{-\frac{1}{2}\left(\frac{\phi}{\Lambda_\Psi}\right)^2} \Psi(x).
\end{equation}
Under this transformation, \(\mathcal{L}_\Psi\) becomes\footnote{Note that if \(\Psi(x)\) transforms nontrivially under a gauge group, the partial derivative \(\partial_\mu\) is replaced by a covariant derivative without modifying the discussion, as the gauge connection commutes with the conformal-like factor.}
\begin{equation}
\mathcal{L}_\Psi =
\frac{i}{2} \, \overline{\Psi} \gamma^\mu \overleftrightarrow{\partial_\mu} \Psi
- y_C[\phi] \, \phi \overline{\Psi}\Psi,
\end{equation}
where the Yukawa coupling acquires an effective \(\phi\)-dependence given by
\begin{equation}
y_C[\phi] \equiv y \, e^{-\left(\frac{\phi}{\Lambda_\Psi}\right)^2}.
\end{equation}
Consequently, a background with \(|\phi| \gg \Lambda_\Psi\) leads to a suppressed coupling \(y_C[\phi]\), reflecting another manifestation of the underlying chameleon screening mechanism. Like \(V(\phi)\), the Yukawa interaction serves as another example of a term that breaks the shift symmetry---otherwise preserved by the UV screener responsible for classicalization---and hinders Vainshtein screening, thereby requiring suppression via chameleon screening.

By analyzing the background fluctuations \(\delta \phi(x) \equiv \phi(x) - \overline{\phi}(r_0)\) using the classicalon solution~\eqref{background_sol}, we obtain the action~\eqref{L_eff_coupl} (retaining only the linear term in \(\delta \phi\)), but with the substitution \(y \mapsto y_C [\overline{\phi}(r_0)]\) in the effective parameters defined in~\eqref{eff_coupl}.
Within the Vainshtein core (\(r_0 \ll R_V\)), the radiative corrections in Eq.~\eqref{delta_m_eff} are then modified to
\begin{equation}
\frac{\delta \overline{m}^2(r_0)}{\Lambda_\ast^2}
\sim \frac{y^4}{\Omega^2} \, e^{-4\left[ \frac{\overline{\phi}(r_0)}{\Lambda_\Psi} \right]^2} \left( \frac{r_0}{\ell_\ast} \right)^2 \ll 1.
\end{equation}
Notably, this exponential suppression extends to the chameleon halo, where \(Z_\phi \sim 1\), since chameleon screening dominates over Vainshtein screening in suppressing the Yukawa coupling. Regarding the nonlinear term in $\delta \phi$ arising from the Taylor expansion of the conformal-like factor within $y_C[\phi]$, the resulting formula closely mirrors that of Eq.~\eqref{Taylor_phi}.
This demonstrates that all couplings inherit the exponential suppression both inside the Vainshtein core and the chameleon halo.
Thus, the classicalon solution remains protected from quantum corrections arising from the fermion coupled to the k-chameleon.

\subsubsection{Kinetic Coupling and Classicalization}
A persistent challenge arises when probing distances smaller than the scale \(\ell_\Psi \sim \ell_\ast\) by localizing a fermion wavepacket constructed from the field \(\Psi(x)\).
Since the Yukawa terms, modulated by the conformal-like factor, lack derivative interactions of both fields, they fail to initiate classicalization at the scale \(\Lambda_\Psi \sim \Lambda_\ast\) and thus cannot self-complete the theory. The second modification to the initial Yukawa theory of Section~\ref{Yukawa_coupling} is therefore to introduce a kinetic coupling between \(\Psi(x)\) and \(\phi(x)\):
\begin{equation}
\mathcal{L}_\Psi =
\frac{i}{2} \, \overline{\Psi} \gamma^\mu \overleftrightarrow{\partial_\mu} \Psi
- y \, e^{-\left(\frac{\phi}{\Lambda_\Psi}\right)^2} \phi \overline{\Psi}\Psi
+ c_2^\prime \left( \frac{\partial^\mu \phi \partial_\mu \phi}{2 \Lambda_\ast^2} \right) \left( \frac{i \overline{\Psi} \gamma^\mu \overleftrightarrow{\partial_\mu} \Psi}{2 \Lambda_\ast^{\prime \, 2}} \right) ,
\label{L_deriv_Psi}
\end{equation}
with a new scale \(\Lambda_\ast^\prime \sim \Lambda_\Psi \sim \Lambda_\ast\) and \(c_2^\prime \equiv \pm 1\). Note that the new term preserves the shift symmetry of the original k-essence model.

\paragraph{Pointlike Source:}
We reconsider the scenario in which \(\phi(x)\) is directly coupled to an external pointlike source. Evidently, the final term of the Lagrangian density \eqref{L_deriv_Psi} is dominated by the kinetic self-coupling of \(\phi(x)\) within the Vainshtein core\footnote{Outside the Vainshtein core, the $\phi$--$\Psi$ kinetic interaction term is suppressed relative to the kinetic term of \(\phi(x)\) by $\Lambda_\ast^2 \Lambda_\ast^{\prime \, 2}$.}, analogous to the kinetic term of \(\phi(x)\) itself. Consequently, the $\phi$--$\Psi$ kinetic interaction is treated perturbatively. Following the procedure outlined in Section~\ref{radiative_stability}, we focus on a point \(\mathbf{r_0}\) inside the classicalon but distant from the source. The effective kinetic term of \(\Psi(x)\) then becomes
\begin{equation}
\frac{i Z_\Psi}{2} \, \overline{\Psi} \gamma^\mu \overleftrightarrow{\partial_\mu} \Psi
\quad \Longrightarrow \quad
\Psi_Z(x) \equiv \sqrt{Z_\Psi(r_0)} \, \Psi(x),
\end{equation}
where, inside the Vainshtein core,
\begin{equation}
Z_\Psi(r_0) \equiv 1 - c_2^\prime \left[ \frac{\overline{\phi}^\prime(r_0)}{\sqrt{2} \Lambda_\ast \Lambda_\ast^\prime} \right]^2
\underset{|\overline{\phi}^\prime| \gg \Lambda_\ast^2}{\simeq}
- c_2^\prime \left[ \frac{\overline{\phi}^\prime(r_0)}{\sqrt{2} \Lambda_\ast \Lambda_\ast^\prime} \right]^2
\quad \Longrightarrow \quad
c_2^\prime = -1.
\end{equation}
Here, we have performed the field strength renormalization of \(\Psi(x)\), as its kinetic term is renormalized by the background field \(\overline{\phi}(r_0)\) in Eq.~\eqref{background_sol}.

The Yukawa coupling modulated by the conformal-like factor---second term of Eq.~\eqref{L_deriv_Psi}---is treated as described in Section~\ref{conformal_coupling}, but with the definition of \(\overline{y}(r_0)\) in Eq.~\eqref{eff_coupl} replaced by
\begin{equation}
\overline{y}(r_0) \equiv \frac{y_C\left[ \overline{\phi}(r_0) \right]}{Z_\Psi(r_0) \sqrt{Z_\phi(r_0)}},
\end{equation}
with the same qualitative physical conclusions.

The $\phi$--$\Psi$ kinetic interaction term introduces couplings between the background fluctuations \(\delta \phi(x)\) and \(\Psi(x)\). This yields the following cubic and quartic interactions:
\begin{equation}
\partial_r \delta\phi_Z \left( \frac{i \overline{\Psi_Z} \gamma^\mu \overleftrightarrow{\partial_\mu} \Psi_Z}{2 \overline{\Lambda_\ast^\prime}^2} \right)
- \left( \frac{\partial^\mu \delta\phi_Z \partial_\mu \delta\phi_Z}{2 \overline{\Lambda_\ast}^2} \right) \left( \frac{i \overline{\Psi_Z} \gamma^\mu \overleftrightarrow{\partial_\mu} \Psi_Z}{2 \overline{\Lambda_\ast^\prime}^2} \right),
\end{equation}
where the blueshifted interaction scales are defined as
\begin{equation}
\overline{\Lambda_\ast} \equiv \sqrt{Z_\phi(r_0)} \, \Lambda_\ast \gg \Lambda_\ast, \quad
\overline{\Lambda_\ast^\prime} \equiv \sqrt{Z_\Psi(r_0)} \, \Lambda_\ast^\prime \gg \Lambda_\ast^\prime.
\end{equation}
Thus, the quantum version of Vainshtein screening extends to the $\phi$--$\Psi$ kinetic interaction, and the corresponding radiative corrections are suppressed accordingly inside the Vainshtein core.

\paragraph{Classicalization:}
With regard to the nonperturbative unitarization of scattering amplitudes at $\sqrt{s} \gg \Lambda_\ast \sim \Lambda_\ast' \sim \Lambda_\Psi$ and $b \sim 1/\sqrt{s}$ via self-sourcing terms, the fermion field $\Psi(x)$ now also contributes to the classicalization dynamics through the $\phi$--$\Psi$ kinetic coupling. By restoring \(\hbar\)-units, one can perform a dimensional analysis\footnote{The dimensionality of the fermion field is $[\Psi] = E^\frac{1}{2} L^{-1}$.} analogous to that in Section~\ref{class_kin_self}, yielding a classicalization radius for both derivative couplings at the same length scale \(R_\circledast\), as defined in Eq.~\eqref{C_rad_1}. This scale thus governs the classicalization dynamics of both the \(\phi(x)\) and \(\Psi(x)\) fields.

Phenomenologically, one anticipates a strongly-coupled fuzzyon dynamics between the 2 fields at the shared scale \(\Lambda_\ast \sim \Lambda_\ast' \sim \Lambda_\Psi\), while the semiclassical regime is expected to exhibit the evaporation of classicalons into \(N_\circledast \gg 1\) k-chameleon bosons and fermion-antifermion pairs. A comprehensive description of the classicalization dynamics would necessitate a dedicated study to capture the nonperturbative collective behavior.

\section{Conclusion and Outlook}
\label{conclusion}
In Section~\ref{K-essence_Model}, we reviewed the classicalization paradigm using the concrete example of a massless k-essence field.
The kinetic self-interaction is widely recognized in the dark energy literature for exhibiting Vainshtein screening around localized sources.
Our objective was to reinterpret these features within the framework of self-UV completion through classicalization and UV/IR mixing. We also summarized the critiques of classicalization in the literature, along with their associated loopholes.

In Section~\ref{from_kessence_kchameleon}, we have explored the conditions under which Vainshtein screening---triggered by the kinetic self-coupling of a gauge singlet scalar---can arise in the presence of either a scalar potential or a Yukawa coupling to matter.
These conditions are critical for the theory to achieve self-UV completion via classicalization.
In addition to the well-established violation of the standard positivity bound for the coefficient of the kinetic self-interaction term, we have shown that a chameleon screening mechanism must be integrated within the classicalon.
This ensures that Vainshtein screening remains robust against perturbations from the potential and the coupling to matter.

This framework offers a fresh perspective on the existence of light scalar bosons, when they appear to suffer from a little hierarchy problem:
the UV/IR mixing intrinsic to the classicalization phenomenon requires such a hierarchy to facilitate a consistent self-UV completion of the theory. Nevertheless, it is important to emphasize that our specific choices of UV screeners---including the scalar potentials and conformal-like couplings---serve merely as illustrative examples of the underlying screening mechanisms.
Exploring alternative possibilities in this direction therefore remains a valuable endeavor.

Several avenues exist for future research in multiple directions, and we outline here a nonexhaustive list of possibilities.
From a theoretical standpoint, potential extensions of this work include:
\begin{itemize}[label=$\spadesuit$]
\item An investigation of the perturbative renormalization of the theory in the presence of a classicalon background.
\item A generalization of our analysis to other UV-screening mechanisms, such as those found in Galileon theories~\cite{Nicolis:2004qq,Nicolis:2008in} or `D-BIon' fields~\cite{Burrage:2014uwa}.
\item The development of a theoretical framework to explore the strongly-coupled fuzzyon regime near the scale \(\Lambda_\ast\).
\end{itemize}
Additionally, classicalization offers several promising applications for concrete model-building in phenomenology, such as:
\begin{itemize}[label=$\spadesuit$]
\item Screening mechanisms were originally proposed to conceal new long-range forces in scalar-tensor theories of modified gravity \cite{Joyce:2014kja,Langlois:2018dxi}; thus, exploring the implications of classicalization in this context represents a natural extension of this work.
\item Extending our work to scalar fields that transform nontrivially under gauge groups would allow the Higgs mechanism to be embedded within a classicalization framework. This approach could offer an elegant resolution to hierarchy problems, such as those encountered in the standard electroweak theory \cite{Hebecker:2020aqr} and in dark Higgs sectors \cite{Cirelli:2024ssz}.
\item The synergy between Vainshtein and chameleon screenings provide a highly effective stabilization mechanism for the shape of a classical potential under quantum corrections. This capability could preserve the plateau of the inflaton potential \cite{Martin:2013tda} within an EFT of inflation~\cite{Cheung:2007st,Weinberg:2008hq,Burgess:2009ea}. Additionally, the potential metastability of the Higgs potential~\cite{Devoto:2022qen} presents another compelling application for this mechanism.
\end{itemize}

As a final comment, we emphasize that classicalization at a scale \(\Lambda_\ast \ll \Lambda_G\) is not expected to be embeddable within string theory (with the gravitational scale\footnote{In scenarios involving extra spatial dimensions, the gravitational scale \(\Lambda_G\) can be significantly lower than the Planck scale \(\Lambda_P\) (see, e.g., Ref.~\cite{Arkani-Hamed:1998sfv}).
} $\Lambda_G \lesssim \Lambda_P$).
Rather, it should be regarded as a hypothetical alternative framework in which a QFT may admit multiple classicalization scales, with \(\Lambda_G\) serving as the ultimate such scale.
While string theory is anticipated to exhibit asymptotic darkness—classicalizing above the scale \(\Lambda_G\) with a spectrum of black hole states~\cite{Aharony:1998tt}—it is also expected to function as a Wilsonian UV completion between the string scale and \(\Lambda_G\)~\cite{Dvali:2010jz,Herraez:2025clp}.
Thus, string theory represents a hybrid UV completion, combining elements of both the Wilsonian approach and self-UV completion, yet it features only a single classicalization scale: \(\Lambda_G\).

\acknowledgments
The author is deeply grateful to E.~Berteaux, R.T.~D'Agnolo, L.~Darm\'e, A.~Deandrea, M.~Ettengruber, A.M.~Iyer, and F.N.~Mahmoudi for stimulating discussions on various aspects of this work. Special thanks go to B.~Bellazzini and G.~Isabella for insightful exchanges on positivity bounds, and to F.~Br\"ummer and G.~Cacciapaglia for pointing out the parallels between classicalization and Higgsplosion. The author also thanks T.~Berthelot for his work during the internship. The author acknowledges the use of \texttt{Wolfram Mathematica} for figure preparation and of \texttt{Mistral AI} for assistance with English editing during drafting.

\bibliographystyle{JHEP}
\input{ms.bbl}

\end{document}

%% file: ms.bbl
\providecommand{\href}[2]{#2}\begingroup\raggedright\endgroup

%% file: ms.bbl
\begin{thebibliography}{100}

\bibitem{Peskin:1995ev}
M.E.~Peskin and D.V.~Schroeder, \emph{{An Introduction to Quantum Field
  Theory}}, CRC Press (1995),
  \href{https://doi.org/10.1201/9780429503559}{10.1201/9780429503559}.

\bibitem{Cirelli:2024ssz}
M.~Cirelli, A.~Strumia and J.~Zupan, \emph{{Dark matter}},
  \href{https://doi.org/10.21468/SciPostPhysRev.1}{\emph{SciPost Phys. Rev.}
  {\bfseries 1} (2026) } [\href{https://arxiv.org/abs/2406.01705}{{\ttfamily
  2406.01705}}].

\bibitem{Joyce:2014kja}
A.~Joyce, B.~Jain, J.~Khoury and M.~Trodden, \emph{{Beyond the cosmological
  standard model}},
  \href{https://doi.org/10.1016/j.physrep.2014.12.002}{\emph{Phys. Rept.}
  {\bfseries 568} (2015) 1} [\href{https://arxiv.org/abs/1407.0059}{{\ttfamily
  1407.0059}}].

\bibitem{Martin:2013tda}
J.~Martin, C.~Ringeval and V.~Vennin, \emph{{Encyclop{\ae}dia Inflationaris}:
  {Opiparous Edition}},
  \href{https://doi.org/10.1016/j.dark.2024.101653}{\emph{Phys. Dark Univ.}
  {\bfseries 5-6} (2014) 75} [\href{https://arxiv.org/abs/1303.3787}{{\ttfamily
  1303.3787}}].

\bibitem{Burgess:2020tbq}
C.P.~Burgess, \emph{{Introduction to Effective Field Theory: Thinking
  Effectively about Hierarchies of Scale}}, Cambridge University Press (2020),
  \href{https://doi.org/10.1017/9781139048040}{10.1017/9781139048040}.

\bibitem{Hebecker:2020aqr}
A.~Hebecker, \emph{{Naturalness, String Landscape and Multiverse}: {A Modern
  Introduction with Exercises}},
  \href{https://doi.org/10.1007/978-3-030-65151-0}{\emph{Lect. Notes Phys.}
  {\bfseries 979} (2021) 1} [\href{https://arxiv.org/abs/2008.10625}{{\ttfamily
  2008.10625}}].

\bibitem{tHooft:1979rat}
G.~'t~Hooft, \emph{{Naturalness, chiral symmetry, and spontaneous chiral
  symmetry breaking}},
  \href{https://doi.org/10.1007/978-1-4684-7571-5_9}{\emph{NATO Sci. Ser. B}
  {\bfseries 59} (1980) 135}.

\bibitem{Cho:2007cb}
A.~Cho, \emph{{Physicists' Nightmare Scenario: The Higgs and Nothing Else}},
  \href{https://doi.org/10.1126/science.315.5819.1657}{\emph{Science}
  {\bfseries 315} (2007) 1657}.

\bibitem{Giudice:2017pzm}
G.F.~Giudice, \emph{{The Dawn of the Post-Naturalness Era}},  in \emph{{From My
  Vast Repertoire ...}: {Guido Altarelli's Legacy}}, A.~Levy, S.~Forte and
  G.~Ridolfi, eds., pp.~267--292 (2019),
  \href{https://doi.org/10.1142/9789813238053_0013}{DOI}
  [\href{https://arxiv.org/abs/1710.07663}{{\ttfamily 1710.07663}}].

\bibitem{Craig:2020ojv}
N.~Craig, \emph{{Naturalness Hits a Snag with Higgs}},
  \href{https://doi.org/10.1103/Physics.13.174}{\emph{APS Physics} {\bfseries
  13} (2020) 174}.

\bibitem{Craig:2022eqo}
N.~Craig, \emph{{Naturalness: past, present, and future}},
  \href{https://doi.org/10.1140/epjc/s10052-023-11928-7}{\emph{Eur. Phys. J. C}
  {\bfseries 83} (2023) 825}
  [\href{https://arxiv.org/abs/2205.05708}{{\ttfamily 2205.05708}}].

\bibitem{McCullough:2024evr}
M.~McCullough, \emph{{In Pursuit of New Paradigms: TASI 2024}},
  \href{https://arxiv.org/abs/2412.15744}{{\ttfamily 2412.15744}}.

\bibitem{DAgnolo:2022mem}
R.T.~D'Agnolo, \emph{{The Higgs boson mass and cosmology}},  HDR thesis,
  Paris-Saclay University, 2022,
  \href{https://cea.hal.science/tel-04444180}{https://cea.hal.science/tel-04444180}.

\bibitem{Brummer:2023znr}
F.~Br{\"u}mmer, G.~Ferrante, M.~Frigerio and T.~Hambye, \emph{{Accidentally
  light scalars from large representations}},
  \href{https://doi.org/10.1007/JHEP01(2024)075}{\emph{JHEP} {\bfseries 01}
  (2024) 075} [\href{https://arxiv.org/abs/2307.10092}{{\ttfamily
  2307.10092}}].

\bibitem{Brummer:2024ejc}
F.~Br{\"u}mmer, G.~Ferrante and M.~Frigerio, \emph{{Hybrid inflation and
  gravitational waves from accidentally light scalars}},
  \href{https://doi.org/10.1103/PhysRevD.110.103506}{\emph{Phys. Rev. D}
  {\bfseries 110} (2024) 103506}
  [\href{https://arxiv.org/abs/2406.02531}{{\ttfamily 2406.02531}}].

\bibitem{Wilson:1970ag}
K.G.~Wilson, \emph{{Renormalization Group and Strong Interactions}},
  \href{https://doi.org/10.1103/PhysRevD.3.1818}{\emph{Phys. Rev. D} {\bfseries
  3} (1971) 1818}.

\bibitem{Duncan:2012aja}
A.~Duncan, \emph{{The Conceptual Framework of Quantum Field Theory}}, Oxford
  University Press (2012),
  \href{https://doi.org/10.1093/acprof:oso/9780199573264.001.0001}{10.1093/acprof:oso/9780199573264.001.0001}.

\bibitem{Aharony:1998tt}
O.~Aharony and T.~Banks, \emph{{Note on the quantum mechanics of M theory}},
  \href{https://doi.org/10.1088/1126-6708/1999/03/016}{\emph{JHEP} {\bfseries
  03} (1999) 016} [\href{https://arxiv.org/abs/hep-th/9812237}{{\ttfamily
  hep-th/9812237}}].

\bibitem{Giddings:2001pt}
S.B.~Giddings and M.~Lippert, \emph{{Precursors, black holes, and a locality
  bound}}, \href{https://doi.org/10.1103/PhysRevD.65.024006}{\emph{Phys. Rev.
  D} {\bfseries 65} (2002) 024006}
  [\href{https://arxiv.org/abs/hep-th/0103231}{{\ttfamily hep-th/0103231}}].

\bibitem{Giddings:2004ud}
S.B.~Giddings and M.~Lippert, \emph{{The information paradox and the locality
  bound}}, \href{https://doi.org/10.1103/PhysRevD.69.124019}{\emph{Phys. Rev.
  D} {\bfseries 69} (2004) 124019}
  [\href{https://arxiv.org/abs/hep-th/0402073}{{\ttfamily hep-th/0402073}}].

\bibitem{Giddings:2005id}
S.B.~Giddings, D.~Marolf and J.B.~Hartle, \emph{{Observables in effective
  gravity}}, \href{https://doi.org/10.1103/PhysRevD.74.064018}{\emph{Phys. Rev.
  D} {\bfseries 74} (2006) 064018}
  [\href{https://arxiv.org/abs/hep-th/0512200}{{\ttfamily hep-th/0512200}}].

\bibitem{Giddings:2006vu}
S.B.~Giddings, \emph{{Locality in quantum gravity and string theory}},
  \href{https://doi.org/10.1103/PhysRevD.74.106006}{\emph{Phys. Rev. D}
  {\bfseries 74} (2006) 106006}
  [\href{https://arxiv.org/abs/hep-th/0604072}{{\ttfamily hep-th/0604072}}].

\bibitem{Giddings:2006sj}
S.B.~Giddings, \emph{{Black hole information, unitarity, and nonlocality}},
  \href{https://doi.org/10.1103/PhysRevD.74.106005}{\emph{Phys. Rev. D}
  {\bfseries 74} (2006) 106005}
  [\href{https://arxiv.org/abs/hep-th/0605196}{{\ttfamily hep-th/0605196}}].

\bibitem{Giddings:2006be}
S.B.~Giddings, \emph{{(Non)perturbative gravity, nonlocality, and nice
  slices}}, \href{https://doi.org/10.1103/PhysRevD.74.106009}{\emph{Phys. Rev.
  D} {\bfseries 74} (2006) 106009}
  [\href{https://arxiv.org/abs/hep-th/0606146}{{\ttfamily hep-th/0606146}}].

\bibitem{Giddings:2007ie}
S.B.~Giddings, \emph{{Quantization in black hole backgrounds}},
  \href{https://doi.org/10.1103/PhysRevD.76.064027}{\emph{Phys. Rev. D}
  {\bfseries 76} (2007) 064027}
  [\href{https://arxiv.org/abs/hep-th/0703116}{{\ttfamily hep-th/0703116}}].

\bibitem{Giddings:2007pj}
S.B.~Giddings, \emph{{Black holes, information, and locality}},
  \href{https://doi.org/10.1142/S0217732307025923}{\emph{Mod. Phys. Lett. A}
  {\bfseries 22} (2007) 2949}
  [\href{https://arxiv.org/abs/0705.2197}{{\ttfamily 0705.2197}}].

\bibitem{Giddings:2021qas}
S.B.~Giddings, \emph{{A {\textquoteleft}black hole theorem,{\textquoteright}
  and its implications}},
  \href{https://doi.org/10.1088/1361-6382/acbe8b}{\emph{Class. Quant. Grav.}
  {\bfseries 40} (2023) 085002}
  [\href{https://arxiv.org/abs/2110.10690}{{\ttfamily 2110.10690}}].

\bibitem{Buoninfante:2023dyd}
L.~Buoninfante, J.~Tokuda and M.~Yamaguchi, \emph{{New lower bounds on
  scattering amplitudes: non-locality constraints}},
  \href{https://doi.org/10.1007/JHEP01(2024)082}{\emph{JHEP} {\bfseries 01}
  (2024) 082} [\href{https://arxiv.org/abs/2305.16422}{{\ttfamily
  2305.16422}}].

\bibitem{Giddings:2024qcf}
S.B.~Giddings, \emph{{The unitarity crisis, nonviolent unitarization, and
  implications for quantum spacetime}},
  \href{https://arxiv.org/abs/2412.18650}{{\ttfamily 2412.18650}}.

\bibitem{Cohen:1998zx}
A.G.~Cohen, D.B.~Kaplan and A.E.~Nelson, \emph{{Effective Field Theory, Black
  Holes, and the Cosmological Constant}},
  \href{https://doi.org/10.1103/PhysRevLett.82.4971}{\emph{Phys. Rev. Lett.}
  {\bfseries 82} (1999) 4971}
  [\href{https://arxiv.org/abs/hep-th/9803132}{{\ttfamily hep-th/9803132}}].

\bibitem{Dienes:2001se}
K.R.~Dienes, \emph{{Solving the hierarchy problem without supersymmetry or
  extra dimensions: an alternative approach}},
  \href{https://doi.org/10.1016/S0550-3213(01)00344-3}{\emph{Nucl. Phys. B}
  {\bfseries 611} (2001) 146}
  [\href{https://arxiv.org/abs/hep-ph/0104274}{{\ttfamily hep-ph/0104274}}].

\bibitem{Cheung:2014vva}
C.~Cheung and G.N.~Remmen, \emph{{Naturalness and the Weak Gravity
  Conjecture}},
  \href{https://doi.org/10.1103/PhysRevLett.113.051601}{\emph{Phys. Rev. Lett.}
  {\bfseries 113} (2014) 051601}
  [\href{https://arxiv.org/abs/1402.2287}{{\ttfamily 1402.2287}}].

\bibitem{Ibanez:2017kvh}
L.E.~Ibanez, V.~Martin-Lozano and I.~Valenzuela, \emph{{Constraining neutrino
  masses, the cosmological constant and BSM physics from the weak gravity
  conjecture}}, \href{https://doi.org/10.1007/JHEP11(2017)066}{\emph{JHEP}
  {\bfseries 11} (2017) 066}
  [\href{https://arxiv.org/abs/1706.05392}{{\ttfamily 1706.05392}}].

\bibitem{Ibanez:2017oqr}
L.E.~Ibanez, V.~Martin-Lozano and I.~Valenzuela, \emph{{Constraining the EW
  Hierarchy from the Weak Gravity Conjecture}},
  \href{https://arxiv.org/abs/1707.05811}{{\ttfamily 1707.05811}}.

\bibitem{Lust:2017wrl}
D.~Lust and E.~Palti, \emph{{Scalar fields, hierarchical UV/IR mixing and the
  Weak Gravity Conjecture}},
  \href{https://doi.org/10.1007/JHEP02(2018)040}{\emph{JHEP} {\bfseries 02}
  (2018) 040} [\href{https://arxiv.org/abs/1709.01790}{{\ttfamily
  1709.01790}}].

\bibitem{Craig:2019fdy}
N.~Craig, I.~Garcia~Garcia and S.~Koren, \emph{{The weak scale from weak
  gravity}}, \href{https://doi.org/10.1007/JHEP09(2019)081}{\emph{JHEP}
  {\bfseries 09} (2019) 081}
  [\href{https://arxiv.org/abs/1904.08426}{{\ttfamily 1904.08426}}].

\bibitem{Craig:2019zbn}
N.~Craig and S.~Koren, \emph{{IR dynamics from UV divergences: UV/IR mixing,
  NCFT, and the hierarchy problem}},
  \href{https://doi.org/10.1007/JHEP03(2020)037}{\emph{JHEP} {\bfseries 03}
  (2020) 037} [\href{https://arxiv.org/abs/1909.01365}{{\ttfamily
  1909.01365}}].

\bibitem{Castellano:2021mmx}
A.~Castellano, A.~Herr\'aez and L.E.~Ib\'a\~nez, \emph{{IR/UV mixing, towers of
  species and swampland conjectures}},
  \href{https://doi.org/10.1007/JHEP08(2022)217}{\emph{JHEP} {\bfseries 08}
  (2022) 217} [\href{https://arxiv.org/abs/2112.10796}{{\ttfamily
  2112.10796}}].

\bibitem{Cribiori:2025oek}
N.~Cribiori and F.~Tonioni, \emph{{Cosmological constraints from UV/IR
  mixing}}, \href{https://doi.org/10.1007/JHEP02(2026)035}{\emph{JHEP}
  {\bfseries 02} (2026) 035}
  [\href{https://arxiv.org/abs/2507.02738}{{\ttfamily 2507.02738}}].

\bibitem{Abel:2021tyt}
S.~Abel and K.R.~Dienes, \emph{{Calculating the Higgs mass in string theory}},
  \href{https://doi.org/10.1103/PhysRevD.104.126032}{\emph{Phys. Rev. D}
  {\bfseries 104} (2021) 126032}
  [\href{https://arxiv.org/abs/2106.04622}{{\ttfamily 2106.04622}}].

\bibitem{Abel:2023hkk}
S.~Abel, K.R.~Dienes and L.A.~Nutricati, \emph{{Running of gauge couplings in
  string theory}},
  \href{https://doi.org/10.1103/PhysRevD.107.126019}{\emph{Phys. Rev. D}
  {\bfseries 107} (2023) 126019}
  [\href{https://arxiv.org/abs/2303.08534}{{\ttfamily 2303.08534}}].

\bibitem{Abel:2024twz}
S.~Abel, K.R.~Dienes and L.A.~Nutricati, \emph{{New nonrenormalization theorem
  from UV/IR mixing}},
  \href{https://doi.org/10.1103/PhysRevD.110.126021}{\emph{Phys. Rev. D}
  {\bfseries 110} (2024) 126021}
  [\href{https://arxiv.org/abs/2407.11160}{{\ttfamily 2407.11160}}].

\bibitem{Shomer:2007vq}
A.~Shomer, \emph{{A Pedagogical explanation for the non-renormalizability of
  gravity}},  \href{https://arxiv.org/abs/0709.3555}{{\ttfamily 0709.3555}}.

\bibitem{Dvali:2010bf}
G.~Dvali and C.~Gomez, \emph{{Self-Completeness of Einstein Gravity}},
  \href{https://arxiv.org/abs/1005.3497}{{\ttfamily 1005.3497}}.

\bibitem{Dvali:2010ue}
G.~Dvali, S.~Folkerts and C.~Germani, \emph{{Physics of trans-Planckian
  gravity}}, \href{https://doi.org/10.1103/PhysRevD.84.024039}{\emph{Phys. Rev.
  D} {\bfseries 84} (2011) 024039}
  [\href{https://arxiv.org/abs/1006.0984}{{\ttfamily 1006.0984}}].

\bibitem{Hawking:1974rv}
S.W.~Hawking, \emph{{Black hole explosions?}},
  \href{https://doi.org/10.1038/248030a0}{\emph{Nature} {\bfseries 248} (1974)
  30}.

\bibitem{Hawking:1975vcx}
S.W.~Hawking, \emph{{Particle creation by black holes}},
  \href{https://doi.org/10.1007/BF02345020}{\emph{Commun. Math. Phys.}
  {\bfseries 43} (1975) 199}.

\bibitem{Giddings:2009gj}
S.B.~Giddings and R.A.~Porto, \emph{{The gravitational $S$ matrix}},
  \href{https://doi.org/10.1103/PhysRevD.81.025002}{\emph{Phys. Rev. D}
  {\bfseries 81} (2010) 025002}
  [\href{https://arxiv.org/abs/0908.0004}{{\ttfamily 0908.0004}}].

\bibitem{Giddings:2011xs}
S.B.~Giddings, \emph{{The gravitational S-matrix: Erice lectures}},
  \href{https://doi.org/10.1142/9789814522489_0005}{\emph{Subnucl. Ser.}
  {\bfseries 48} (2013) 93} [\href{https://arxiv.org/abs/1105.2036}{{\ttfamily
  1105.2036}}].

\bibitem{Amati:1987wq}
D.~Amati, M.~Ciafaloni and G.~Veneziano, \emph{{Superstring collisions at
  planckian energies}},
  \href{https://doi.org/10.1016/0370-2693(87)90346-7}{\emph{Phys. Lett. B}
  {\bfseries 197} (1987) 81}.

\bibitem{Gross:1987kza}
D.J.~Gross and P.F.~Mende, \emph{{The high-energy behavior of string scattering
  amplitudes}}, \href{https://doi.org/10.1016/0370-2693(87)90355-8}{\emph{Phys.
  Lett. B} {\bfseries 197} (1987) 129}.

\bibitem{tHooft:1987vrq}
G.~'t~Hooft, \emph{{Graviton dominance in ultra-high-energy scattering}},
  \href{https://doi.org/10.1016/0370-2693(87)90159-6}{\emph{Phys. Lett. B}
  {\bfseries 198} (1987) 61}.

\bibitem{Gross:1987ar}
D.J.~Gross and P.F.~Mende, \emph{{String theory beyond the Planck scale}},
  \href{https://doi.org/10.1016/0550-3213(88)90390-2}{\emph{Nucl. Phys. B}
  {\bfseries 303} (1988) 407}.

\bibitem{Amati:1987uf}
D.~Amati, M.~Ciafaloni and G.~Veneziano, \emph{{Classical and Quantum Gravity
  Effects from Planckian Energy Superstring Collisions}},
  \href{https://doi.org/10.1142/S0217751X88000710}{\emph{Int. J. Mod. Phys. A}
  {\bfseries 3} (1988) 1615}.

\bibitem{Amati:1988tn}
D.~Amati, M.~Ciafaloni and G.~Veneziano, \emph{{Can spacetime be probed below
  the string size?}},
  \href{https://doi.org/10.1016/0370-2693(89)91366-X}{\emph{Phys. Lett. B}
  {\bfseries 216} (1989) 41}.

\bibitem{Mende:1989wt}
P.F.~Mende and H.~Ooguri, \emph{{Borel summation of string theory for Planck
  scale scattering}},
  \href{https://doi.org/10.1016/0550-3213(90)90202-O}{\emph{Nucl. Phys. B}
  {\bfseries 339} (1990) 641}.

\bibitem{Amati:1990xe}
D.~Amati, M.~Ciafaloni and G.~Veneziano, \emph{{Higher-order gravitational
  deflection and soft bremsstrahlung in planckian energy superstring
  collisions}}, \href{https://doi.org/10.1016/0550-3213(90)90375-N}{\emph{Nucl.
  Phys. B} {\bfseries 347} (1990) 550}.

\bibitem{Amati:1993tb}
D.~Amati, M.~Ciafaloni and G.~Veneziano, \emph{{Effective action and all-order
  gravitational eikonal at planckian energies}},
  \href{https://doi.org/10.1016/0550-3213(93)90367-X}{\emph{Nucl. Phys. B}
  {\bfseries 403} (1993) 707}.

\bibitem{Banks:1999gd}
T.~Banks and W.~Fischler, \emph{{A Model for high-energy scattering in quantum
  gravity}},  \href{https://arxiv.org/abs/hep-th/9906038}{{\ttfamily
  hep-th/9906038}}.

\bibitem{Eardley:2002re}
D.M.~Eardley and S.B.~Giddings, \emph{{Classical black hole production in
  high-energy collisions}},
  \href{https://doi.org/10.1103/PhysRevD.66.044011}{\emph{Phys. Rev. D}
  {\bfseries 66} (2002) 044011}
  [\href{https://arxiv.org/abs/gr-qc/0201034}{{\ttfamily gr-qc/0201034}}].

\bibitem{Kohlprath:2002yh}
E.~Kohlprath and G.~Veneziano, \emph{{Black holes from high-energy beam-beam
  collisions}},
  \href{https://doi.org/10.1088/1126-6708/2002/06/057}{\emph{JHEP} {\bfseries
  06} (2002) 057} [\href{https://arxiv.org/abs/gr-qc/0203093}{{\ttfamily
  gr-qc/0203093}}].

\bibitem{Giddings:2004xy}
S.B.~Giddings and V.S.~Rychkov, \emph{{Black holes from colliding
  wavepackets}}, \href{https://doi.org/10.1103/PhysRevD.70.104026}{\emph{Phys.
  Rev. D} {\bfseries 70} (2004) 104026}
  [\href{https://arxiv.org/abs/hep-th/0409131}{{\ttfamily hep-th/0409131}}].

\bibitem{Giddings:2007bw}
S.B.~Giddings, D.J.~Gross and A.~Maharana, \emph{{Gravitational effects in
  ultrahigh-energy string scattering}},
  \href{https://doi.org/10.1103/PhysRevD.77.046001}{\emph{Phys. Rev. D}
  {\bfseries 77} (2008) 046001}
  [\href{https://arxiv.org/abs/0705.1816}{{\ttfamily 0705.1816}}].

\bibitem{Giddings:2007qq}
S.B.~Giddings and M.~Srednicki, \emph{{High-energy gravitational scattering and
  black hole resonances}},
  \href{https://doi.org/10.1103/PhysRevD.77.085025}{\emph{Phys. Rev. D}
  {\bfseries 77} (2008) 085025}
  [\href{https://arxiv.org/abs/0711.5012}{{\ttfamily 0711.5012}}].

\bibitem{Dvali:2014ila}
G.~Dvali, C.~Gomez, R.S.~Isermann, D.~L\"ust and S.~Stieberger, \emph{{Black
  hole formation and classicalization in ultra-Planckian 2\textrightarrow{}N
  scattering}},
  \href{https://doi.org/10.1016/j.nuclphysb.2015.02.004}{\emph{Nucl. Phys. B}
  {\bfseries 893} (2015) 187}
  [\href{https://arxiv.org/abs/1409.7405}{{\ttfamily 1409.7405}}].

\bibitem{Addazi:2016ksu}
A.~Addazi, M.~Bianchi and G.~Veneziano, \emph{{Glimpses of black hole
  formation/evaporation in highly inelastic, ultra-planckian string
  collisions}}, \href{https://doi.org/10.1007/JHEP02(2017)111}{\emph{JHEP}
  {\bfseries 02} (2017) 111}
  [\href{https://arxiv.org/abs/1611.03643}{{\ttfamily 1611.03643}}].

\bibitem{Hossenfelder:2012jw}
S.~Hossenfelder, \emph{{Minimal Length Scale Scenarios for Quantum Gravity}},
  \href{https://doi.org/10.12942/lrr-2013-2}{\emph{Living Rev. Rel.} {\bfseries
  16} (2013) 2} [\href{https://arxiv.org/abs/1203.6191}{{\ttfamily
  1203.6191}}].

\bibitem{Marshakov:2002ff}
A.~Marshakov, \emph{{String theory or field theory?}},
  \href{https://doi.org/10.1070/PU2002v045n09ABEH001148}{\emph{Phys. Usp.}
  {\bfseries 45} (2002) 915}
  [\href{https://arxiv.org/abs/hep-th/0212114}{{\ttfamily hep-th/0212114}}].

\bibitem{Boos:2020qgg}
J.~Boos, \emph{{Effects of Non-locality in Gravity and Quantum Theory}},
  Springer Theses, University of Alberta (2020),
  \href{https://doi.org/10.1007/978-3-030-82910-0}{10.1007/978-3-030-82910-0},
  [\href{https://arxiv.org/abs/2009.10856}{{\ttfamily 2009.10856}}].

\bibitem{Donoghue:1994dn}
J.F.~Donoghue, \emph{{General relativity as an effective field theory: The
  leading quantum corrections}},
  \href{https://doi.org/10.1103/PhysRevD.50.3874}{\emph{Phys. Rev. D}
  {\bfseries 50} (1994) 3874}
  [\href{https://arxiv.org/abs/gr-qc/9405057}{{\ttfamily gr-qc/9405057}}].

\bibitem{Burgess:2003jk}
C.P.~Burgess, \emph{{Quantum Gravity in Everyday Life: General Relativity as an
  Effective Field Theory}},
  \href{https://doi.org/10.12942/lrr-2004-5}{\emph{Living Rev. Rel.} {\bfseries
  7} (2004) 5} [\href{https://arxiv.org/abs/gr-qc/0311082}{{\ttfamily
  gr-qc/0311082}}].

\bibitem{Donoghue:2017pgk}
J.F.~Donoghue, M.M.~Ivanov and A.~Shkerin, \emph{{EPFL Lectures on General
  Relativity as a Quantum Field Theory}},
  \href{https://arxiv.org/abs/1702.00319}{{\ttfamily 1702.00319}}.

\bibitem{Donoghue:2022eay}
J.F.~Donoghue, \emph{{Quantum General Relativity and Effective Field Theory}},
  in \emph{{Handbook of Quantum Gravity}}, C.~Bambi, L.~Modesto and I.~Shapiro,
  eds., Springer (2023),
  \href{https://doi.org/10.1007/978-981-19-3079-9_1-1}{DOI}
  [\href{https://arxiv.org/abs/2211.09902}{{\ttfamily 2211.09902}}].

\bibitem{Dvali:2011th}
G.~Dvali, C.~Gomez and A.~Kehagias, \emph{{Classicalization of gravitons and
  Goldstones}}, \href{https://doi.org/10.1007/JHEP11(2011)070}{\emph{JHEP}
  {\bfseries 11} (2011) 070} [\href{https://arxiv.org/abs/1103.5963}{{\ttfamily
  1103.5963}}].

\bibitem{Dvali:2011aa}
G.~Dvali and C.~Gomez, \emph{{Black hole's quantum N-portrait}},
  \href{https://doi.org/10.1002/prop.201300001}{\emph{Fortsch. Phys.}
  {\bfseries 61} (2013) 742} [\href{https://arxiv.org/abs/1112.3359}{{\ttfamily
  1112.3359}}].

\bibitem{Dvali:2012rt}
G.~Dvali and C.~Gomez, \emph{{Black hole's $1/N$ hair}},
  \href{https://doi.org/10.1016/j.physletb.2013.01.020}{\emph{Phys. Lett. B}
  {\bfseries 719} (2013) 419}
  [\href{https://arxiv.org/abs/1203.6575}{{\ttfamily 1203.6575}}].

\bibitem{Dvali:2012gb}
G.~Dvali and C.~Gomez, \emph{{Landau–Ginzburg limit of black hole's quantum
  portrait: Self-similarity and critical exponent}},
  \href{https://doi.org/10.1016/j.physletb.2012.08.019}{\emph{Phys. Lett. B}
  {\bfseries 716} (2012) 240}
  [\href{https://arxiv.org/abs/1203.3372}{{\ttfamily 1203.3372}}].

\bibitem{Dvali:2012uq}
G.~Dvali, C.~Gomez and D.~Lust, \emph{{Black hole quantum mechanics in the
  presence of species}},
  \href{https://doi.org/10.1002/prop.201300002}{\emph{Fortsch. Phys.}
  {\bfseries 61} (2013) 768} [\href{https://arxiv.org/abs/1206.2365}{{\ttfamily
  1206.2365}}].

\bibitem{Dvali:2012wq}
G.~Dvali and C.~Gomez, \emph{{Black Hole Macro-Quantumness}},
  \href{https://arxiv.org/abs/1212.0765}{{\ttfamily 1212.0765}}.

\bibitem{Dvali:2013vxa}
G.~Dvali, D.~Flassig, C.~Gomez, A.~Pritzel and N.~Wintergerst,
  \emph{{Scrambling in the black hole portrait}},
  \href{https://doi.org/10.1103/PhysRevD.88.124041}{\emph{Phys. Rev. D}
  {\bfseries 88} (2013) 124041}
  [\href{https://arxiv.org/abs/1307.3458}{{\ttfamily 1307.3458}}].

\bibitem{Casadio:2015lis}
R.~Casadio, A.~Giugno, O.~Micu and A.~Orlandi, \emph{{Thermal BEC Black
  Holes}}, \href{https://doi.org/10.3390/e17106893}{\emph{Entropy} {\bfseries
  17} (2015) 6893} [\href{https://arxiv.org/abs/1511.01279}{{\ttfamily
  1511.01279}}].

\bibitem{Giusti:2019wdx}
A.~Giusti, \emph{{On the corpuscular theory of gravity}},
  \href{https://doi.org/10.1142/S0219887819300010}{\emph{Int. J. Geom. Meth.
  Mod. Phys.} {\bfseries 16} (2019) 1930001}.

\bibitem{Buoninfante:2019swn}
L.~Buoninfante and A.~Mazumdar, \emph{{Nonlocal star as a blackhole mimicker}},
  \href{https://doi.org/10.1103/PhysRevD.100.024031}{\emph{Phys. Rev. D}
  {\bfseries 100} (2019) 024031}
  [\href{https://arxiv.org/abs/1903.01542}{{\ttfamily 1903.01542}}].

\bibitem{Dvali:2020wft}
G.~Dvali, L.~Eisemann, M.~Michel and S.~Zell, \emph{{Black hole metamorphosis
  and stabilization by memory burden}},
  \href{https://doi.org/10.1103/PhysRevD.102.103523}{\emph{Phys. Rev. D}
  {\bfseries 102} (2020) 103523}
  [\href{https://arxiv.org/abs/2006.00011}{{\ttfamily 2006.00011}}].

\bibitem{Casadio:2021eio}
R.~Casadio, \emph{{Geometry and thermodynamics of coherent quantum black
  holes}}, \href{https://doi.org/10.1142/S0218271822501280}{\emph{Int. J. Mod.
  Phys. D} {\bfseries 31} (2022) 2250128}
  [\href{https://arxiv.org/abs/2103.00183}{{\ttfamily 2103.00183}}].

\bibitem{Dvali:2021ooc}
G.~Dvali and R.~Venugopalan, \emph{{Classicalization and unitarization of wee
  partons in QCD and gravity: The CGC-black hole correspondence}},
  \href{https://doi.org/10.1103/PhysRevD.105.056026}{\emph{Phys. Rev. D}
  {\bfseries 105} (2022) 056026}
  [\href{https://arxiv.org/abs/2106.11989}{{\ttfamily 2106.11989}}].

\bibitem{Dvali:2021tez}
G.~Dvali, O.~Kaikov and J.S.V.~Berm\'udez, \emph{{How special are black holes?
  Correspondence with objects saturating unitarity bounds in generic
  theories}}, \href{https://doi.org/10.1103/PhysRevD.105.056013}{\emph{Phys.
  Rev. D} {\bfseries 105} (2022) 056013}
  [\href{https://arxiv.org/abs/2112.00551}{{\ttfamily 2112.00551}}].

\bibitem{Dvali:2021ofp}
G.~Dvali, F.~K\"uhnel and M.~Zantedeschi, \emph{{Vortices in Black Holes}},
  \href{https://doi.org/10.1103/PhysRevLett.129.061302}{\emph{Phys. Rev. Lett.}
  {\bfseries 129} (2022) 061302}
  [\href{https://arxiv.org/abs/2112.08354}{{\ttfamily 2112.08354}}].

\bibitem{Dvali:2023qlk}
G.~Dvali, O.~Kaikov, F.~K{\"u}hnel, J.S.~Valbuena-Bermudez and M.~Zantedeschi,
  \emph{{Vortex Effects in Merging Black Holes and Saturons}},
  \href{https://doi.org/10.1103/PhysRevLett.132.151402}{\emph{Phys. Rev. Lett.}
  {\bfseries 132} (2024) 151402}
  [\href{https://arxiv.org/abs/2310.02288}{{\ttfamily 2310.02288}}].

\bibitem{Raj:2023irr}
H.~Raj and R.~Venugopalan, \emph{{Universal features of 2\textrightarrow{}N
  scattering in QCD and gravity from shockwave collisions}},
  \href{https://doi.org/10.1103/PhysRevD.109.044064}{\emph{Phys. Rev. D}
  {\bfseries 109} (2024) 044064}
  [\href{https://arxiv.org/abs/2311.03463}{{\ttfamily 2311.03463}}].

\bibitem{Feng:2025nai}
W.~Feng, A.~Giusti and R.~Casadio, \emph{{Horizon quantum mechanics for
  coherent quantum black holes}},
  \href{https://doi.org/10.1140/epjp/s13360-025-06065-x}{\emph{Eur. Phys. J.
  Plus} {\bfseries 140} (2025) 145}
  [\href{https://arxiv.org/abs/2408.17091}{{\ttfamily 2408.17091}}].

\bibitem{Dvali:2025gvd}
G.~Dvali, \emph{{Occam{\textquoteright}s Razor for~Black Holes}},
  \href{https://doi.org/10.1007/978-3-031-76066-2_5}{\emph{Fundam. Theor.
  Phys.} {\bfseries 219} (2025) 141}.

\bibitem{Dvali:2025sog}
G.~Dvali, \emph{{Swift Memory Burden in Merging Black Holes: how information
  load affects black hole's classical dynamics}},
  \href{https://arxiv.org/abs/2509.22540}{{\ttfamily 2509.22540}}.

\bibitem{Zhang:1999is}
W.-M.~Zhang, \emph{{Coherent states in field theory}},
  \href{https://arxiv.org/abs/hep-th/9908117}{{\ttfamily hep-th/9908117}}.

\bibitem{Binetruy:2012kx}
P.~Binetruy, \emph{{Vacuum energy, holography and a quantum portrait of the
  visible Universe}},  \href{https://arxiv.org/abs/1208.4645}{{\ttfamily
  1208.4645}}.

\bibitem{Dvali:2013eja}
G.~Dvali and C.~Gomez, \emph{{Quantum compositeness of gravity: black holes,
  AdS and inflation}},
  \href{https://doi.org/10.1088/1475-7516/2014/01/023}{\emph{JCAP} {\bfseries
  01} (2014) 023} [\href{https://arxiv.org/abs/1312.4795}{{\ttfamily
  1312.4795}}].

\bibitem{Dvali:2014gua}
G.~Dvali and C.~Gomez, \emph{{Quantum exclusion of positive cosmological
  constant?}}, \href{https://doi.org/10.1002/andp.201500216}{\emph{Annalen
  Phys.} {\bfseries 528} (2016) 68}
  [\href{https://arxiv.org/abs/1412.8077}{{\ttfamily 1412.8077}}].

\bibitem{Casadio:2015xva}
R.~Casadio, F.~Kuhnel and A.~Orlandi, \emph{{Consistent cosmic microwave
  background spectra from quantum depletion}},
  \href{https://doi.org/10.1088/1475-7516/2015/09/002}{\emph{JCAP} {\bfseries
  09} (2015) 002} [\href{https://arxiv.org/abs/1502.04703}{{\ttfamily
  1502.04703}}].

\bibitem{Kuhnel:2015yka}
F.~Kuhnel and M.~Sandstad, \emph{{Corpuscular consideration of eternal
  inflation}}, \href{https://doi.org/10.1140/epjc/s10052-015-3736-7}{\emph{Eur.
  Phys. J. C} {\bfseries 75} (2015) 505}
  [\href{https://arxiv.org/abs/1504.02377}{{\ttfamily 1504.02377}}].

\bibitem{Dvali:2017eba}
G.~Dvali, C.~Gomez and S.~Zell, \emph{{Quantum break-time of de Sitter}},
  \href{https://doi.org/10.1088/1475-7516/2017/06/028}{\emph{JCAP} {\bfseries
  06} (2017) 028} [\href{https://arxiv.org/abs/1701.08776}{{\ttfamily
  1701.08776}}].

\bibitem{Dvali:2018jhn}
G.~Dvali, C.~Gomez and S.~Zell, \emph{{Quantum Breaking Bound on de Sitter and
  Swampland}}, \href{https://doi.org/10.1002/prop.201800094}{\emph{Fortsch.
  Phys.} {\bfseries 67} (2019) 1800094}
  [\href{https://arxiv.org/abs/1810.11002}{{\ttfamily 1810.11002}}].

\bibitem{Dvali:2020etd}
G.~Dvali, \emph{{$S$-Matrix and Anomaly of de Sitter}},
  \href{https://doi.org/10.3390/sym13010003}{\emph{Symmetry} {\bfseries 13}
  (2020) 3} [\href{https://arxiv.org/abs/2012.02133}{{\ttfamily 2012.02133}}].

\bibitem{Giusti:2021shf}
A.~Giusti, S.~Buffa, L.~Heisenberg and R.~Casadio, \emph{{A quantum state for
  the late Universe}},
  \href{https://doi.org/10.1016/j.physletb.2022.136900}{\emph{Phys. Lett. B}
  {\bfseries 826} (2022) 136900}
  [\href{https://arxiv.org/abs/2108.05111}{{\ttfamily 2108.05111}}].

\bibitem{Berezhiani:2021zst}
L.~Berezhiani, G.~Dvali and O.~Sakhelashvili, \emph{{de Sitter space as a BRST
  invariant coherent state of gravitons}},
  \href{https://doi.org/10.1103/PhysRevD.105.025022}{\emph{Phys. Rev. D}
  {\bfseries 105} (2022) 025022}
  [\href{https://arxiv.org/abs/2111.12022}{{\ttfamily 2111.12022}}].

\bibitem{Berezhiani:2024boz}
L.~Berezhiani, G.~Dvali and O.~Sakhelashvili, \emph{{Consistent canonical
  quantization of gravity: Recovery of classical GR from BRST-invariant
  coherent states}}, \href{https://doi.org/10.1103/652w-ym62}{\emph{Phys. Rev.
  D} {\bfseries 113} (2026) 125032}
  [\href{https://arxiv.org/abs/2409.18777}{{\ttfamily 2409.18777}}].

\bibitem{Meade:2007sz}
P.~Meade and L.~Randall, \emph{{Black holes and quantum gravity at the LHC}},
  \href{https://doi.org/10.1088/1126-6708/2008/05/003}{\emph{JHEP} {\bfseries
  05} (2008) 003} [\href{https://arxiv.org/abs/0708.3017}{{\ttfamily
  0708.3017}}].

\bibitem{Calmet:2008dg}
X.~Calmet, W.~Gong and S.D.H.~Hsu, \emph{{Colorful quantum black holes at the
  LHC}}, \href{https://doi.org/10.1016/j.physletb.2008.08.011}{\emph{Phys.
  Lett. B} {\bfseries 668} (2008) 20}
  [\href{https://arxiv.org/abs/0806.4605}{{\ttfamily 0806.4605}}].

\bibitem{Gingrich:2009hj}
D.M.~Gingrich, \emph{{Quantum black holes with charge, color and spin at the
  LHC}}, \href{https://doi.org/10.1088/0954-3899/37/10/105008}{\emph{J. Phys.
  G} {\bfseries 37} (2010) 105008}
  [\href{https://arxiv.org/abs/0912.0826}{{\ttfamily 0912.0826}}].

\bibitem{Calmet:2012fv}
X.~Calmet and N.~Gausmann, \emph{{Non-thermal quantum black holes with
  quantized masses}},
  \href{https://doi.org/10.1142/S0217751X13500450}{\emph{Int. J. Mod. Phys. A}
  {\bfseries 28} (2013) 1350045}
  [\href{https://arxiv.org/abs/1209.4618}{{\ttfamily 1209.4618}}].

\bibitem{Treder:1985kb}
H.J.~Treder, \emph{{The planckions as largest elementary particles and as
  smallest test bodies}},
  \href{https://doi.org/10.1007/BF00735287}{\emph{Found. Phys.} {\bfseries 15}
  (1985) 161}.

\bibitem{Dvali:2016ovn}
G.~Dvali, \emph{{Strong Coupling and Classicalization}},
  \href{https://doi.org/10.1142/9789813208292_0005}{\emph{Subnucl. Ser.}
  {\bfseries 53} (2017) 189}
  [\href{https://arxiv.org/abs/1607.07422}{{\ttfamily 1607.07422}}].

\bibitem{Calmet:2014gya}
X.~Calmet, \emph{{The Lightest of Black Holes}},
  \href{https://doi.org/10.1142/S0217732314502046}{\emph{Mod. Phys. Lett. A}
  {\bfseries 29} (2014) 1450204}
  [\href{https://arxiv.org/abs/1410.2807}{{\ttfamily 1410.2807}}].

\bibitem{Calmet:2017omb}
X.~Calmet, R.~Casadio, A.Y.~Kamenshchik and O.V.~Teryaev, \emph{{Graviton
  propagator, renormalization scale and black-hole like states}},
  \href{https://doi.org/10.1016/j.physletb.2017.09.080}{\emph{Phys. Lett. B}
  {\bfseries 774} (2017) 332}
  [\href{https://arxiv.org/abs/1708.01485}{{\ttfamily 1708.01485}}].

\bibitem{Narison:2002woh}
S.~Narison, \emph{{QCD as a Theory of Hadrons}: {From Partons to Confinement}},
  Cambridge Monographs on Particle Physics, Nuclear Physics and Cosmology,
  Cambridge University Press (2007),
  \href{https://doi.org/10.1017/CBO9780511535000.}{10.1017/CBO9780511535000.}

\bibitem{Loll:2019rdj}
R.~Loll, \emph{{Quantum Gravity from Causal Dynamical Triangulations: A
  Review}}, \href{https://doi.org/10.1088/1361-6382/ab57c7}{\emph{Class. Quant.
  Grav.} {\bfseries 37} (2020) 013002}
  [\href{https://arxiv.org/abs/1905.08669}{{\ttfamily 1905.08669}}].

\bibitem{Dvali:2010jz}
G.~Dvali, G.F.~Giudice, C.~Gomez and A.~Kehagias, \emph{{UV-completion by
  classicalization}},
  \href{https://doi.org/10.1007/JHEP08(2011)108}{\emph{JHEP} {\bfseries 08}
  (2011) 108} [\href{https://arxiv.org/abs/1010.1415}{{\ttfamily 1010.1415}}].

\bibitem{Dvali:2010ns}
G.~Dvali and D.~Pirtskhalava, \emph{{Dynamics of unitarization by
  classicalization}},
  \href{https://doi.org/10.1016/j.physletb.2011.03.054}{\emph{Phys. Lett. B}
  {\bfseries 699} (2011) 78} [\href{https://arxiv.org/abs/1011.0114}{{\ttfamily
  1011.0114}}].

\bibitem{Dvali:2011nj}
G.~Dvali, \emph{{Classicalize or not to Classicalize?}},
  \href{https://arxiv.org/abs/1101.2661}{{\ttfamily 1101.2661}}.

\bibitem{Bajc:2011ey}
B.~Bajc, A.~Momen and G.~Senjanovic, \emph{{Classicalization via Path
  Integral}},  \href{https://arxiv.org/abs/1102.3679}{{\ttfamily 1102.3679}}.

\bibitem{Dvali:2011nh}
G.~Dvali, C.~Gomez and S.~Mukhanov, \emph{{Black Hole Masses are Quantized}},
  \href{https://arxiv.org/abs/1106.5894}{{\ttfamily 1106.5894}}.

\bibitem{Brouzakis:2011zs}
N.~Brouzakis, J.~Rizos and N.~Tetradis, \emph{{On the dynamics of
  classicalization}},
  \href{https://doi.org/10.1016/j.physletb.2012.01.011}{\emph{Phys. Lett. B}
  {\bfseries 708} (2012) 170}
  [\href{https://arxiv.org/abs/1109.6174}{{\ttfamily 1109.6174}}].

\bibitem{Grojean:2011bq}
C.~Grojean and R.S.~Gupta, \emph{{Theory and LHC phenomenology of classicalon
  decays}}, \href{https://doi.org/10.1007/JHEP05(2012)114}{\emph{JHEP}
  {\bfseries 05} (2012) 114} [\href{https://arxiv.org/abs/1110.5317}{{\ttfamily
  1110.5317}}].

\bibitem{Rizos:2011wj}
J.~Rizos and N.~Tetradis, \emph{{Dynamical classicalization}},
  \href{https://doi.org/10.1007/JHEP04(2012)110}{\emph{JHEP} {\bfseries 04}
  (2012) 110} [\href{https://arxiv.org/abs/1112.5546}{{\ttfamily 1112.5546}}].

\bibitem{Dvali:2012zc}
G.~Dvali, A.~Franca and C.~Gomez, \emph{{Road Signs for UV-Completion}},
  \href{https://arxiv.org/abs/1204.6388}{{\ttfamily 1204.6388}}.

\bibitem{Dvali:2012mx}
G.~Dvali and C.~Gomez, \emph{{Ultra-high energy probes of classicalization}},
  \href{https://doi.org/10.1088/1475-7516/2012/07/015}{\emph{JCAP} {\bfseries
  07} (2012) 015} [\href{https://arxiv.org/abs/1205.2540}{{\ttfamily
  1205.2540}}].

\bibitem{Rizos:2012qs}
J.~Rizos, N.~Tetradis and G.~Tsolias, \emph{{Classicalization as a tunnelling
  phenomenon}}, \href{https://doi.org/10.1007/JHEP08(2012)054}{\emph{JHEP}
  {\bfseries 08} (2012) 054} [\href{https://arxiv.org/abs/1206.3785}{{\ttfamily
  1206.3785}}].

\bibitem{Alberte:2012is}
L.~Alberte and F.~Bezrukov, \emph{{Semiclassical calculation of multiparticle
  scattering cross sections in classicalizing theories}},
  \href{https://doi.org/10.1103/PhysRevD.86.105008}{\emph{Phys. Rev. D}
  {\bfseries 86} (2012) 105008}
  [\href{https://arxiv.org/abs/1206.5311}{{\ttfamily 1206.5311}}].

\bibitem{Vikman:2012bx}
A.~Vikman, \emph{{Suppressing Quantum Fluctuations in Classicalization}},
  \href{https://doi.org/10.1209/0295-5075/101/34001}{\emph{EPL} {\bfseries 101}
  (2013) 34001} [\href{https://arxiv.org/abs/1208.3647}{{\ttfamily
  1208.3647}}].

\bibitem{Berkhahn:2013woa}
F.~Berkhahn, S.~Muller, F.~Niedermann and R.~Schneider, \emph{{Microscopic
  picture of non-relativistic classicalons}},
  \href{https://doi.org/10.1088/1475-7516/2013/08/028}{\emph{JCAP} {\bfseries
  08} (2013) 028} [\href{https://arxiv.org/abs/1302.6581}{{\ttfamily
  1302.6581}}].

\bibitem{Brouzakis:2014bwa}
N.~Brouzakis and N.~Tetradis, \emph{{Suppression of quantum corrections by
  classical backgrounds}},
  \href{https://doi.org/10.1103/PhysRevD.89.125004}{\emph{Phys. Rev. D}
  {\bfseries 89} (2014) 125004}
  [\href{https://arxiv.org/abs/1401.2775}{{\ttfamily 1401.2775}}].

\bibitem{Asimakis:2014bza}
P.~Asimakis, N.~Brouzakis, A.~Katsis and N.~Tetradis, \emph{{Quantum
  corrections in classicalon theories}},
  \href{https://doi.org/10.1016/j.physletb.2015.02.031}{\emph{Phys. Lett. B}
  {\bfseries 743} (2015) 75} [\href{https://arxiv.org/abs/1412.4275}{{\ttfamily
  1412.4275}}].

\bibitem{Dvali:2015ywa}
G.~Dvali, A.~Franca, C.~Gomez and N.~Wintergerst, \emph{{Nambu-Goldstone
  effective theory of information at quantum criticality}},
  \href{https://doi.org/10.1103/PhysRevD.92.125002}{\emph{Phys. Rev. D}
  {\bfseries 92} (2015) 125002}
  [\href{https://arxiv.org/abs/1507.02948}{{\ttfamily 1507.02948}}].

\bibitem{Dvali:2018xoc}
G.~Dvali, \emph{{Classicalization Clearly: Quantum Transition into States of
  Maximal Memory Storage Capacity}},
  \href{https://arxiv.org/abs/1804.06154}{{\ttfamily 1804.06154}}.

\bibitem{Dvali:2022vzz}
G.~Dvali and L.~Eisemann, \emph{{Perturbative understanding of nonperturbative
  processes and quantumization versus classicalization}},
  \href{https://doi.org/10.1103/PhysRevD.106.125019}{\emph{Phys. Rev. D}
  {\bfseries 106} (2022) 125019}
  [\href{https://arxiv.org/abs/2211.02618}{{\ttfamily 2211.02618}}].

\bibitem{Aydemir:2012nz}
U.~Aydemir, M.M.~Anber and J.F.~Donoghue, \emph{{Self-healing of unitarity in
  effective field theories and the onset of new physics}},
  \href{https://doi.org/10.1103/PhysRevD.86.014025}{\emph{Phys. Rev. D}
  {\bfseries 86} (2012) 014025}
  [\href{https://arxiv.org/abs/1203.5153}{{\ttfamily 1203.5153}}].

\bibitem{Chattopadhyay:2023nbj}
P.~Chattopadhyay and F.~Nortier, \emph{{Ghost-free Electroweak Symmetry
  Breaking with Weakly Nonlocal Interactions}},
  \href{https://doi.org/10.5506/APhysPolB.55.8-A2}{\emph{Acta Phys. Polon. B}
  {\bfseries 55} (2024) 8} [\href{https://arxiv.org/abs/2311.08311}{{\ttfamily
  2311.08311}}].

\bibitem{Nortier:2023dkq}
F.~Nortier, \emph{{Exorcizing Ghosts from the Vacuum Spectra in String-inspired
  Nonlocal Tachyon Condensation}},
  \href{https://doi.org/10.5506/APhysPolB.54.9-A4}{\emph{Acta Phys. Polon. B}
  {\bfseries 54} (2023) 9 A4}
  [\href{https://arxiv.org/abs/2307.11741}{{\ttfamily 2307.11741}}].

\bibitem{Addazi:2015ppa}
A.~Addazi, \emph{{Unitarization and Causalization of Non-local quantum field
  theories by Classicalization}},
  \href{https://doi.org/10.1142/S0217751X16500093}{\emph{Int. J. Mod. Phys. A}
  {\bfseries 31} (2016) 1650009}
  [\href{https://arxiv.org/abs/1505.07357}{{\ttfamily 1505.07357}}].

\bibitem{Buoninfante:2018gce}
L.~Buoninfante, A.~Ghoshal, G.~Lambiase and A.~Mazumdar, \emph{{Transmutation
  of nonlocal scale in infinite derivative field theories}},
  \href{https://doi.org/10.1103/PhysRevD.99.044032}{\emph{Phys. Rev. D}
  {\bfseries 99} (2019) 044032}
  [\href{https://arxiv.org/abs/1812.01441}{{\ttfamily 1812.01441}}].

\bibitem{Addazi:2020nkm}
A.~Addazi, \emph{{Hidden non-locality and self-superrenormalization of quantum
  gravity}}, \href{https://doi.org/10.1142/S0217732320502880}{\emph{Mod. Phys.
  Lett. A} {\bfseries 35} (2020) 2050288}
  [\href{https://arxiv.org/abs/2005.01961}{{\ttfamily 2005.01961}}].

\bibitem{Keltner:2015qqb}
L.A.~Keltner, \emph{{UV Properties of Galileons}},  PhD thesis, Case Western
  Reserve University, 2015,
  \href{https://inspirehep.net/literature/2627162}{https://inspirehep.net/literature/2627162}.

\bibitem{Keltner:2015xda}
L.~Keltner and A.J.~Tolley, \emph{{UV properties of Galileons: Spectral
  Densities}},  \href{https://arxiv.org/abs/1502.05706}{{\ttfamily
  1502.05706}}.

\bibitem{Brax:2012jr}
P.~Brax, C.~Burrage and A.-C.~Davis, \emph{{Screening fifth forces in k-essence
  and DBI models}},
  \href{https://doi.org/10.1088/1475-7516/2013/01/020}{\emph{JCAP} {\bfseries
  01} (2013) 020} [\href{https://arxiv.org/abs/1209.1293}{{\ttfamily
  1209.1293}}].

\bibitem{Vainshtein:1972sx}
A.I.~Vainshtein, \emph{{To the problem of nonvanishing gravitation mass}},
  \href{https://doi.org/10.1016/0370-2693(72)90147-5}{\emph{Phys. Lett. B}
  {\bfseries 39} (1972) 393}.

\bibitem{Nicolis:2004qq}
A.~Nicolis and R.~Rattazzi, \emph{{Classical and quantum consistency of the DGP
  model}}, \href{https://doi.org/10.1088/1126-6708/2004/06/059}{\emph{JHEP}
  {\bfseries 06} (2004) 059}
  [\href{https://arxiv.org/abs/hep-th/0404159}{{\ttfamily hep-th/0404159}}].

\bibitem{Armendariz-Picon:1999hyi}
C.~Armendariz-Picon, T.~Damour and V.F.~Mukhanov, \emph{{$k$-Inflation}},
  \href{https://doi.org/10.1016/S0370-2693(99)00603-6}{\emph{Phys. Lett. B}
  {\bfseries 458} (1999) 209}
  [\href{https://arxiv.org/abs/hep-th/9904075}{{\ttfamily hep-th/9904075}}].

\bibitem{Chiba:1999ka}
T.~Chiba, T.~Okabe and M.~Yamaguchi, \emph{{Kinetically driven quintessence}},
  \href{https://doi.org/10.1103/PhysRevD.62.023511}{\emph{Phys. Rev. D}
  {\bfseries 62} (2000) 023511}
  [\href{https://arxiv.org/abs/astro-ph/9912463}{{\ttfamily
  astro-ph/9912463}}].

\bibitem{Armendariz-Picon:2000nqq}
C.~Armendariz-Picon, V.F.~Mukhanov and P.J.~Steinhardt, \emph{{Dynamical
  Solution to the Problem of a Small Cosmological Constant and Late-Time Cosmic
  Acceleration}},
  \href{https://doi.org/10.1103/PhysRevLett.85.4438}{\emph{Phys. Rev. Lett.}
  {\bfseries 85} (2000) 4438}
  [\href{https://arxiv.org/abs/astro-ph/0004134}{{\ttfamily
  astro-ph/0004134}}].

\bibitem{Armendariz-Picon:2000ulo}
C.~Armendariz-Picon, V.F.~Mukhanov and P.J.~Steinhardt, \emph{{Essentials of
  $k$-essence}}, \href{https://doi.org/10.1103/PhysRevD.63.103510}{\emph{Phys.
  Rev. D} {\bfseries 63} (2001) 103510}
  [\href{https://arxiv.org/abs/astro-ph/0006373}{{\ttfamily
  astro-ph/0006373}}].

\bibitem{Brax:2014wla}
P.~Brax and P.~Valageas, \emph{{K-mouflage cosmology: The background
  evolution}}, \href{https://doi.org/10.1103/PhysRevD.90.023507}{\emph{Phys.
  Rev. D} {\bfseries 90} (2014) 023507}
  [\href{https://arxiv.org/abs/1403.5420}{{\ttfamily 1403.5420}}].

\bibitem{Luty:2003vm}
M.A.~Luty, M.~Porrati and R.~Rattazzi, \emph{{Strong interactions and stability
  in the DGP model}},
  \href{https://doi.org/10.1088/1126-6708/2003/09/029}{\emph{JHEP} {\bfseries
  09} (2003) 029} [\href{https://arxiv.org/abs/hep-th/0303116}{{\ttfamily
  hep-th/0303116}}].

\bibitem{Nicolis:2008in}
A.~Nicolis, R.~Rattazzi and E.~Trincherini, \emph{{Galileon as a local
  modification of gravity}},
  \href{https://doi.org/10.1103/PhysRevD.79.064036}{\emph{Phys. Rev. D}
  {\bfseries 79} (2009) 064036}
  [\href{https://arxiv.org/abs/0811.2197}{{\ttfamily 0811.2197}}].

\bibitem{Deffayet:2009wt}
C.~Deffayet, G.~Esposito-Farese and A.~Vikman, \emph{{Covariant Galileon}},
  \href{https://doi.org/10.1103/PhysRevD.79.084003}{\emph{Phys. Rev. D}
  {\bfseries 79} (2009) 084003}
  [\href{https://arxiv.org/abs/0901.1314}{{\ttfamily 0901.1314}}].

\bibitem{Deffayet:2009mn}
C.~Deffayet, S.~Deser and G.~Esposito-Farese, \emph{{Generalized Galileons: All
  scalar models whose curved background extensions maintain second-order field
  equations and stress tensors}},
  \href{https://doi.org/10.1103/PhysRevD.80.064015}{\emph{Phys. Rev. D}
  {\bfseries 80} (2009) 064015}
  [\href{https://arxiv.org/abs/0906.1967}{{\ttfamily 0906.1967}}].

\bibitem{Adams:2006sv}
A.~Adams, N.~Arkani-Hamed, S.~Dubovsky, A.~Nicolis and R.~Rattazzi,
  \emph{{Causality, analyticity and an IR obstruction to UV completion}},
  \href{https://doi.org/10.1088/1126-6708/2006/10/014}{\emph{JHEP} {\bfseries
  10} (2006) 014} [\href{https://arxiv.org/abs/hep-th/0602178}{{\ttfamily
  hep-th/0602178}}].

\bibitem{Kovner:2012yi}
A.~Kovner and M.~Lublinsky, \emph{{Classicalization and unitarity}},
  \href{https://doi.org/10.1007/JHEP11(2012)030}{\emph{JHEP} {\bfseries 11}
  (2012) 030} [\href{https://arxiv.org/abs/1207.5037}{{\ttfamily 1207.5037}}].

\bibitem{Kaloper:2014vqa}
N.~Kaloper, A.~Padilla, P.~Saffin and D.~Stefanyszyn, \emph{{Unitarity and the
  Vainshtein mechanism}},
  \href{https://doi.org/10.1103/PhysRevD.91.045017}{\emph{Phys. Rev. D}
  {\bfseries 91} (2015) 045017}
  [\href{https://arxiv.org/abs/1409.3243}{{\ttfamily 1409.3243}}].

\bibitem{Burrage:2020bxp}
C.~Burrage, B.~Coltman, A.~Padilla, D.~Saadeh and T.~Wilson, \emph{{Massive
  Galileons and Vainshtein screening}},
  \href{https://doi.org/10.1088/1475-7516/2021/02/050}{\emph{JCAP} {\bfseries
  02} (2021) 050} [\href{https://arxiv.org/abs/2008.01456}{{\ttfamily
  2008.01456}}].

\bibitem{Padilla:2017wth}
A.~Padilla and I.D.~Saltas, \emph{{Vainshtein in the UV and a Wilsonian
  analysis of derivatively coupled scalars}},
  \href{https://doi.org/10.1088/1475-7516/2018/06/039}{\emph{JCAP} {\bfseries
  06} (2018) 039} [\href{https://arxiv.org/abs/1712.04019}{{\ttfamily
  1712.04019}}].

\bibitem{Dvali:2007kt}
G.~Dvali, S.~Hofmann and J.~Khoury, \emph{{Degravitation of the cosmological
  constant and graviton width}},
  \href{https://doi.org/10.1103/PhysRevD.76.084006}{\emph{Phys. Rev. D}
  {\bfseries 76} (2007) 084006}
  [\href{https://arxiv.org/abs/hep-th/0703027}{{\ttfamily hep-th/0703027}}].

\bibitem{Brax:2016jjt}
P.~Brax and P.~Valageas, \emph{{Quantum field theory of K-mouflage}},
  \href{https://doi.org/10.1103/PhysRevD.94.043529}{\emph{Phys. Rev. D}
  {\bfseries 94} (2016) 043529}
  [\href{https://arxiv.org/abs/1607.01129}{{\ttfamily 1607.01129}}].

\bibitem{Dvali:2010gv}
G.~Dvali, C.~Gomez and S.~Mukhanov, \emph{{Probing quantum geometry at LHC}},
  \href{https://doi.org/10.1007/JHEP02(2011)012}{\emph{JHEP} {\bfseries 02}
  (2011) 012} [\href{https://arxiv.org/abs/1006.2466}{{\ttfamily 1006.2466}}].

\bibitem{Lee:1991ax}
T.D.~Lee and Y.~Pang, \emph{{Nontopological solitons}},
  \href{https://doi.org/10.1016/0370-1573(92)90064-7}{\emph{Phys. Rept.}
  {\bfseries 221} (1992) 251}.

\bibitem{Dvali:2019jjw}
G.~Dvali, \emph{{Area Law Saturation of Entropy Bound from Perturbative
  Unitarity in Renormalizable Theories}},
  \href{https://doi.org/10.1002/prop.202000090}{\emph{Fortsch. Phys.}
  {\bfseries 69} (2021) 2000090}
  [\href{https://arxiv.org/abs/1906.03530}{{\ttfamily 1906.03530}}].

\bibitem{Dvali:2019ulr}
G.~Dvali, \emph{{Unitarity Entropy Bound: Solitons and Instantons}},
  \href{https://doi.org/10.1002/prop.202000091}{\emph{Fortsch. Phys.}
  {\bfseries 69} (2021) 2000091}
  [\href{https://arxiv.org/abs/1907.07332}{{\ttfamily 1907.07332}}].

\bibitem{Dvali:2020wqi}
G.~Dvali, \emph{{Entropy bound and unitarity of scattering amplitudes}},
  \href{https://doi.org/10.1007/JHEP03(2021)126}{\emph{JHEP} {\bfseries 03}
  (2021) 126} [\href{https://arxiv.org/abs/2003.05546}{{\ttfamily
  2003.05546}}].

\bibitem{Dvali:2021rlf}
G.~Dvali and O.~Sakhelashvili, \emph{{Black-hole-like saturons in
  Gross-Neveu}}, \href{https://doi.org/10.1103/PhysRevD.105.065014}{\emph{Phys.
  Rev. D} {\bfseries 105} (2022) 065014}
  [\href{https://arxiv.org/abs/2111.03620}{{\ttfamily 2111.03620}}].

\bibitem{Dvali:2023xfz}
G.~Dvali, \emph{{Saturon Dark Matter}},
  \href{https://arxiv.org/abs/2302.08353}{{\ttfamily 2302.08353}}.

\bibitem{Contri:2025eod}
G.~Contri, G.~Dvali and O.~Sakhelashvili, \emph{{Similarities in the
  evaporation of saturated solitons and black holes}},
  \href{https://doi.org/10.1103/6q41-f43p}{\emph{Phys. Rev. D} {\bfseries 112}
  (2025) 125027} [\href{https://arxiv.org/abs/2509.08049}{{\ttfamily
  2509.08049}}].

\bibitem{Hook:2023pba}
A.~Hook and R.~Rattazzi, \emph{{Softening the UV without new particles}},
  \href{https://doi.org/10.1103/PhysRevD.108.115019}{\emph{Phys. Rev. D}
  {\bfseries 108} (2023) 115019}
  [\href{https://arxiv.org/abs/2306.12489}{{\ttfamily 2306.12489}}].

\bibitem{Cheung:2024wme}
C.~Cheung and I.Z.~Rothstein, \emph{{Hiding new physics at the end of field
  space}}, \href{https://doi.org/10.1103/ycb8-sj4l}{\emph{Phys. Rev. D}
  {\bfseries 113} (2026) 016002}
  [\href{https://arxiv.org/abs/2411.07380}{{\ttfamily 2411.07380}}].

\bibitem{Cheung:2007st}
C.~Cheung, P.~Creminelli, A.L.~Fitzpatrick, J.~Kaplan and L.~Senatore,
  \emph{{The effective field theory of inflation}},
  \href{https://doi.org/10.1088/1126-6708/2008/03/014}{\emph{JHEP} {\bfseries
  03} (2008) 014} [\href{https://arxiv.org/abs/0709.0293}{{\ttfamily
  0709.0293}}].

\bibitem{Pham:1985cr}
T.N.~Pham and T.N.~Truong, \emph{{Evaluation of the derivative quartic terms of
  the meson chiral Lagrangian from forward dispersion relations}},
  \href{https://doi.org/10.1103/PhysRevD.31.3027}{\emph{Phys. Rev. D}
  {\bfseries 31} (1985) 3027}.

\bibitem{Bellazzini:2020cot}
B.~Bellazzini, J.~Elias~Mir{\'o}, R.~Rattazzi, M.~Riembau and F.~Riva,
  \emph{{Positive moments for scattering amplitudes}},
  \href{https://doi.org/10.1103/PhysRevD.104.036006}{\emph{Phys. Rev. D}
  {\bfseries 104} (2021) 036006}
  [\href{https://arxiv.org/abs/2011.00037}{{\ttfamily 2011.00037}}].

\bibitem{Arkani-Hamed:2020blm}
N.~Arkani-Hamed, T.-C.~Huang and Y.-t.~Huang, \emph{{The EFT-Hedron}},
  \href{https://doi.org/10.1007/JHEP05(2021)259}{\emph{JHEP} {\bfseries 05}
  (2021) 259} [\href{https://arxiv.org/abs/2012.15849}{{\ttfamily
  2012.15849}}].

\bibitem{Weinberg:2008hq}
S.~Weinberg, \emph{{Effective field theory for inflation}},
  \href{https://doi.org/10.1103/PhysRevD.77.123541}{\emph{Phys. Rev. D}
  {\bfseries 77} (2008) 123541}
  [\href{https://arxiv.org/abs/0804.4291}{{\ttfamily 0804.4291}}].

\bibitem{Burgess:2009ea}
C.P.~Burgess, H.M.~Lee and M.~Trott, \emph{{Power-counting and the validity of
  the classical approximation during inflation}},
  \href{https://doi.org/10.1088/1126-6708/2009/09/103}{\emph{JHEP} {\bfseries
  09} (2009) 103} [\href{https://arxiv.org/abs/0902.4465}{{\ttfamily
  0902.4465}}].

\bibitem{deRham:2014wfa}
C.~de~Rham and R.H.~Ribeiro, \emph{{Riding on irrelevant operators}},
  \href{https://doi.org/10.1088/1475-7516/2014/11/016}{\emph{JCAP} {\bfseries
  11} (2014) 016} [\href{https://arxiv.org/abs/1405.5213}{{\ttfamily
  1405.5213}}].

\bibitem{Giudice:2016yja}
G.F.~Giudice and M.~McCullough, \emph{{A clockwork theory}},
  \href{https://doi.org/10.1007/JHEP02(2017)036}{\emph{JHEP} {\bfseries 02}
  (2017) 036} [\href{https://arxiv.org/abs/1610.07962}{{\ttfamily
  1610.07962}}].

\bibitem{Buoninfante:2024yth}
L.~Buoninfante et~al., \emph{{Visions in quantum gravity}},
  \href{https://doi.org/10.21468/SciPostPhysCommRep.11}{\emph{SciPost Phys.
  Comm. Rep.} {\bfseries 11} (2025) }
  [\href{https://arxiv.org/abs/2412.08696}{{\ttfamily 2412.08696}}].

\bibitem{Arkani-Hamed:1998sfv}
N.~Arkani-Hamed, S.~Dimopoulos and G.R.~Dvali, \emph{{Phenomenology,
  astrophysics, and cosmology of theories with submillimeter dimensions and TeV
  scale quantum gravity}},
  \href{https://doi.org/10.1103/PhysRevD.59.086004}{\emph{Phys. Rev. D}
  {\bfseries 59} (1999) 086004}
  [\href{https://arxiv.org/abs/hep-ph/9807344}{{\ttfamily hep-ph/9807344}}].

\bibitem{Akhoury:2011en}
R.~Akhoury, S.~Mukohyama and R.~Saotome, \emph{{No Classicalization Beyond
  Spherical Symmetry}},  \href{https://arxiv.org/abs/1109.3820}{{\ttfamily
  1109.3820}}.

\bibitem{Brax:2011sv}
P.~Brax, C.~Burrage and A.-C.~Davis, \emph{{Laboratory tests of the Galileon}},
  \href{https://doi.org/10.1088/1475-7516/2011/09/020}{\emph{JCAP} {\bfseries
  09} (2011) 020} [\href{https://arxiv.org/abs/1106.1573}{{\ttfamily
  1106.1573}}].

\bibitem{Bloomfield:2014zfa}
J.K.~Bloomfield, C.~Burrage and A.-C.~Davis, \emph{{Shape dependence of
  Vainshtein screening}},
  \href{https://doi.org/10.1103/PhysRevD.91.083510}{\emph{Phys. Rev. D}
  {\bfseries 91} (2015) 083510}
  [\href{https://arxiv.org/abs/1408.4759}{{\ttfamily 1408.4759}}].

\bibitem{Davis:2021oce}
A.-C.~Davis and S.~Melville, \emph{{Scalar fields near compact objects:
  resummation versus UV completion}},
  \href{https://doi.org/10.1088/1475-7516/2021/11/012}{\emph{JCAP} {\bfseries
  11} (2021) 012} [\href{https://arxiv.org/abs/2107.00010}{{\ttfamily
  2107.00010}}].

\bibitem{Tolley:2020gtv}
A.J.~Tolley, Z.-Y.~Wang and S.-Y.~Zhou, \emph{{New positivity bounds from full
  crossing symmetry}},
  \href{https://doi.org/10.1007/JHEP05(2021)255}{\emph{JHEP} {\bfseries 05}
  (2021) 255} [\href{https://arxiv.org/abs/2011.02400}{{\ttfamily
  2011.02400}}].

\bibitem{Caron-Huot:2020cmc}
S.~Caron-Huot and V.~Van~Duong, \emph{{Extremal effective field theories}},
  \href{https://doi.org/10.1007/JHEP05(2021)280}{\emph{JHEP} {\bfseries 05}
  (2021) 280} [\href{https://arxiv.org/abs/2011.02957}{{\ttfamily
  2011.02957}}].

\bibitem{Bellazzini:2021oaj}
B.~Bellazzini, M.~Riembau and F.~Riva, \emph{{IR side of positivity bounds}},
  \href{https://doi.org/10.1103/PhysRevD.106.105008}{\emph{Phys. Rev. D}
  {\bfseries 106} (2022) 105008}
  [\href{https://arxiv.org/abs/2112.12561}{{\ttfamily 2112.12561}}].

\bibitem{Serra:2023nrn}
F.~Serra and L.G.~Trombetta, \emph{{Five-point superluminality bounds}},
  \href{https://doi.org/10.1007/JHEP06(2024)117}{\emph{JHEP} {\bfseries 06}
  (2024) 117} [\href{https://arxiv.org/abs/2312.06759}{{\ttfamily
  2312.06759}}].

\bibitem{Bonvin:2006vc}
C.~Bonvin, C.~Caprini and R.~Durrer, \emph{{No-Go Theorem for $k$-Essence Dark
  Energy}}, \href{https://doi.org/10.1103/PhysRevLett.97.081303}{\emph{Phys.
  Rev. Lett.} {\bfseries 97} (2006) 081303}
  [\href{https://arxiv.org/abs/astro-ph/0606584}{{\ttfamily
  astro-ph/0606584}}].

\bibitem{Bonvin:2007mw}
C.~Bonvin, C.~Caprini and R.~Durrer, \emph{{Superluminal motion and closed
  signal curves}},  \href{https://arxiv.org/abs/0706.1538}{{\ttfamily
  0706.1538}}.

\bibitem{Hinterbichler:2009kq}
K.~Hinterbichler, A.~Nicolis and M.~Porrati, \emph{{Superluminality in DGP}},
  \href{https://doi.org/10.1088/1126-6708/2009/09/089}{\emph{JHEP} {\bfseries
  09} (2009) 089} [\href{https://arxiv.org/abs/0905.2359}{{\ttfamily
  0905.2359}}].

\bibitem{Goon:2010xh}
G.L.~Goon, K.~Hinterbichler and M.~Trodden, \emph{{Stability and
  superluminality of spherical DBI Galileon solutions}},
  \href{https://doi.org/10.1103/PhysRevD.83.085015}{\emph{Phys. Rev. D}
  {\bfseries 83} (2011) 085015}
  [\href{https://arxiv.org/abs/1008.4580}{{\ttfamily 1008.4580}}].

\bibitem{Evslin:2011vh}
J.~Evslin and T.~Qiu, \emph{{Closed timelike curves in the Galileon model}},
  \href{https://doi.org/10.1007/JHEP11(2011)032}{\emph{JHEP} {\bfseries 11}
  (2011) 032} [\href{https://arxiv.org/abs/1106.0570}{{\ttfamily 1106.0570}}].

\bibitem{Evslin:2011rj}
J.~Evslin, \emph{{Stability of closed timelike curves in a Galileon model}},
  \href{https://doi.org/10.1007/JHEP03(2012)009}{\emph{JHEP} {\bfseries 03}
  (2012) 009} [\href{https://arxiv.org/abs/1112.1349}{{\ttfamily 1112.1349}}].

\bibitem{Hawking:1991nk}
S.W.~Hawking, \emph{{Chronology protection conjecture}},
  \href{https://doi.org/10.1103/PhysRevD.46.603}{\emph{Phys. Rev. D} {\bfseries
  46} (1992) 603}.

\bibitem{Kim:1991mc}
S.W.~Kim and K.S.~Thorne, \emph{{Do vacuum fluctuations prevent the creation of
  closed timelike curves?}},
  \href{https://doi.org/10.1103/PhysRevD.43.3929}{\emph{Phys. Rev. D}
  {\bfseries 43} (1991) 3929}.

\bibitem{Babichev:2007dw}
E.~Babichev, V.~Mukhanov and A.~Vikman, \emph{{$k$-Essence, superluminal
  propagation, causality and emergent geometry}},
  \href{https://doi.org/10.1088/1126-6708/2008/02/101}{\emph{JHEP} {\bfseries
  02} (2008) 101} [\href{https://arxiv.org/abs/0708.0561}{{\ttfamily
  0708.0561}}].

\bibitem{Burrage:2011cr}
C.~Burrage, C.~de~Rham, L.~Heisenberg and A.J.~Tolley, \emph{{Chronology
  protection in Galileon models and massive gravity}},
  \href{https://doi.org/10.1088/1475-7516/2012/07/004}{\emph{JCAP} {\bfseries
  07} (2012) 004} [\href{https://arxiv.org/abs/1111.5549}{{\ttfamily
  1111.5549}}].

\bibitem{Kaplan:2024qtf}
D.E.~Kaplan, S.~Rajendran and F.~Serra, \emph{{Wrong signs are alright}},
  \href{https://doi.org/10.1007/JHEP03(2025)031}{\emph{JHEP} {\bfseries 03}
  (2025) 031} [\href{https://arxiv.org/abs/2406.06681}{{\ttfamily
  2406.06681}}].

\bibitem{Cintia:2025fzn}
G.~Cintia, F.~Piazza and S.~Ramos, \emph{{Modified microcausality from
  perturbation theory}}, \href{https://doi.org/10.1103/y5cg-1c7g}{\emph{Phys.
  Rev. D} {\bfseries 112} (2025) 085015}
  [\href{https://arxiv.org/abs/2504.16992}{{\ttfamily 2504.16992}}].

\bibitem{Meiman:1964vmk}
N.N.~Meiman, \emph{{The Causality Principle and the Asymptotic Behavior of the
  Scattering Amplitude}}, {\emph{Sov. Phys. JETP} {\bfseries 20} (1965) 1320}.

\bibitem{Jaffe:1966an}
A.M.~Jaffe, \emph{{High-Energy Behavior of Local Quantum Fields}},  1966,
  \href{https://inspirehep.net/literature/49928}{https://inspirehep.net/literature/49928}.

\bibitem{Jaffe:1967nb}
A.M.~Jaffe, \emph{{High-Energy Behavior in Quantum Field Theory. I. Strictly
  Localizable Fields}},
  \href{https://doi.org/10.1103/PhysRev.158.1454}{\emph{Phys. Rev.} {\bfseries
  158} (1967) 1454}.

\bibitem{Tokuda:2019nqb}
J.~Tokuda, \emph{{Extension of positivity bounds to non-local theories: IR
  obstructions to Lorentz invariant UV completions}},
  \href{https://doi.org/10.1007/JHEP05(2019)216}{\emph{JHEP} {\bfseries 05}
  (2019) 216} [\href{https://arxiv.org/abs/1902.10039}{{\ttfamily
  1902.10039}}].

\bibitem{Iofa:1969fj}
M.Z.~Iofa and V.Y.~Fainberg, \emph{{Wightman Formulation for a Nonlocalizable
  Field Theory. I.}}, {\emph{Sov. Phys. JETP} {\bfseries 29} (1969) 880}.

\bibitem{Iofa:1969ex}
M.Z.~Iofa and V.Y.~Fainberg, \emph{{Wightman formulation for nonlocalizable
  field theories. II. Theory of asymptotic fields and particles}},
  \href{https://doi.org/10.1007/BF01028040}{\emph{Teor. Mat. Fiz.} {\bfseries
  1} (1969) 187}.

\bibitem{Steinmann:1970cm}
O.~Steinmann, \emph{{Scattering formalism for non-localizable fields}},
  \href{https://doi.org/10.1007/BF01649431}{\emph{Commun. Math. Phys.}
  {\bfseries 18} (1970) 179}.

\bibitem{Fainberg:1977wp}
V.Y.~Fainberg and M.A.~Solovev, \emph{{Causality, localizability, and
  holomorphically convex hulls}},
  \href{https://doi.org/10.1007/BF01625773}{\emph{Commun. Math. Phys.}
  {\bfseries 57} (1977) 149}.

\bibitem{Fainberg:1978cc}
V.Y.~Fainberg and M.A.~Solovev, \emph{{How can local properties be described in
  field theories without strict locality?}},
  \href{https://doi.org/10.1016/0003-4916(78)90211-7}{\emph{Annals Phys.}
  {\bfseries 113} (1978) 421}.

\bibitem{Solovev:1980tle}
M.A.~Solovev, \emph{{Relativistically invariant formulation of causality in a
  nonlocal theory of exponential growth}},
  \href{https://doi.org/10.1007/bf01018393}{\emph{Theor. Math. Phys.}
  {\bfseries 43} (1980) 412}.

\bibitem{Fainberg:1992jt}
V.Y.~Fainberg and M.A.~Solovev, \emph{{Nonlocalizability and asymptotical
  commutativity}}, \href{https://doi.org/10.1007/BF01016400}{\emph{Theor. Math.
  Phys.} {\bfseries 93} (1992) 1438}
  [\href{https://arxiv.org/abs/hep-th/9211099}{{\ttfamily hep-th/9211099}}].

\bibitem{Soloviev:1999rv}
M.A.~Soloviev, \emph{{PCT, spin and statistics, and analytic wave front set}},
  \href{https://doi.org/10.1007/BF02557234}{\emph{Theor. Math. Phys.}
  {\bfseries 121} (1999) 1377}
  [\href{https://arxiv.org/abs/hep-th/0605243}{{\ttfamily hep-th/0605243}}].

\bibitem{Soloviev:2001qe}
M.A.~Soloviev, \emph{{Nonlocal extension of the Borchers classes of quantum
  fields}},  \href{https://arxiv.org/abs/math-ph/0112053}{{\ttfamily
  math-ph/0112053}}.

\bibitem{Soloviev:2001qd}
M.A.~Soloviev, \emph{{Lorentz-Covariant Ultradistributions, Hyperfunctions, and
  Analytic Functionals}},
  \href{https://doi.org/10.1023/A:1012368004774}{\emph{Theor. Math. Phys.}
  {\bfseries 128} (2001) 1252}
  [\href{https://arxiv.org/abs/math-ph/0112052}{{\ttfamily math-ph/0112052}}].

\bibitem{Bruning:2004jr}
E.~Bruning and S.~Nagamachi, \emph{{Relativistic quantum field theory with a
  fundamental length}}, \href{https://doi.org/10.1063/1.1737055}{\emph{J. Math.
  Phys.} {\bfseries 45} (2004) 2199}.

\bibitem{Soloviev:2005qd}
M.A.~Soloviev, \emph{{Two classes of generalized functions used in nonlocal
  field theory}}, \href{https://doi.org/10.1007/s11232-005-0096-8}{\emph{Theor.
  Math. Phys.} {\bfseries 143} (2005) 651}
  [\href{https://arxiv.org/abs/math-ph/0605065}{{\ttfamily math-ph/0605065}}].

\bibitem{Soloviev:2006ah}
M.A.~Soloviev, \emph{{Axiomatic formulations of nonlocal and noncommutative
  field theories}},
  \href{https://doi.org/10.1007/s11232-006-0068-7}{\emph{Theor. Math. Phys.}
  {\bfseries 147} (2006) 660}
  [\href{https://arxiv.org/abs/hep-th/0605249}{{\ttfamily hep-th/0605249}}].

\bibitem{Soloviev:2009cy}
M.A.~Soloviev, \emph{{Quantum field theory with a fundamental length: A general
  mathematical framework}}, \href{https://doi.org/10.1063/1.3269595}{\emph{J.
  Math. Phys.} {\bfseries 50} (2009) 123519}
  [\href{https://arxiv.org/abs/0912.0595}{{\ttfamily 0912.0595}}].

\bibitem{Soloviev:2010bg}
M.A.~Soloviev, \emph{{Reconstruction in quantum field theory with a fundamental
  length}}, \href{https://doi.org/10.1063/1.3483691}{\emph{J. Math. Phys.}
  {\bfseries 51} (2010) 093520}
  [\href{https://arxiv.org/abs/1012.3546}{{\ttfamily 1012.3546}}].

\bibitem{Fainberg:1971ia}
V.Y.~Fainberg and M.Z.~Iofa, \emph{{Bounds on the elastic amplitude in
  nonlocalizable field theories}},
  \href{https://doi.org/10.1007/BF02723604}{\emph{Nuovo Cim. A} {\bfseries 5}
  (1971) 275}.

\bibitem{Buoninfante:2024ibt}
L.~Buoninfante, L.-Q.~Shao and A.~Tokareva, \emph{{Nonlocal positivity bounds:
  Islands in terra incognita}},
  \href{https://doi.org/10.1103/37kc-25s3}{\emph{Phys. Rev. D} {\bfseries 112}
  (2025) L021904} [\href{https://arxiv.org/abs/2412.08634}{{\ttfamily
  2412.08634}}].

\bibitem{Khoze:2017tjt}
V.V.~Khoze and M.~Spannowsky, \emph{{Higgsplosion: Solving the hierarchy
  problem via rapid decays of heavy states into multiple Higgs bosons}},
  \href{https://doi.org/10.1016/j.nuclphysb.2017.11.002}{\emph{Nucl. Phys. B}
  {\bfseries 926} (2018) 95}
  [\href{https://arxiv.org/abs/1704.03447}{{\ttfamily 1704.03447}}].

\bibitem{Khoze:2017ifq}
V.V.~Khoze, \emph{{Multiparticle production in the large
  {\ensuremath{\lambda}}n limit: realising Higgsplosion in a scalar QFT}},
  \href{https://doi.org/10.1007/JHEP06(2017)148}{\emph{JHEP} {\bfseries 06}
  (2017) 148} [\href{https://arxiv.org/abs/1705.04365}{{\ttfamily
  1705.04365}}].

\bibitem{Khoze:2017lft}
V.V.~Khoze and M.~Spannowsky, \emph{{Higgsploding universe}},
  \href{https://doi.org/10.1103/PhysRevD.96.075042}{\emph{Phys. Rev. D}
  {\bfseries 96} (2017) 075042}
  [\href{https://arxiv.org/abs/1707.01531}{{\ttfamily 1707.01531}}].

\bibitem{Khoze:2017uga}
V.V.~Khoze, J.~Reiness, M.~Spannowsky and P.~Waite, \emph{{Precision
  measurements for the Higgsploding standard model}},
  \href{https://doi.org/10.1088/1361-6471/ab1a70}{\emph{J. Phys. G} {\bfseries
  46} (2019) 065004} [\href{https://arxiv.org/abs/1709.08655}{{\ttfamily
  1709.08655}}].

\bibitem{Belyaev:2018mtd}
A.~Belyaev, F.~Bezrukov, C.~Shepherd and D.~Ross, \emph{{Problems with
  Higgsplosion}}, \href{https://doi.org/10.1103/PhysRevD.98.113001}{\emph{Phys.
  Rev. D} {\bfseries 98} (2018) 113001}
  [\href{https://arxiv.org/abs/1808.05641}{{\ttfamily 1808.05641}}].

\bibitem{Monin:2018cbi}
A.~Monin, \emph{{Inconsistencies of higgsplosion}},
  \href{https://arxiv.org/abs/1808.05810}{{\ttfamily 1808.05810}}.

\bibitem{Khoze:2018qhz}
V.V.~Khoze and M.~Spannowsky, \emph{{Consistency of Higgsplosion in localizable
  QFT}}, \href{https://doi.org/10.1016/j.physletb.2019.01.052}{\emph{Phys.
  Lett. B} {\bfseries 790} (2019) 466}
  [\href{https://arxiv.org/abs/1809.11141}{{\ttfamily 1809.11141}}].

\bibitem{Zhang:2016pja}
Z.~Zhang, \emph{{Covariant diagrams for one-loop matching}},
  \href{https://doi.org/10.1007/JHEP05(2017)152}{\emph{JHEP} {\bfseries 05}
  (2017) 152} [\href{https://arxiv.org/abs/1610.00710}{{\ttfamily
  1610.00710}}].

\bibitem{Simon:1990ic}
J.Z.~Simon, \emph{{Higher-derivative Lagrangians, nonlocality, problems, and
  solutions}}, \href{https://doi.org/10.1103/PhysRevD.41.3720}{\emph{Phys. Rev.
  D} {\bfseries 41} (1990) 3720}.

\bibitem{Burgess:2014lwa}
C.P.~Burgess and M.~Williams, \emph{{Who you gonna call? Runaway ghosts, higher
  derivatives and time-dependence in EFTs}},
  \href{https://doi.org/10.1007/JHEP08(2014)074}{\emph{JHEP} {\bfseries 08}
  (2014) 074} [\href{https://arxiv.org/abs/1404.2236}{{\ttfamily 1404.2236}}].

\bibitem{Solomon:2017nlh}
A.R.~Solomon and M.~Trodden, \emph{{Higher-derivative operators and effective
  field theory for general scalar-tensor theories}},
  \href{https://doi.org/10.1088/1475-7516/2018/02/031}{\emph{JCAP} {\bfseries
  02} (2018) 031} [\href{https://arxiv.org/abs/1709.09695}{{\ttfamily
  1709.09695}}].

\bibitem{Langlois:2018dxi}
D.~Langlois, \emph{{Dark energy and modified gravity in degenerate higher-order
  scalar{\textendash}tensor (DHOST) theories: A review}},
  \href{https://doi.org/10.1142/S0218271819420069}{\emph{Int. J. Mod. Phys. D}
  {\bfseries 28} (2019) 1942006}
  [\href{https://arxiv.org/abs/1811.06271}{{\ttfamily 1811.06271}}].

\bibitem{Goon:2016ihr}
G.~Goon, K.~Hinterbichler, A.~Joyce and M.~Trodden, \emph{{Aspects of Galileon
  non-renormalization}},
  \href{https://doi.org/10.1007/JHEP11(2016)100}{\emph{JHEP} {\bfseries 11}
  (2016) 100} [\href{https://arxiv.org/abs/1606.02295}{{\ttfamily
  1606.02295}}].

\bibitem{Heisenberg:2020cyi}
L.~Heisenberg, J.~Noller and J.~Zosso, \emph{{Horndeski under the quantum
  loupe}}, \href{https://doi.org/10.1088/1475-7516/2020/10/010}{\emph{JCAP}
  {\bfseries 10} (2020) 010}
  [\href{https://arxiv.org/abs/2004.11655}{{\ttfamily 2004.11655}}].

\bibitem{Yukawa:1935xg}
H.~Yukawa, \emph{{On the Interaction of Elementary Particles. I}},
  \href{https://doi.org/10.1143/PTPS.1.1}{\emph{Proc. Phys. Math. Soc. Jap.}
  {\bfseries 17} (1935) 48}.

\bibitem{Khoury:2003aq}
J.~Khoury and A.~Weltman, \emph{{Chameleon Fields: Awaiting Surprises for Tests
  of Gravity in Space}},
  \href{https://doi.org/10.1103/PhysRevLett.93.171104}{\emph{Phys. Rev. Lett.}
  {\bfseries 93} (2004) 171104}
  [\href{https://arxiv.org/abs/astro-ph/0309300}{{\ttfamily
  astro-ph/0309300}}].

\bibitem{Khoury:2003rn}
J.~Khoury and A.~Weltman, \emph{{Chameleon cosmology}},
  \href{https://doi.org/10.1103/PhysRevD.69.044026}{\emph{Phys. Rev. D}
  {\bfseries 69} (2004) 044026}
  [\href{https://arxiv.org/abs/astro-ph/0309411}{{\ttfamily
  astro-ph/0309411}}].

\bibitem{Damour:1994zq}
T.~Damour and A.M.~Polyakov, \emph{{The string dilaton and a least coupling
  principle}}, \href{https://doi.org/10.1016/0550-3213(94)90143-0}{\emph{Nucl.
  Phys. B} {\bfseries 423} (1994) 532}
  [\href{https://arxiv.org/abs/hep-th/9401069}{{\ttfamily hep-th/9401069}}].

\bibitem{Brax:2011ja}
P.~Brax, C.~van~de Bruck, A.-C.~Davis, B.~Li and D.J.~Shaw, \emph{{Nonlinear
  structure formation with the environmentally dependent dilaton}},
  \href{https://doi.org/10.1103/PhysRevD.83.104026}{\emph{Phys. Rev. D}
  {\bfseries 83} (2011) 104026}
  [\href{https://arxiv.org/abs/1102.3692}{{\ttfamily 1102.3692}}].

\bibitem{Olive:2007aj}
K.A.~Olive and M.~Pospelov, \emph{{Environmental dependence of masses and
  coupling constants}},
  \href{https://doi.org/10.1103/PhysRevD.77.043524}{\emph{Phys. Rev. D}
  {\bfseries 77} (2008) 043524}
  [\href{https://arxiv.org/abs/0709.3825}{{\ttfamily 0709.3825}}].

\bibitem{Hinterbichler:2010es}
K.~Hinterbichler and J.~Khoury, \emph{{Symmetron Fields: Screening Long-Range
  Forces Through Local Symmetry Restoration}},
  \href{https://doi.org/10.1103/PhysRevLett.104.231301}{\emph{Phys. Rev. Lett.}
  {\bfseries 104} (2010) 231301}
  [\href{https://arxiv.org/abs/1001.4525}{{\ttfamily 1001.4525}}].

\bibitem{Hinterbichler:2011ca}
K.~Hinterbichler, J.~Khoury, A.~Levy and A.~Matas, \emph{{Symmetron
  cosmology}}, \href{https://doi.org/10.1103/PhysRevD.84.103521}{\emph{Phys.
  Rev. D} {\bfseries 84} (2011) 103521}
  [\href{https://arxiv.org/abs/1107.2112}{{\ttfamily 1107.2112}}].

\bibitem{Wei:2004rw}
H.~Wei and R.-G.~Cai, \emph{{$K$-chameleon and the coincidence problem}},
  \href{https://doi.org/10.1103/PhysRevD.71.043504}{\emph{Phys. Rev. D}
  {\bfseries 71} (2005) 043504}
  [\href{https://arxiv.org/abs/hep-th/0412045}{{\ttfamily hep-th/0412045}}].

\bibitem{Devoto:2022qen}
F.~Devoto, S.~Devoto, L.~Di~Luzio and G.~Ridolfi, \emph{{False vacuum decay: an
  introductory review}},
  \href{https://doi.org/10.1088/1361-6471/ac7f24}{\emph{J. Phys. G} {\bfseries
  49} (2022) 103001} [\href{https://arxiv.org/abs/2205.03140}{{\ttfamily
  2205.03140}}].

\bibitem{Ghoshal:2017egr}
A.~Ghoshal, A.~Mazumdar, N.~Okada and D.~Villalba, \emph{{Stability of infinite
  derivative Abelian Higgs models}},
  \href{https://doi.org/10.1103/PhysRevD.97.076011}{\emph{Phys. Rev. D}
  {\bfseries 97} (2018) 076011}
  [\href{https://arxiv.org/abs/1709.09222}{{\ttfamily 1709.09222}}].

\bibitem{Ghoshal:2022mnj}
A.~Ghoshal and F.~Nortier, \emph{{Fate of the false vacuum in string-inspired
  nonlocal field theory}},
  \href{https://doi.org/10.1088/1475-7516/2022/08/047}{\emph{JCAP} {\bfseries
  08} (2022) 047} [\href{https://arxiv.org/abs/2203.04438}{{\ttfamily
  2203.04438}}].

\bibitem{Barreira:2014gwa}
A.~Barreira, P.~Brax, S.~Clesse, B.~Li and P.~Valageas, \emph{{Linear
  perturbations in K-mouflage cosmologies with massive neutrinos}},
  \href{https://doi.org/10.1103/PhysRevD.91.063528}{\emph{Phys. Rev. D}
  {\bfseries 91} (2015) 063528}
  [\href{https://arxiv.org/abs/1411.5965}{{\ttfamily 1411.5965}}].

\bibitem{Burrage:2014uwa}
C.~Burrage and J.~Khoury, \emph{{Screening of scalar fields in
  Dirac-Born-Infeld theory}},
  \href{https://doi.org/10.1103/PhysRevD.90.024001}{\emph{Phys. Rev. D}
  {\bfseries 90} (2014) 024001}
  [\href{https://arxiv.org/abs/1403.6120}{{\ttfamily 1403.6120}}].

\bibitem{Herraez:2025clp}
A.~Herr{\'a}ez, D.~L{\"u}st, J.~Masias and C.~Montella, \emph{{A short overview
  on the Black Hole-Tower Correspondence and Species Thermodynamics}},
  \href{https://doi.org/10.22323/1.490.0161}{\emph{PoS} {\bfseries CORFU2024}
  (2025) 161} [\href{https://arxiv.org/abs/2506.02335}{{\ttfamily
  2506.02335}}].

\end{thebibliography}
